\documentclass[a4paper,twoside,twocolumn,english,pra,aps]{revtex4-2}
\usepackage[T1]{fontenc}
\usepackage[utf8]{inputenc}
\usepackage{amsmath}
\usepackage{amssymb}
\usepackage{graphicx}
\makeatletter

\pdfpageheight\paperheight
\pdfpagewidth\paperwidth

\newcommand*\LyXZeroWidthSpace{\hspace{0pt}}
\newcommand{\lyxmathsym}[1]{\ifmmode\begingroup\def\b@ld{bold}
  \text{\ifx\math@version\b@ld\bfseries\fi#1}\endgroup\else#1\fi}

\usepackage[colorlinks = true,
            linkcolor =red,
            urlcolor  =magenta,
            citecolor = blue,
            anchorcolor = blue]{hyperref}

\usepackage{times}      

\makeatother

\usepackage{babel}
\begin{document}
\title{Quantum-Classical Boundary Engineering in Weak-to-Strong Measurements
via Squeezed Vacua}
\author{Janarbek Yuanbek$^{1,2}$ ,Wen-Long Ma$^{1,2}$}
\email{wenlongma@semi.ac.cn}

\author{Yusuf Turek$^{3}$}
\email{yusuftu1984@hotmail.com}

\affiliation{$^{1}$State Key Laboratory of Semiconductor Physics and Chip Technologies,
Institute of Semiconductors, Chinese Academy of Sciences, Beijing
100083, China}
\affiliation{$^{2}$Center of Materials Science and Opto-Electronic Technology,
University of Chinese Academy of Sciences, Beijing 100049, China}
\address{$^{3}$School of Physics, Liaoning University, Shenyang, Liaoning
110036, China}
\date{\today}
\begin{abstract}
This study establishes a post-selected von Neumann framework to regulate
non-classical features of single-photon-subtracted squeezed vacuum
(SPSSV) and two-mode squeezed vacuum (TMSV) states during weak-to-strong
measurement transitions. By synergizing Wigner-Yanase skew information,
Amplitude Squared (AS) squeezing, sum squeezing, and photon statistics,
we demonstrate weak value amplification as a unified control mechanism
for quantum properties. Phase-space analysis via the Husimi Kano $\mathrm{Q}$
function reveals a critical transition: as coupling strength increases,
SPSSV and TMSV states evolve from quantum non-Gaussianity to classical
single-peak separability, marking a quantum-classical boundary crossing.
This critical point is validated as the optimal threshold for noise
suppression and signal enhancement in quantum metrology. The work
provides a tunable platform for quantum sensing and weak-signal detection
technologies.
\end{abstract}
\maketitle

\section{INTRODUCTION\label{sec:1 }}

Quantum measurement fundamentally shapes our understanding of the
microscopic world. Unlike classical measurement, which merely disturbs
a system, quantum measurement can reconstruct system states but irreversibly
collapses quantum superpositions, this destruction of coherence underpins
the central challenge for quantum technologies\citep{briegel2009measurement,taylor2013biological,RevModPhys.75.715}.
Overcoming decoherence and achieving precise quantum state control
under experimental constraints are therefore critical for harnessing
quantum advantages. In order to address this inherent limitation,
Aharonov, Albert, and Vaidman introduced the concept of weak measurement
in 1980\citep{PhysRevLett.60.1351}. Weak measurement is a refined
measurement technique that has been developed to exert minimal impact
on the quantum system during the measurement process. The calibration
of the sensitivity of the measuring apparatus is pivotal in this regard,
as it enables the preservation of quantum coherence, thereby facilitating
the acquisition of information about the quantum system without inducing
significant changes to its quantum state\citep{2014Quantum}. Consequently,
the outcomes of weak measurements reflect the state of the system
in a manner that is less intrusive compared to strong measurements\citep{PhysRevA.92.012120,PhysRevLett.127.180401,PhysRevLett.104.080503,PhysRevA.86.040102}.
Weak measurements address key limitations of projective quantum measurements\citep{PhysRevA.91.062107,PhysRevLett.111.023604}.
Utilizing weak system-probe coupling, they enable non-destructive
observation, allowing continuous monitoring while preserving quantum
states and mitigating uncertainty constraints\citep{kim2012protecting}.
A key innovation is post-selection, which dramatically enhances measurement
efficiency by selectively amplifying the useful signal relative to
noise. Crucially, leveraging quantum superposition and erasure principles
enables significant signal amplification. This boosts single qubit
measurement precision by orders of magnitude\citep{PhysRevA.90.012108,PhysRevLett.66.1107},
translating to vastly enhanced experimental sensitivity. Such amplification
allows observation of subtle effects (e.g., electron spin deflection)
in single measurements, overcoming the need for extensive averaging.
These capabilities hold profound implications for advancing quantum
metrology\citep{PhysRevLett.104.103602,Ouyang:16}, single-photon
detection\citep{RN1927,Zhang2014}, quantum computing\citep{RN1928,Chuang2010},
teleportation\citep{PhysRevA.54.2614,PhysRevA.54.2629,PhysRevA.60.937,PhysRevA.67.033802,PhysRevA.67.042314,PhysRevA.68.052308,PhysRevD.23.1693},
atom-light entanglement manipulation\citep{Hacker2018DeterministicCO},
precision measurement, and quantum error correction\citep{Meng:12,PhysRevX.7.031012}. 

The quantum weak-to-strong measurement transition is fundamentally
governed by the system-pointer coupling strength\citep{PhysRevA.100.062111,pan2020weak,PhysRevLett.114.210801}.
Weak coupling preserves system coherence, yielding complex weak values
accessible only through statistical averaging over many trials combined
with pre- and post-selection. Strong coupling induces instantaneous
wavefunction collapse to an eigenstate, producing a deterministic
pointer shift reflecting the corresponding eigenvalue in a single
measurement. Understanding this transition mechanism provides insights
for optimizing quantum metrology (balancing noise suppression and
signal amplification) and elucidates the quantum-classical boundary
under controlled parameter variation\citep{PhysRevLett.114.210801,Jebli_2020,TUREK2023128663}.

The actualization of such protocols, nevertheless, is contingent upon
the synthesis, scrutiny, and refinement of pertinent quantum states,
such as the coherent state \citep{PhysRev.131.2766,PhysRev.140.B676,PhysRevD.4.2309},
squeezing state \citep{Carranza:12,Andersen201530YO,RN1932}, photon
number states \citep{PhysRevLett.56.58,PhysRevA.36.4547,PhysRevA.39.3414,RN1930,Liu_2004,Waks2006,RN1931},
even and odd coherent states \citep{PhysRevLett.75.4011,PhysRevA.13.2226}.
Existing quantum states are increasingly inadequate for practical
quantum information applications. This has motivated intensive research
into generating novel quantum states and exploring their properties\citep{doi:10.1126/science.1146204,2011JMOp...58..890A}.

The increasing reliance on squeezed states has propelled the study
of squeezing operators and squeezed states to the forefront of research
in quantum optics and quantum information\citep{Andersen_2016}. The
single-photon-subtracted squeezed vacuum (SPSSV) state\citep{Agarwal2013,PhysRevA.75.032104,PhysRevA.43.492}
and the two-mode squeezed vacuum (TMSV) state \citep{Riabinin_2021,Daoming2015QuantumPO}
is a notable quantum technology advancement, offering distinct advantages
such as heightened sensitivity\citep{PhysRevA.107.052614,doi:10.1126/sciadv.adl1814},
diminished noise\citep{PhysRevResearch.7.013130,xu2020conditional},
and pronounced correlations\citep{liu2017entanglement}. This is achieved
by integrating the deterministic nature of the single photon with
the noise suppression capabilities inherent in the squeezed state.
The applications of this technology extend to a number of frontier
fields, ranging from basic science to engineering technology. It is
a significant instrument that fosters the advancement of quantum precision
measurement\citep{PhysRevResearch.6.033292}. In the future, with
the advancement of squeezed light source technology, its application
scenarios are poised to be further expanded to deep space exploration\citep{gao2024generation,Oelker:14}
and other fields\citep{PhysRevLett.80.869,RevModPhys.58.1001,1056132,PhysRevA.72.053806,PhysRevA.76.011804,PhysRevA.61.042302,PhysRevA.72.053812,PhysRevA.60.910}.

In order to address these issues, the present study elucidates how
post-selected von Neumann measurements affect the quantum properties
of SPSSV and TMSV states. The text provides a comprehensive analysis
of the discrepancy between weak and strong measurement regimes, systematically
examining their impact on single-mode radiated fields. It places particular
emphasis on the role of post-selection and weak value characterisation
in this context\citep{yuanbek2024single,PhysRevA.75.032104,PhysRevA.43.492}. 

The present paper establishes the precision metrology advantages of
the SPSSV and TMSVS states, and elucidates the weak-to-strong measurement
transition mechanism. The SPSSV and TMSV polarization degrees are
utilised as the measured system, with a quantitative analysis of the
radiation field's Wigner-Yanase skew information, AS squeezing, photon
statistics, sum squeezing, measurement transition and Husimi-Kano
$\mathrm{Q}$ function being conducted. A systematic comparison of
the results with the initial state is also performed, and a comprehensive
assessment of the impact of post-selected von Neumann measurements
is conducted. In order to corroborate the measurement transition,
an offset metric for observable measurements is hereby introduced.
Precise modulation of the dimensionless coupling parameter $\mathrm{s}$
has been shown to consistently control this transition, characterised
by pointer position and momentum offsets. This process yields the
final, normalised post-measurement SPSSV and TMSV states. The findings
of this study provide a novel framework for theoretical exploration,
and the realisation of this transition using SPSSV and TMSV pointer
states establishes a foundation for its application in quantum information
processing and metrology.

The structure of this paper is outlined as follows. In Sec.\ref{sec:2 },
we describe constructs a theoretical model based on the von Neumann
post-selection measurement framework, obtains terminal pointer states
through the post-selection protocol, analyzes the weak-to-strong measurement
conversion mechanism, and establishes the theoretical foundation for
subsequent analysis. In Sec. \ref{sec:3}, we employ the Wigner-Yanase
skew information, AS squeezing, sum squeezing, and photon statistics
as metrics to validate their superior performance in precision measurement
protocols. In Sec. \ref{sec:4}, we proposed universal expressions
for pointer position and momentum displacement in the SPSSV and TMSV
states, achieved weak-to-strong measurement transitions through coupling
strength modulation, and validated theoretical results by combining
numerical simulation comparison with Husimi-Kano $\mathrm{Q}$ function
analysis. In Sec. \ref{sec:5}, We systematically summarize the key
findings of this study and provide an outlook on potential future
research directions. All quantities are expressed in units where $\mathrm{\hbar=1}$,
unless explicitly stated otherwise.

\section{Fundamental principles \label{sec:2 }}

The total Hamiltonian in measurement theory is typically decomposed
into three fundamental components, each governing distinct physical
processes during the measurement interaction. These contributions
are formally expressed as
\begin{equation}
\mathrm{\hat{H}=\hat{H}_{s}+\hat{H}_{p}+\hat{H}_{int}}.
\end{equation}

Here, $\mathrm{\hat{H}_{s}}$ denotes the Hamiltonian of the measured
system, $\mathrm{\hat{H}_{p}}$represents the Hamiltonian of the measuring
apparatus (pointer), and $\mathrm{\hat{H}_{int}}$ characterizes the
interaction between the system and the apparatus.

In the framework of ideal measurement theory\citep{vonNeumann+2018},
the specific forms of the Hamiltonian describing the pointer and the
system under measurement do not influence the measurement outcomes.
The interaction Hamiltonian, which encodes the essential information
about the pointer and the measured system, serves as the foundation
for our analysis. In this study, we adopt the interaction Hamiltonian
as

\begin{equation}
\mathrm{\hat{H}_{int}=g(t)\hat{A}\otimes\hat{P}},\label{eq:introduction hamli}
\end{equation}
Here, $\mathrm{\hat{A}}$ is the observable to be measured and $\mathrm{\hat{P}}$
denotes the momentum operator of the pointer, which is the conjugate
variable of the position operator $\mathrm{\hat{X}}$, satisfying
the canonical commutation relation $\mathrm{[\hat{X},\hat{P}]=i}$,
The momentum operator $\mathrm{\hat{P}}$ and the position operator
$\mathrm{\hat{X}}$ can be expressed in terms of the annihilation
operator ($\mathrm{\hat{a}}$) and creation operator ($\mathrm{\hat{a}^{\dagger}}$)
as\citep{PhysRevA.41.1526}

\begin{align}
\mathrm{\hat{P}} & \mathrm{=\frac{i}{2\sigma}(\hat{a}^{\dagger}-\hat{a}),}\\
\mathrm{\hat{X}} & =\mathrm{\sigma(\hat{a}^{\dagger}+\hat{a})},
\end{align}
where $\mathrm{\sigma=\sqrt{1/2m\omega}}$ represents the width of
the Gaussian ground state of the pointer, which depends on its mass
$\mathrm{m}$ and the oscillation frequency $\mathrm{\omega}$. The
parameter $\mathrm{g(t)}$ quantifies the coupling strength between
the measured system and the pointer, and $\mathrm{g(t)}$ is non-vanishing
over a finite interval, with its time-integrated value representing

\begin{equation}
\mathrm{\int_{t_{0}}^{t}g(\tau)d\tau=g\delta(t-t_{0})}.
\end{equation}

\begin{figure}
\begin{centering}
\includegraphics[scale=0.04]{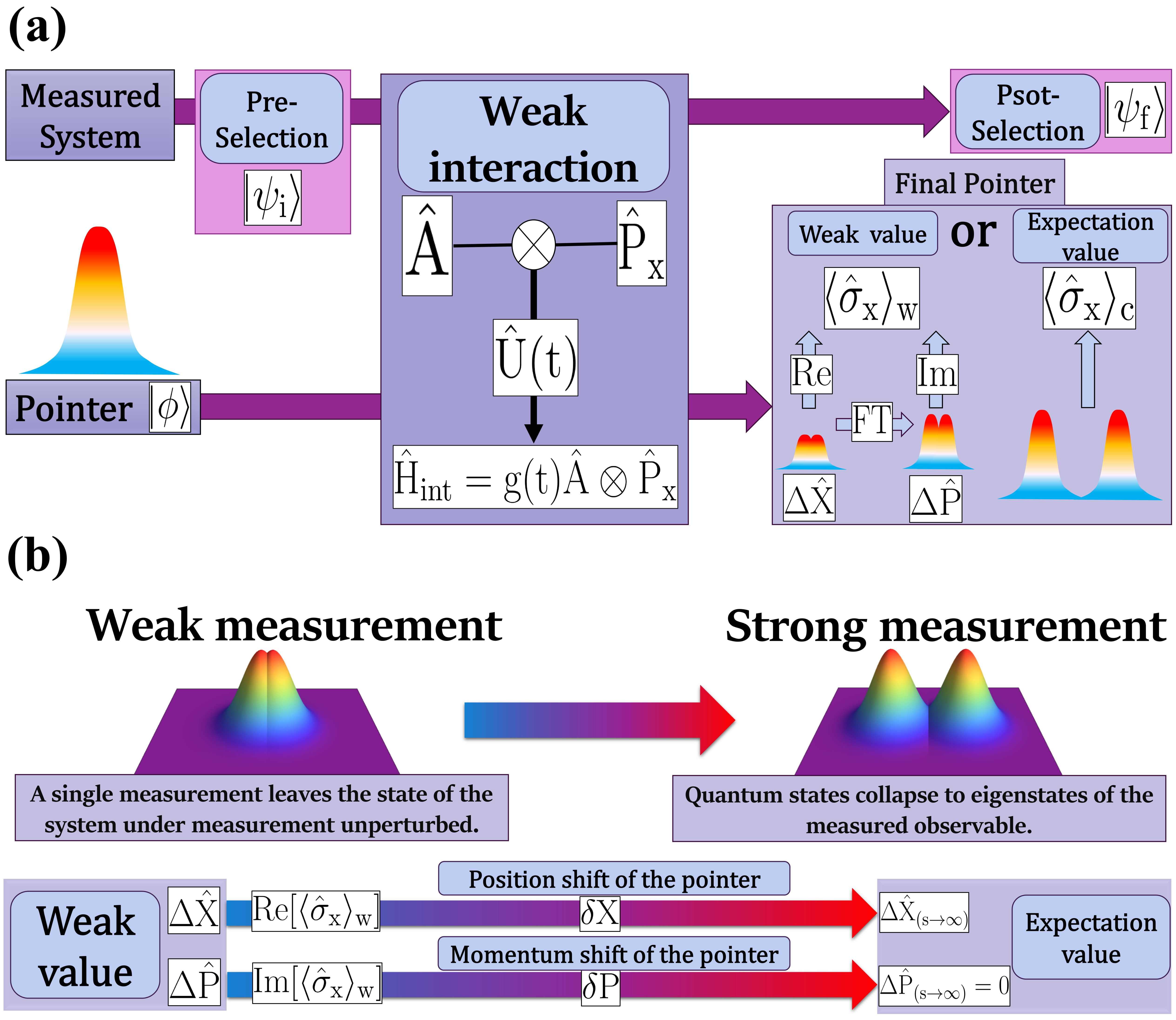}
\par\end{centering}
\caption{Conceptual framework of the measurement transition model. (a) Schematic
representation of the post-selected Von Neumann measurement model
and the standard protocol of weak measurement theory involves four
steps: $\mathrm{(i})$ Prepare the system in the initial state $\mathrm{\vert\psi_{i}\rangle}$
and the measuring device in $\mathrm{\vert\phi\rangle}$. $\mathrm{(ii})$
Induce weak interaction to drive joint evolution of the system and
device. $\mathrm{(iii})$ Perform postselection projection of the
system onto a specific final state $\mathrm{\vert\psi_{f}\rangle}$.
$\mathrm{(iv})$ Extract the weak value of the system's observable
through shifts in the position space (real part $\mathrm{\operatorname{Re}}$)
and the Fourier-transformed momentum space (imaginary part $\mathrm{\operatorname{Im}}$)
of the measuring device. (b) Schematic representation of the relationship
between pointer-induced the weak-to-strong measurement transition
model. \label{fig:weak-to-strong}}
\end{figure}

To achieve this, we designate the polarization and spatial degrees
of freedom of the SPSSV and TMSV states as the measured system and
the pointer, respectively. In Fig. \ref{fig:weak-to-strong}, we assume
that the initial state of the entire system is set to 
\begin{equation}
\mathrm{\vert\Psi_{in}\rangle=\vert\psi_{i}\rangle\otimes\vert\phi\rangle,}\label{eq:inital state}
\end{equation}
Here, the initial state 
\begin{equation}
\mathrm{\vert\psi_{i}\rangle=\cos\frac{\alpha}{2}\vert H\rangle+e^{i\delta}\sin\frac{\alpha}{2}\vert V\rangle}
\end{equation}
can be prepared in the optical lab by using quarter and half wave
plates, with $\mathrm{\delta\in[0,2\pi]}$ and $\mathrm{\alpha\in[0,\pi)}$.
The joint state in Eq. (\ref{eq:inital state}) is transformed by
the system's time-evolution operator, defined as
\begin{equation}
\mathrm{\hat{U}(t)=\exp\left[-i\int_{0}^{t}\hat{H}_{int}d\tau\right],}\label{eq:time-eve}
\end{equation}
the initial state evolves to

\begin{align}
\mathrm{\vert\Psi\rangle} & =\mathrm{\exp\left[-i\int_{0}^{t}\hat{H}_{int}d\tau\right]\vert\psi_{i}\rangle\otimes\vert\phi\rangle}\nonumber \\
 & =\mathrm{e^{-igt\hat{A}\otimes\hat{P}}\vert\psi_{i}\rangle\otimes\vert\phi\rangle}.\label{eq:time-eve-post}
\end{align}

A core postulate of quantum theory posits that measurement inherently
induces an irreversible disturbance in a quantum system, we now investigate
the solutions to Eq. (\ref{eq:time-eve-post}) under diverse conditions.
If $\mathrm{\vert\alpha_{i}\rangle}$ is an eigenstate of observable
$\mathrm{\hat{A}}$with eigenvalue $\mathrm{\alpha_{i}}$($\mathrm{\hat{A}\vert\alpha_{i}\rangle=a_{i}\vert\alpha_{i}\rangle}$),
then $\mathrm{\hat{A}}$can be expressed as

\begin{equation}
\mathrm{\hat{A}=\sum_{i}a_{i}\vert\alpha_{i}\rangle\langle\alpha_{i}\vert},
\end{equation}
observable $\mathrm{\hat{A}}$ has three distinct values: eigenvalues,
expectation values, and weak values. These are accessible via pointer
shifts in measurements. Since eigenvalues are special cases of the
other two values, we later sections will detail the readout procedures
for (conditional) expectation values and weak values in relevant measurements.

1.Expectation Value, let the pointer's initial state be $\mathrm{\vert\phi\rangle}$,
with wave function $\mathrm{\vert\phi(x)\rangle}$, the measured system
is prepared in superposition state $\mathrm{|\psi_{i}\rangle=\sum_{i}\alpha_{i}|a_{i}\rangle}$$\mathrm{(\sum_{i}|\alpha_{i}|^{2}=1)}$,
of observable $\mathrm{\hat{A}}$, Under Eq. (\ref{eq:time-eve}),
the total system (subnormalized) evolves into the state:

\begin{align}
\mathrm{\left|\Psi\right\rangle } & =\mathrm{e^{-igt\hat{A}\otimes\hat{P}}|\psi_{i}\rangle\otimes|\phi(x)\rangle}\nonumber \\
 & =\mathrm{\sum_{i}a_{i}|\alpha_{i}\rangle\otimes|\phi(x-ga_{i})\rangle},
\end{align}
strong measurement shifts the pointer wave function to $\mathrm{|\phi(x-ga_{i})\rangle}$,
displacing its center by $\mathrm{ga_{i}}$, the pointer displacement
gives

\begin{equation}
\mathrm{\delta X_{(s\to\infty)}=\frac{\langle\Psi|\hat{X}|\Psi\rangle}{\langle\Psi|\Psi\rangle}-\langle\phi|\hat{X}|\phi\rangle=g\langle\hat{A}\rangle},
\end{equation}
where $\mathrm{\langle A\rangle}$is the expectation value of $\mathrm{\hat{A}}$in
state $\mathrm{\vert\psi_{i}\rangle}$, written as

\begin{equation}
\mathrm{\langle\hat{A}\rangle=\langle\psi_{i}|\hat{A}|\psi_{i}\rangle=\sum_{i}a_{i}|\alpha_{i}|^{2}},\label{eq:expectation}
\end{equation}
this formula represents the expected value obtained from performing
a strong measurement on observable $\mathrm{\hat{A}}$ in a quantum
system prepared in the initial state $\mathrm{|\psi_{i}\rangle}$.
As introduced in the introduction, Aharonov and his collaborators
proposed weak value , which establishes another fundamental framework
for quantum measurement theory. Within the scope of this theoretical
framework, the expected value is extended to a more universally applicable
weak value expression, the definition of which is provided below.

2.Weak value. Contrary to the previously analyzed scenario, in the
context of weak system-pointer coupling, a solitary measurement is
inadequate in yielding meaningful information. It is evident that
a first-order approximation of the unitary operator is employed, therefore,
the first line of Eq. (\ref{eq:time-eve-post}) takes the following
form

\begin{equation}
\mathrm{|\Psi\rangle\approx(1-ig\hat{A}\otimes\hat{P})\vert\psi_{i}\rangle\otimes|\phi(x)\rangle},
\end{equation}
as shown in Fig. \ref{fig:weak-to-strong}, after performing post-selection
on the state$\mathrm{|\psi_{f}\rangle}$, the (subnormalized) system
state becomes

\begin{align}
\mathrm{\vert\Psi^{\prime}\rangle} & \approx\mathrm{\langle\psi_{i}\vert(1-ig\hat{A}\otimes\hat{P})|\psi_{i}\rangle\otimes|\phi(x)\rangle}\nonumber \\
 & =\mathrm{\langle\psi_{f}|\psi_{i}\rangle(1-ig\langle\hat{A}\rangle_{w}\hat{P})|\phi(x)\rangle}\nonumber \\
 & \approx\mathrm{\langle\psi_{f}|\psi_{i}\rangle e^{-ig\langle\hat{A}\rangle_{w}\hat{P}}|\phi(x)\rangle}\nonumber \\
 & \approx\mathrm{\langle\psi_{f}|\psi_{i}\rangle\otimes\vert\phi(x-gt\operatorname{Re}[\langle\hat{A}\rangle_{w}])\rangle},
\end{align}
where $\mathrm{\langle\hat{A}\rangle_{w}}$ represents the weak value,
defined as

\begin{align}
\mathrm{\langle\hat{A}\rangle_{w}} & \mathrm{=\frac{\langle\psi_{i}|\hat{A}|\psi_{f}\rangle}{\langle\psi_{i}|\psi_{f}\rangle}=\operatorname{Re}[\langle\hat{A}\rangle_{w}]+i\operatorname{Im}[\langle\hat{A}\rangle_{w}]}.\label{eq:weak-valu}
\end{align}

In the present Eq. (\ref{eq:weak-valu}), the weak value is to be
decomposed into real and imaginary components, denoted here as $\mathrm{\operatorname{Re}[\langle\hat{A}\rangle_{w}]}$
and $\mathrm{\operatorname{Im}[\langle\hat{A}\rangle_{w}]}$ respectively\citep{PhysRevA.76.044103},
and $\mathrm{\langle\psi_{f}|\psi_{i}\rangle=\delta_{f,i}}$ is the
Kronecker delta defined as

\begin{equation}
\mathrm{\delta_{f,i}=\begin{cases}
1, & f=i\\
0, & f\neq i
\end{cases}}
\end{equation}

satisfaction of $\mathrm{f=i}$ causes Eq. (\ref{eq:weak-valu}) to
collapse to the standard expectation value in Eq. (\ref{eq:expectation}),
with both observables manifesting in the post-measurement pointer
displacement. By decomposing Eq. (\ref{eq:weak-valu}) into its real
and imaginary components of the weak value, the pointers position
and momentum displacements after measurement satisfy

\begin{align}
\mathrm{\delta X} & \propto\mathrm{g\operatorname{Re}[\langle\hat{A}\rangle_{w}]},\label{eq:weak-approach}\\
\mathrm{\delta P} & =\mathrm{2g\operatorname{Im}[\langle\hat{A}\rangle_{w}]Var(P)},\label{eq:strong-approach}
\end{align}
where $\mathrm{Var(P)=\langle\phi|P^{2}|\phi\rangle-\langle\phi|P|\phi\rangle^{2}}$
denotes the variance of the momentum operator $\mathrm{P}$ in the
initial pointer state $\mathrm{|\phi\rangle}$. Then, we develop Eq.
(\ref{eq:time-eve-post}) further through expansion, for the total
system governed by the interaction Hamiltonian {[}see Eq.(\ref{eq:introduction hamli}){]}
, the time evolution is

\begin{align}
\mathrm{\vert\Psi\rangle} & \mathrm{=e^{-igt\hat{A}\otimes\hat{P}}\vert\psi_{i}\rangle\otimes\vert\phi\rangle}\nonumber \\
 & =\mathrm{\sum_{n}\frac{1}{n!}(\hat{\sigma}_{x})^{n}\left[\frac{gt(\hat{a}^{\dagger}-\hat{a})}{2\sigma}\right]^{n}\vert\psi_{i}\rangle\otimes\vert\phi\rangle}\nonumber \\
 & =\mathrm{\frac{1}{2}\left[r_{+}\hat{D}\left(\frac{s}{2}\right)+r_{-}\hat{D}^{\dagger}\left(\frac{s}{2}\right)\right]\mathrm{\vert\psi_{i}\rangle\otimes\vert\phi\rangle}},\label{eq:3-1}
\end{align}
where $\mathrm{r_{\pm}=\mathbb{I}\pm\hat{\sigma}_{x}}$ and the coupling
strength parameter $\mathrm{s=gt/\sigma}$ is a dimensionless, continuous
variable employed to characterize the measurement strength. When $\mathrm{0<s\ll1}$
(or $\mathrm{s\gg1}$), the measurement is classified as weak (strong),
respectively. Experimental control over the parameter $\mathrm{s}$
can be achieved through modulation of three factors: the coupling
coefficient $\mathrm{g}$, the interaction time $\mathrm{t}$, and
the spatial distribution parameter $\mathrm{\sigma}$. Among these,
experimental studies \citep{pan2020weak} demonstrate that adjusting
$\mathrm{t}$ offers the most direct and efficient means to manipulate
$\mathrm{s}$. For the subsequent analysis, we adopt the working assumption
that variations in s arise solely from changes in $\mathrm{t}$, with
$\mathrm{g}$ and $\mathrm{\sigma}$ maintained at constant values.
$\mathbb{\mathrm{\mathbb{I}}}$ is $\mathrm{2\times2}$ unit matrix
operator and $\mathrm{\hat{D}(s/2)=e^{s(\hat{a}^{\dagger}-\hat{a})/2}}$
is the displacement operator, satisfying the following transformation
relations

\begin{align}
\mathrm{\hat{D}^{\dagger}(\alpha)\hat{a}\hat{D}(\alpha)} & =\mathrm{\hat{a}+\alpha},\\
\mathrm{\hat{D}(\alpha)\hat{a}\hat{D}^{\dagger}(\alpha)} & =\mathrm{\hat{a}-\alpha}.
\end{align}

The diagonal and anti-aligned polarization states can be expanded
in the horizontal ($\mathrm{\vert H\rangle}$) and vertical ($\mathrm{\vert V\rangle}$)
polarization bases of the optical beam, and are given by
\begin{align}
\mathrm{\vert D\rangle} & \mathrm{=\frac{1}{\sqrt{2}}\left(\vert H\rangle+\vert V\rangle\right)},\\
\mathrm{\vert A\rangle} & \mathrm{=\frac{1}{\sqrt{2}}\left(\vert H\rangle-\vert V\rangle\right)},
\end{align}
using the above expression, we choose the Pauli operators as the observables

\begin{equation}
\mathrm{\hat{A}=\hat{\sigma}_{x}=\vert D\rangle\langle D\vert-\vert A\rangle\langle A\vert}.
\end{equation}

For the implementation of post-selected von Neumann measurement the
post-selected state $\mathrm{\vert\psi_{f}\rangle=\vert H\rangle}$
is taken over on the state $\mathrm{\vert\Psi\rangle}$ given in Eq.
(\ref{eq:3-1}), then the pointer state reads $\mathrm{\vert\tilde{\Phi}\rangle=\langle\psi_{f}\vert\Psi\rangle}$.
The state $\mathrm{\vert\tilde{\Phi}\rangle}$ is not normalized,
the above state $\mathrm{\vert\tilde{\Phi}\rangle}$ becomes as 
\begin{align}
\mathrm{\vert\tilde{\Phi}\rangle} & =\mathrm{\langle\psi_{f}\vert\Psi\rangle}\nonumber \\
 & =\mathrm{\frac{\langle\psi_{f}\vert\psi_{i}\rangle}{2}\mathrm{\bigg[t_{+}\hat{D}\left(\frac{s}{2}\right)}+t_{-}\hat{D}^{\dagger}\left(\frac{s}{2}\right)\bigg]\otimes\vert\phi\rangle},\label{eq:not-normalization}
\end{align}

where $\mathrm{t_{\pm}=\mathbb{I}\pm\langle\hat{\sigma}_{x}\rangle_{w}}$,
by imposing the normalisation condition $\mathrm{\langle\psi_{i}\vert\psi_{i}\rangle=\langle\psi_{f}\vert\psi_{f}\rangle=1}$,
the overlap $\mathrm{\langle\psi_{f}\vert\psi_{i}\rangle=\cos(\alpha/2)}$
naturally satisfies the constraints on the weak value parameter $\mathrm{\alpha}$.
The primary objective of this study is to examine the influence of
post-selected measurement on the intrinsic properties of the SPSSV
and TMSV states. This process will be investigated first by analysing
the different pointer states.

\subsection{Pointer States in Single Mode}

The choice of the pointer state ($\mathrm{\vert\phi\rangle}$) as
a SPSSV state ($\mathrm{\vert\phi_{1}\rangle}$) for a single mode
radiation field can be formally represented by the following expression

\begin{equation}
\mathrm{\vert\phi_{1}\rangle}\mathrm{=\frac{\hat{a}}{\sinh(r)}\hat{S}(\xi)\vert0\rangle=\sum_{n=0}^{\infty}C_{n}|2n+1\rangle},
\end{equation}
where 
\begin{equation}
\mathrm{\hat{S}(\xi)=\exp\left[\frac{1}{2}\left(\xi\hat{a}^{\dagger2}-\xi^{\ast}\hat{a}^{2}\right)\right]},
\end{equation}
is the squeezing operator and $\mathrm{\xi}$ is a complex number,
and
\begin{equation}
\mathrm{C_{n}=\frac{\mathrm{e}^{\mathrm{i}(n+1)\theta}(\tanh r)^{n}\sqrt{(2n+1)!}}{(\cosh r)^{3/2}n!2^{n}}},
\end{equation}
we can defined in the punctured complex plane as $\mathrm{\xi=re^{i\theta}}$,
with parameters satisfying $\mathrm{0<r<\infty}$ and $\mathrm{0<\theta<2\pi}$.
In the Fock basis $\mathrm{\hat{S}(\xi)|0\rangle=|\xi\rangle}$ has
the representation\citep{Int}

\begin{equation}
\mathrm{|\xi\rangle=\sum_{m=0}^{\infty}\frac{(-1)^{m}\sqrt{(2m)!}}{2^{m}m!\sqrt{\cosh(r)}}e^{im\theta}\tanh^{m}r|2m\rangle,}
\end{equation}

The following typical commutation relations were employed in the subsequent
derivation process

\begin{align}
\mathrm{\hat{S}^{\dagger}(\xi)\hat{a}(\xi)\hat{S}(\xi)} & \mathrm{=\hat{a}\cosh r+\hat{a}^{\dagger}\mathrm{e}^{i\theta}\sinh r,}\\
\mathrm{\hat{S}^{\dagger}(\xi)\hat{a}^{\dagger}(\xi)\hat{S}(\xi)} & \mathrm{=\hat{a}^{\dagger}\cosh r+\hat{a}\mathrm{e}^{-i\theta}\sinh r,}
\end{align}
we proceed to impose normalization on the derived equation

\begin{align}
\mathrm{\mathrm{\vert\Phi\rangle_{S}}} & \mathrm{=\frac{\vert\tilde{\Phi}\rangle_{S}}{\sqrt{P_{S}}}}\nonumber \\
 & =\mathrm{\lambda\bigg[t_{+}\hat{D}\left(\frac{s}{2}\right)+t_{-}\hat{D}^{\dagger}\left(\frac{s}{2}\right)\bigg]\otimes\vert\phi_{1}\rangle},\label{eq:normalization}
\end{align}
Here, $\mathrm{\lambda=1/\sqrt{P_{S}}}$ and $\mathrm{P_{S}}$ is
also can characterize the probability of successful post-selection
of the final pointer state $\vert\Phi\rangle$ and we can define as

\begin{equation}
\mathrm{P_{s}}=\mathrm{_{S}\langle\tilde{\Phi}|\tilde{\Phi}\rangle_{S}},
\end{equation}
Where the transformation relation is invoked to establish the equivalence\citep{Agarwal2013}

\begin{align}
\mathrm{\hat{S}^{\dagger}(\xi)\hat{D}^{\dagger}(s)\hat{S}(\xi)} & \mathrm{=\hat{D}(\beta)},
\end{align}
with $\mathrm{\beta}=\mathrm{-s\left[\cosh(r)-e^{i\theta}\sinh(r)\right]}$.
From this relation,
\begin{align}
\mathrm{P} & \mathrm{=\langle\phi_{1}\vert\hat{D}\left(\pm s\right)\vert\phi_{1}\rangle\mathrm{=\left(1-\vert\beta\vert^{2}\right)\exp\left[-\frac{1}{2}\vert\beta\vert^{2}\right]},}
\end{align}
Consequently, the normalization coefficient $\mathrm{\lambda}$ is
defined as

\begin{align}
\mathrm{\lambda} & \mathrm{=\frac{1}{\sqrt{2}}\left[1+\vert\langle\hat{\sigma}_{x}\rangle_{w}\vert^{2}+(1-\vert\langle\hat{\sigma}_{x}\rangle_{w}\vert^{2})P\right]^{-\frac{1}{2}}}.
\end{align}

\begin{figure}
\begin{centering}
\includegraphics[scale=0.355]{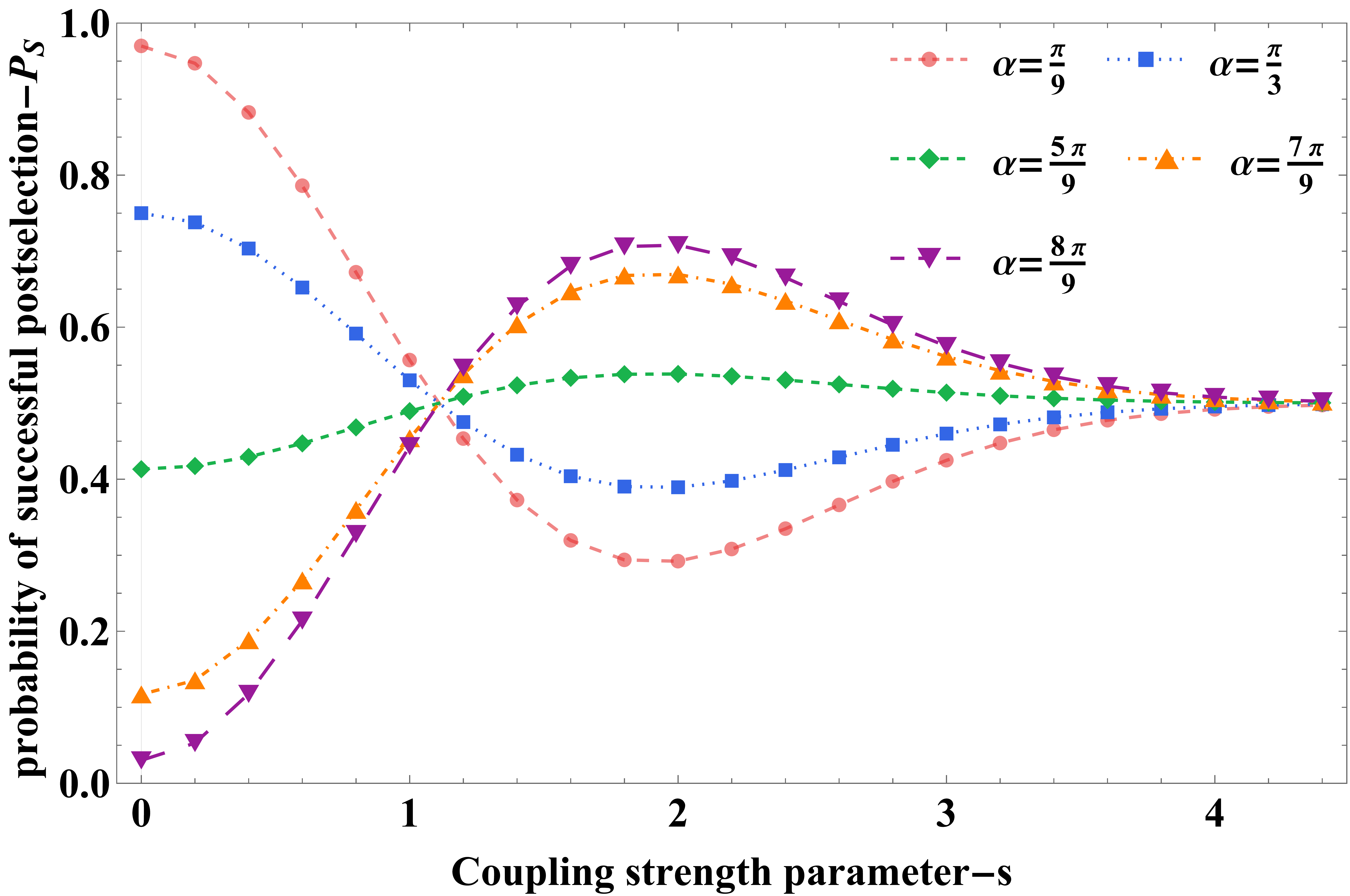}
\par\end{centering}
\caption{The post-selection success probability of the final pointer state
$\mathrm{\vert\Phi\rangle_{s}}$, $\mathrm{P_{S}}$ as a dependent
variable on $\mathrm{s}$ for different $\mathrm{\alpha}$, with $\mathrm{r=0.1}$.
Here $\mathrm{\ensuremath{\theta}=\delta=0}$.\label{fig:Probility}}
\end{figure}

As a result of post-selected von Neumann measurement, the weak value
of the system observable $\mathrm{\hat{\sigma}_{x}}$is given by
\begin{align}
\mathrm{\langle\hat{\sigma}_{x}\rangle_{w}} & \mathrm{=\frac{\langle\psi_{f}\vert\hat{\sigma}_{x}\vert\psi_{i}\rangle}{\langle\psi_{f}\vert\psi_{i}\rangle}=e^{i\delta}\tan\frac{\alpha}{2}},\label{eq:weak values}
\end{align}
the Eq.(\ref{eq:normalization}) is the final state of the pointer
after post-selected von Neumann measurement, which will be used throughout
our work. A concise analysis of the anomalies in weak values (as shown
in Eq. (\ref{eq:weak values})) reveals that when the initial and
final states are nearly orthogonal, weak measurements can exceed the
range of standard deviations $\mathrm{\langle\hat{\sigma}_{x}\rangle_{w}}$
typically observed in classical measurements. Furthermore, when the
parameter $\mathrm{\mathrm{\delta}}$ is non-zero$\mathrm{\left(\mathrm{\delta\neq0}\right)}$,
the behavior of the weak values may become even more complex, exhibiting
non-intuitive or non-classical numerical characteristics. 

As shown in Fig. \ref{fig:Probility}, the final state of the pointer
depends on both the weak value and the coupling strength parameter
$\mathrm{s}$, with its success probability denoted as $\mathrm{P_{s}}$.
The figure reveals two distinct trends in the weak measurement regime:

(a). For weak coupling coefficients ($\mathrm{0<s\leq1}$), the success
probability $\mathrm{P_{s}}$ decreases as the weak value increases.

(b). For stronger couplings ($\mathrm{s>1}$), higher weak values
lead to greater success probabilities ($\mathrm{1.1<s\leq4}$), indicating
that larger weak values are more advantageous in this range. However,
above the critical coupling strength $\mathrm{s>4}$, success probability
$\mathrm{P_{s}}$ converges to a constant value of $\mathrm{0.5}$,
independent of the weak value parameters $\mathrm{\alpha}$ and $\mathrm{\delta}$.

Moreover, if we consider a larger coupling strength parameter $\mathrm{s}$,
even when the weak value is large and would otherwise be associated
with a low success probability, we can still obtain a successful final
pointer state with a probability that is not too low. This behavior
has significant implications for quantum measurement theory, highlighting
the distinctive nature of quantum systems under weak coupling conditions
and their departure from classical physical predictions.As indicated
in the introductory section, the unusually large weak values can serve
not only to enhance minute system details but also to optimize quantum
states. 

In order to elucidate the physical mechanism underlying the evolution
from weak to strong measurements in subsequent chapters, this paper
introduces a transition parameter characterising the values of the
system's observable under weak and strong measurement regimes. This
is based on the theoretical framework of Ref. \citep{PhysRevA.103.052215,turek2023single,10.1063/5.0230512}.
The transition value within the SPSSV state exhibits the following
properties 
\begin{equation}
\mathrm{\sigma_{x}^{S}=\frac{\langle\tilde{\Phi}\vert\Psi^{\prime}\rangle}{\langle\tilde{\Phi}\vert\tilde{\Phi}\rangle}=\frac{\langle\tilde{\Phi}\vert\Psi^{\prime}\rangle}{P_{S}}},\label{eq:T-Value}
\end{equation}
wherein

\begin{align}
\mathrm{\vert\Psi^{\prime}\rangle} & \mathrm{=\langle\psi_{f}\vert\hat{\sigma}_{x}\vert\Psi\rangle}\nonumber \\
 & =\mathrm{\frac{1}{2}\langle\psi_{f}\vert\left[r_{+}\hat{D}\left(\frac{s}{2}\right)-r_{-}\hat{D}^{\dagger}\left(\frac{s}{2}\right)\right]\vert\psi_{i}\rangle\otimes\vert\phi_{1}\rangle}\nonumber \\
 & =\mathrm{\frac{\cos\frac{\alpha}{2}}{2}}\mathrm{\bigg[t_{+}\hat{D}\left(\frac{s}{2}\right)-t_{-}\hat{D}^{\dagger}\left(\frac{s}{2}\right)\bigg]\vert\phi_{1}\rangle.}
\end{align}

Therefore, we derive 

\begin{align}
\mathrm{\langle\tilde{\Phi}\vert\Psi^{\prime}\rangle} & =\mathrm{\cos^{2}\frac{\alpha}{2}\left(\operatorname{Re}\left[\langle\hat{\sigma}_{x}\rangle_{w}\right]+i\operatorname{Im}\left[\langle\hat{\sigma}_{x}\rangle_{w}\right]P\right)}.
\end{align}

Following the substitution of the corresponding functional forms of
$\mathrm{\tilde{\vert\Phi\rangle}}$ and $\mathrm{\vert\Psi^{\prime}\rangle}$
into Eq. (\ref{eq:T-Value}), the $\mathrm{\sigma_{x}^{S}}$ parameter
is able to be determined analytically as such

\begin{align}
\mathrm{\sigma_{x}^{S}} & \mathrm{=\frac{2\left[\operatorname{Re}\left[\langle\hat{\sigma}_{x}\rangle_{w}\right]+i\operatorname{Im}\left[\langle\hat{\sigma}_{x}\rangle_{w}\right]P\right]}{1+\left|\langle\sigma_{x}\rangle_{w}^{\ast}\right|^{2}+\left(1-\left|\langle\sigma_{x}\rangle_{w}^{\ast}\right|^{2}\right)P}.}
\end{align}

Within the framework of quantum measurement theory, Fig. (\ref{fig:weak-to-strong})
demonstrates that, the fundamental concept underlying the transition
from weak to strong measurements manifests through continuous modulation
of interaction strength between the system and measurement apparatus,
as well as the dynamic interplay between quantum state information
extraction and measurement-induced disturbance. This transition is
mathematically characterized by a parameterized measurement operator
formalism, where the coupling strength $\mathrm{s}$ serves as the
critical control parameter.

1. Weak measurement, implemented via weak coupling ($\mathrm{s\rightarrow0}$),
i.e.,

\begin{align}
\mathrm{\left(\sigma_{x}^{S}\right)_{s\rightarrow0}} & \mathrm{=\operatorname{Re}\left[\langle\hat{\sigma}_{x}\rangle_{w}\right]+i\operatorname{Im}\left[\langle\hat{\sigma}_{x}\rangle_{w}\right]}\nonumber \\
 & \mathrm{=\langle\hat{\sigma}_{x}\rangle_{w}}.
\end{align}

the measurement outcome is governed by the weak value $\mathrm{\langle\hat{\sigma}_{x}\rangle_{w}}$.
Under this regime, the pointer states exhibit significant wavepacket
overlap, directly reflecting quantum coherence between pre-selected
($\mathrm{\vert\psi_{i}\rangle}$) and post-selected ($\mathrm{\vert\psi_{f}\rangle}$)
states.

2. Strong measurement, achieved through strong coupling ($\mathrm{s\rightarrow\infty}$),
i.e.,

\begin{align}
\mathrm{\left(\sigma_{x}^{S}\right)_{s\rightarrow\infty}} & \mathrm{=\frac{2\operatorname{Re}\left[\langle\hat{\sigma}_{x}\rangle_{w}\right]}{1+\left|\langle\sigma_{x}\rangle_{w}^{\ast}\right|^{2}}}\nonumber \\
 & \mathrm{=2\frac{\cos\delta\tan\frac{\alpha}{2}}{\sec^{2}\frac{\alpha}{2}}}\nonumber \\
 & \mathrm{=\cos\delta\sin\alpha=\sigma_{x}^{c}}.
\end{align}

In this framework, $\mathrm{\sigma_{x}^{c}}$ corresponds to the conditional
expectation value of the system observable $\mathrm{\hat{\sigma}_{x}}$
under strong measurement protocols. The evaluation of $\mathrm{\sigma_{x}^{c}}$
adheres to the Aharonov--Bergmann--Lebowitz (ABL) rule \citep{PhysRev.134.B1410},
mathematically formulated as

\begin{align}
\mathrm{\sigma_{x}^{c}} & =\mathrm{\sum_{j}a_{j}\frac{\vert\langle\psi_{f}\vert a_{j}\rangle\langle a_{j}\vert\psi_{i}\rangle\vert^{2}}{\sum_{i}\vert\langle\psi_{f}\vert a_{i}\rangle\langle a_{i}\vert\psi_{i}\rangle\vert^{2}}}\nonumber \\
 & =\mathrm{\frac{\vert\langle\psi_{f}\vert D\rangle\langle D\vert\psi_{i}\rangle\vert^{2}-\vert\langle\psi_{f}\vert A\rangle\langle A\vert\psi_{i}\rangle\vert^{2}}{\vert\langle\psi_{f}\vert D\rangle\langle D\vert\psi_{i}\rangle\vert^{2}+\vert\langle\psi_{f}\vert A\rangle\langle A\vert\psi_{i}\rangle\vert^{2}}}\nonumber \\
 & =\mathrm{\cos\delta\sin\alpha}.
\end{align}

this regime induces complete wavepacket separation of pointer states.
The measurement outcome converges to the expectation value $\mathrm{\mathrm{\langle\hat{\sigma}_{x}\rangle_{s}=\langle\psi_{f}\vert\hat{\sigma}_{x}\vert\psi_{f}\rangle}}$
of operator $\mathrm{\hat{\sigma}_{x}}$, accompanied by prominent
decoherence effects. This process ultimately reduces to quantum state
collapse under standard projective measurement protocols.

\subsection{Pointer States in Double Mode }

For a two mode radiation field, let us consider the selection of the
pointer state $\mathrm{\vert\phi\rangle}$ as a TMSV state $\mathrm{\vert\phi_{2}\rangle}$,
which can be formally represented through the following expression
\begin{align}
\mathrm{\vert\phi_{2}\rangle} & =\mathrm{\hat{S}(\chi)\vert0,0\rangle_{a,b}}\nonumber \\
 & =\mathrm{\frac{1}{\cosh\eta}\sum_{n=0}^{\infty}\left(-\mathrm{e}^{\mathrm{i}\zeta}\tanh\eta\right)^{n}|n,n\rangle}
\end{align}
where 
\begin{equation}
\mathrm{\hat{S}(\chi)=e^{\chi\hat{a}^{\dagger}\hat{b}^{\dagger}-\chi^{\ast}\hat{a}\hat{b}}},
\end{equation}
is the two mode squeezing operator. Where, the operators $\mathrm{\hat{a}}$
($\mathrm{\hat{a}^{\dagger}}$) and $\mathrm{\hat{b}}$ ($\mathrm{\hat{b}^{\dagger}}$)
represent the annihilation (creation) operators corresponding to the
two bosonic modes, with their commutation relation defined as$\mathrm{[\hat{a},\hat{a}^{\dagger}]=[\hat{b},\hat{b}^{\dagger}]=1}$
and $\mathrm{[\hat{a},\hat{b}]=0}$. Here, $\mathrm{\chi=\eta e^{i\zeta}}$
and $\mathrm{\lambda}$ represents the squeezing parameter, with $\mathrm{0\leq\eta<\infty}$
and $\mathrm{0\leq\zeta\leq2\pi}$. 

\begin{figure}[h]
\begin{centering}
\includegraphics[scale=0.04]{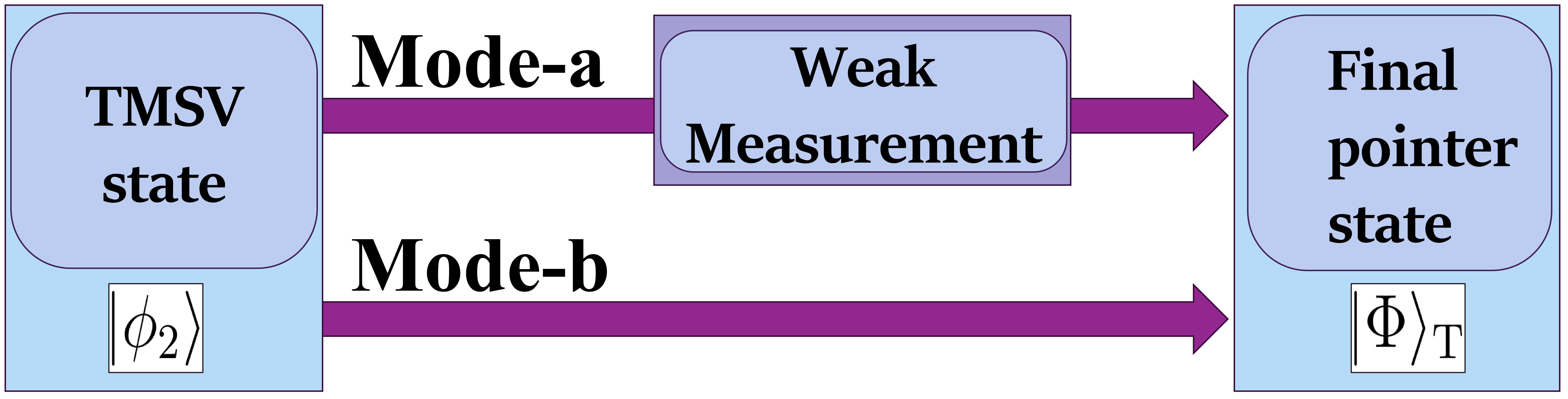}
\par\end{centering}
\caption{Schematic diagram for generating $\mathrm{\vert\Phi\rangle_{T}}$
using von Neumann measurement with post-selection.\label{fig:TMSVS-a-mode}}
\end{figure}

Then we emphasize that, since the TMSV state is a bimodal system,
Fig. \ref{fig:TMSVS-a-mode} thus demonstrates the implementation
of weak measurement solely on its individual mode (mode a) and the
relevant properties of the two mode squeezing operator are employed
in the calculation to obtain\citep{Agarwal2013}

\begin{align}
\mathrm{\hat{S}^{\dagger}(\chi)\hat{a}\hat{S}(\chi)} & =\mathrm{\hat{a}\cosh\eta+\hat{b}^{\dagger}e^{i\zeta}\sinh\eta},\\
\mathrm{\hat{S}^{\dagger}(\chi)\hat{b}\hat{S}(\chi)} & =\mathrm{\hat{b}\cosh\eta+\hat{a}^{\dagger}e^{i\zeta}\sinh\eta},
\end{align}
and 
\begin{equation}
\mathrm{\hat{D}(t)\hat{S}(\chi)=\hat{S}(\chi)\hat{D}(c)},\label{eq:11}
\end{equation}
where $\mathrm{c=t\cosh\lambda+t^{\ast}e^{i\zeta}\sinh\lambda}$.
Here, $\mathrm{\hat{D}(t)}$ and $\mathrm{\hat{D}(c)}$ represent
the displacement operators as mentioned above, consistent with the
preceding steps, normalization is subsequently imposed on the derived
equation, expressed as

\begin{align}
\mathrm{\mathrm{\vert\Phi\rangle_{T}}} & \mathrm{=\frac{\vert\tilde{\Phi}\rangle_{T}}{\sqrt{P_{T}}}}\nonumber \\
 & =\mathrm{\kappa\bigg[r_{+}\hat{D}\left(\frac{s}{2}\right)+r_{-}\hat{D}^{\dagger}\left(\frac{s}{2}\right)\bigg]\vert\phi_{2}\rangle},\label{eq:normalization-1}
\end{align}
Here, $\mathrm{\kappa=1/\sqrt{P_{T}}}$ is the normalization coefficient
given by

\begin{equation}
\mathrm{\kappa=\sqrt{2}\left[1+\vert\langle\hat{\sigma}_{x}\rangle_{w}\vert^{2}+\left(1-\vert\langle\hat{\sigma}_{x}\rangle_{w}\vert^{2}\right)K\right]^{-\frac{1}{2}}},
\end{equation}
and $\mathrm{P_{T}}$ further quantifies the post-selection success
probability of $\mathrm{\vert\Phi\rangle_{T}}$, defined as

\begin{align}
\mathrm{P_{T}} & =\mathrm{_{T}\langle\tilde{\Phi}|\tilde{\Phi}\rangle_{T}}\nonumber \\
 & =\mathrm{\frac{\cos^{2}\frac{\alpha}{2}}{2}\left[1+\left|\langle\hat{\sigma}_{x}\rangle_{w}\right|^{2}+\left(1-\left|\langle\hat{\sigma}_{x}\rangle_{w}\right|^{2}\right)K\right]},
\end{align}
with 
\begin{equation}
\mathrm{K=\langle\phi_{2}\vert\hat{D}(\pm s)\vert\phi_{2}\rangle=e^{-\frac{s^{2}\cosh(2\eta)}{2}}}.
\end{equation}

\begin{figure}
\begin{centering}
\includegraphics[scale=0.355]{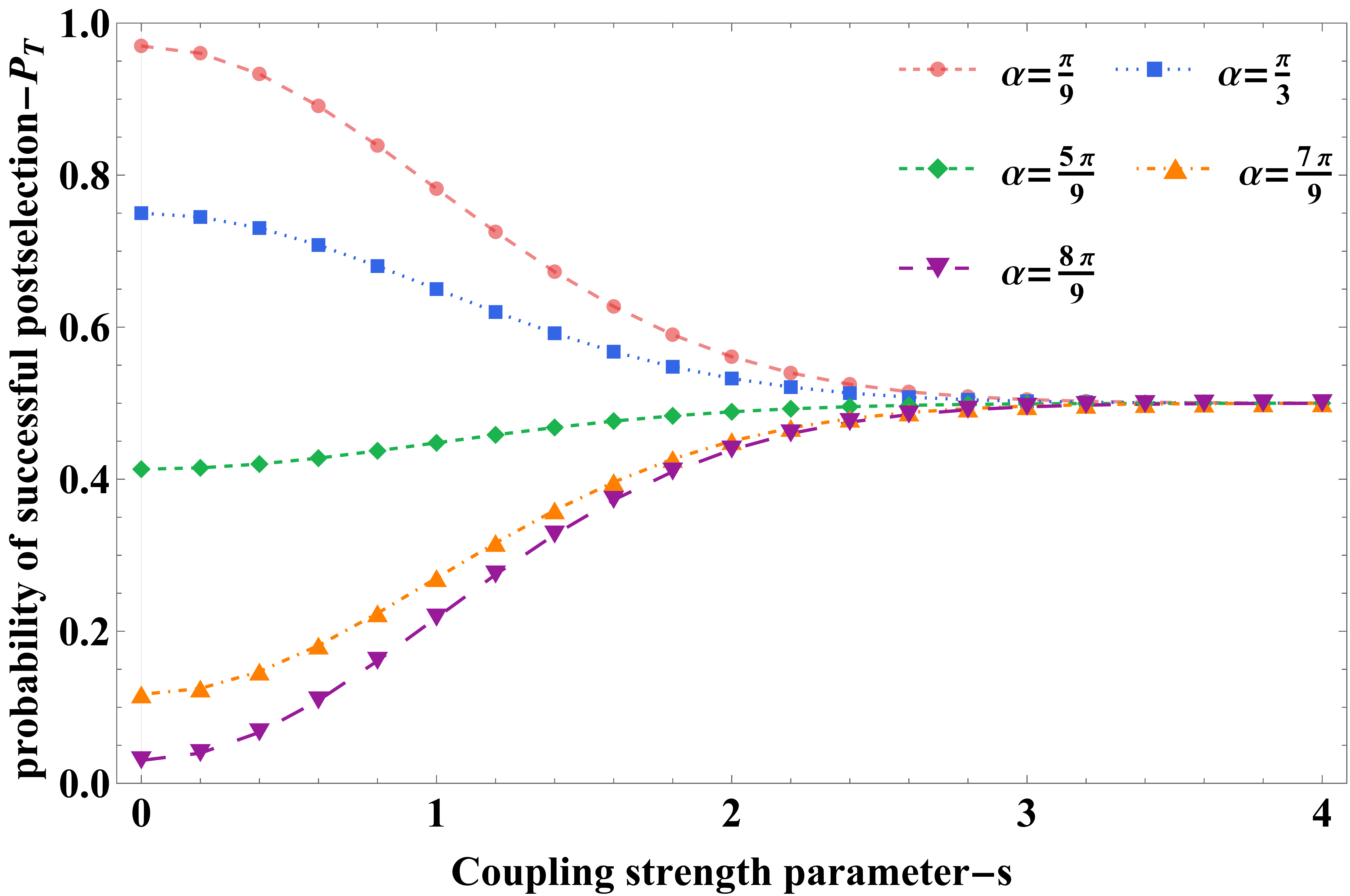}
\par\end{centering}
\caption{The postselection success probability of the final pointer state $\mathrm{\vert\Phi\rangle_{T}}$,
$\mathrm{P_{T}}$ as a dependent variable on $\mathrm{s}$ for different
$\mathrm{\alpha}$, with $\mathrm{\eta=0.1}$. Here $\mathrm{\ensuremath{\zeta}=\delta=0}$.\label{fig:Probility-Tmsvs}}
\end{figure}

Fig. \ref{fig:Probility-Tmsvs} also illustrates the success probability
$\mathrm{P_{T}}$ of the TMSV state $\mathrm{\vert\phi_{2}\rangle}$
as a function of the coupling strength $\mathrm{s}$ and the weak
value parameter $\mathrm{\alpha}$. It is seen that within the weak
coupling regime ($\mathrm{0<s<1}$), increasing the coupling strength
$\mathrm{s}$ causes the success probability $\mathrm{P_{T}}$ to
increase with the weak value $\mathrm{\alpha}$. Conversely, in the
strong coupling regime ($\mathrm{s>1}$), pa becomes independent of
$\mathrm{s}$. Specifically, for $\mathrm{s>2.5}$, $\mathrm{P_{T}}$
remains constant at $\mathrm{0.5}$ and is independent of the weak
value parameters $\mathrm{\alpha}$ and $\mathrm{\delta}$.

Similarly, the TMSV state transition value satisfies 
\begin{align}
\mathrm{\sigma_{x}^{T}} & =\mathrm{\frac{_{T}\langle\tilde{\Phi}\vert\Psi^{\prime\prime}\rangle}{P_{T}}}\nonumber \\
 & =\mathrm{\frac{2\langle\hat{\sigma}_{x}\rangle_{w}K}{1+\left|\langle\hat{\sigma}_{x}\rangle_{w}\right|^{2}+\left(1-\left|\langle\hat{\sigma}_{x}\rangle_{w}\right|^{2}\right)K}}.\label{eq:T-value-TMSVS}
\end{align}
Consistent with the aforementioned analysis

\begin{align}
\mathrm{\left(\sigma_{x}^{T}\right)_{s\rightarrow0}} & \mathrm{=\langle\hat{\sigma}_{x}\rangle_{w}},\\
\mathrm{\left(\sigma_{x}^{T}\right)_{s\rightarrow\infty}} & =\mathrm{\cos\delta\sin\alpha=\sigma_{x}^{c}}.
\end{align}

Experimental realization of continuous transition from weak values
to expectation values has been demonstrated in single-ion trap systems,
providing critical validation for theoretical unification of quantum
measurement frameworks and enabling novel technological applications\citep{pan2020weak}.
We next investigate the influence of anomalous weak values associated
with the measured system observable on the intrinsic properties of
the SPSSV state and TMSV state.

\section{The effects of post-selected measurement on the properties of SPSSV
and TMSV states\label{sec:3}}

Within this analytical framework, we rigorously examine the alterations
imparted to the essential quantum signatures characterizing SPSSV
and TMSV states through post-selected von Neumann measurement protocols.\LyXZeroWidthSpace{}

\subsection{Wigner-Yanase skew information of SPSSV state}

Recent advancements in quantum information theory have extended the
foundational framework of the Wigner-Yanase skew information ($\mathrm{W}$)
to develop a novel metric for quantifying non-classical features in
optical fields. Contemporary research efforts \citep{Turek_2023,doi:10.1073/pnas.49.6.910,PhysRevA.100.032116}
have formulated this information-theoretic quantity as a particularly
advantageous measure, demonstrating attributes such as conceptual
elegance, tractability in mathematical analysis, and intuitive physical
interpretability. Specifically, for pure single-mode quantum states
within the radiation field formalism, the skew information assumes
the following analytically expressible form

\begin{figure*}
\begin{centering}
\includegraphics[scale=0.25]{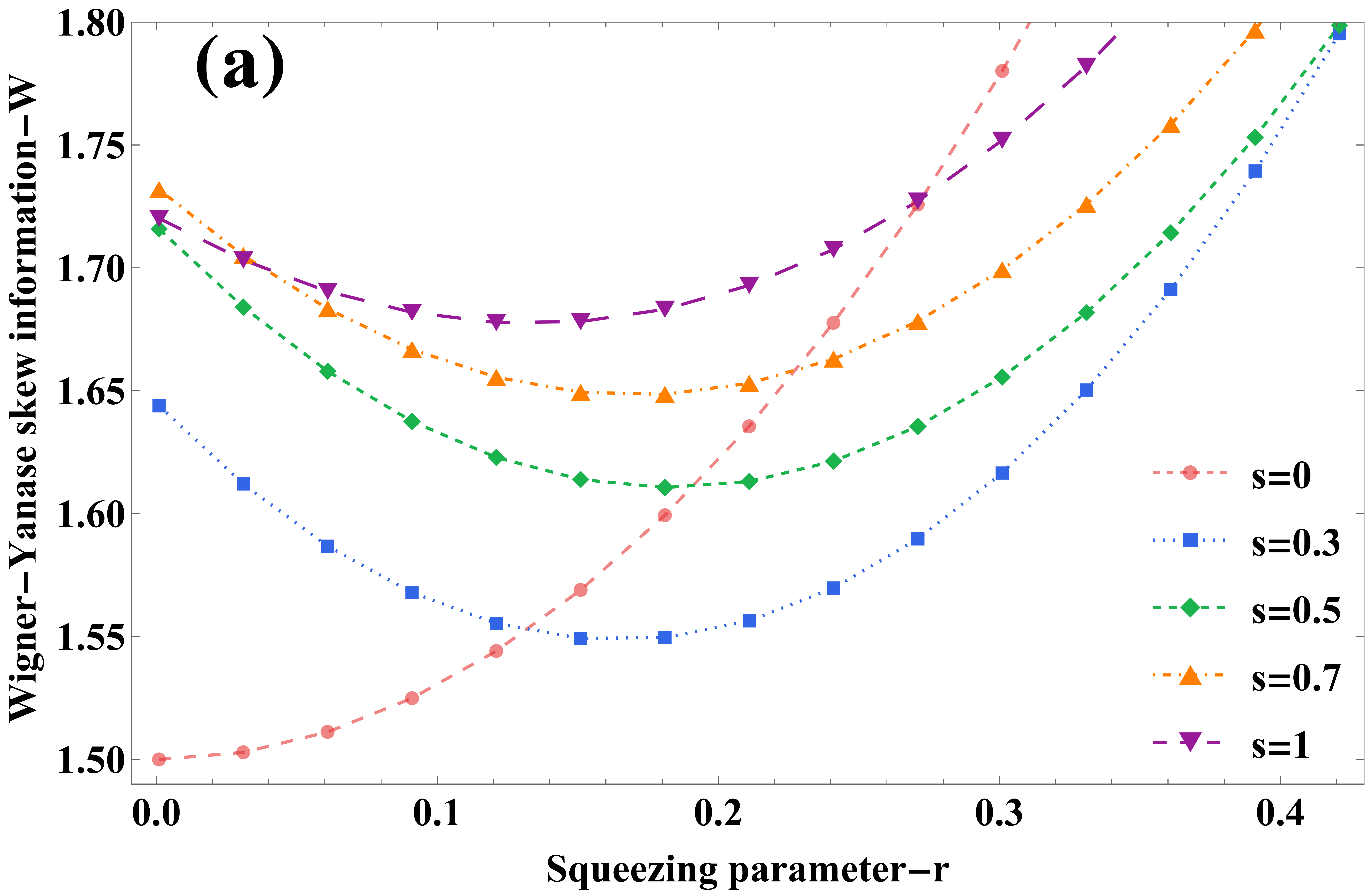}\includegraphics[scale=0.25]{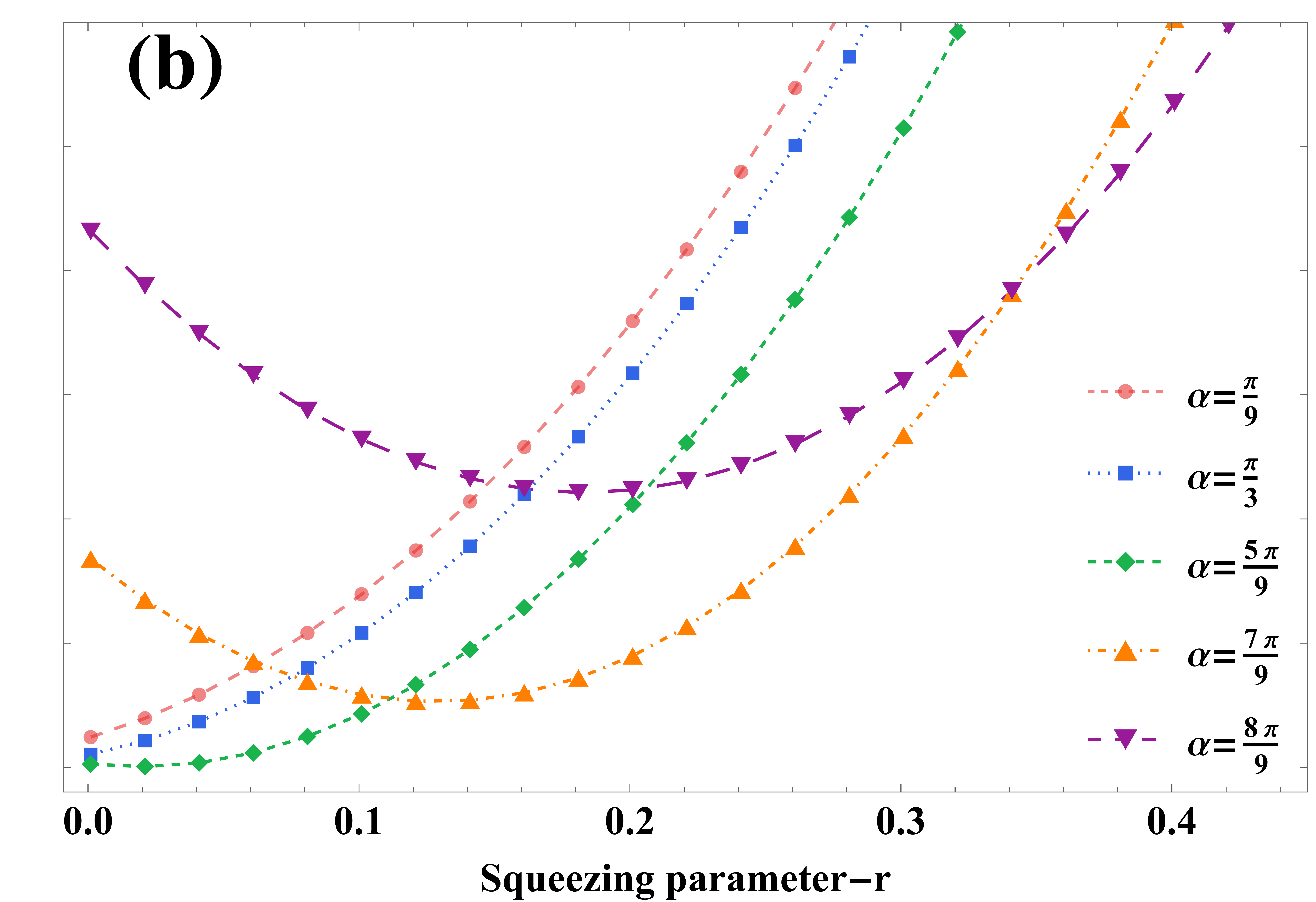}\includegraphics[scale=0.25]{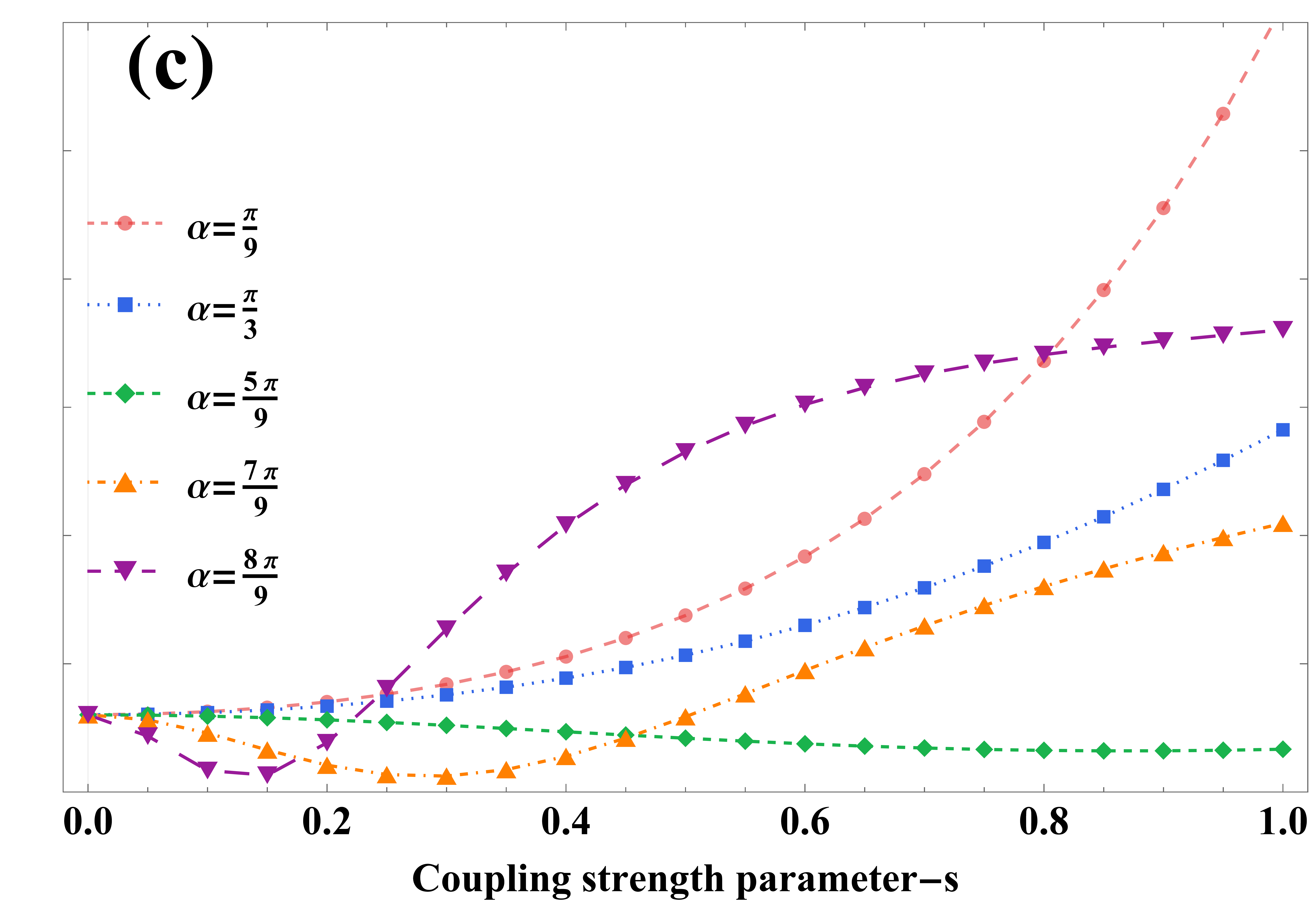}
\par\end{centering}
\caption{Wigner-Yanase skew informatio of the SPSSV state after postselected
measurement. (a) $\mathrm{W}$ as a dependent variable on $\mathrm{r}$
for different $\mathrm{s}$, with $\mathrm{\alpha=8\pi/9}$. (b) $\mathrm{W}$
as a dependent variable on $\mathrm{r}$ for different $\mathrm{\alpha}$,
with $\mathrm{s=0.5}$. (c) $\mathrm{W}$ as a dependent variable
on $\mathrm{s}$ for different $\mathrm{\alpha}$, with $\mathrm{r=0.1}$.
Here $\mathrm{\ensuremath{\theta}=\delta=0}$.\label{fig:WYS-I}}
\end{figure*}

\begin{align}
\mathrm{W} & =\mathrm{\frac{1}{2}+\langle\hat{a}^{\dagger}\hat{a}\rangle-\left|\langle\hat{a}\rangle\right|^{2}},\label{eq:WISI}
\end{align}
Here, $\langle\cdots\rangle$ denotes the expectation values of the
corresponding quantum operators for the state $\mathrm{\mathrm{\vert\Phi_{S}\rangle}}$,
with their analytic expressions rigorously derived in Appendix \ref{sec:A1}.
Setting the coupling coefficient $\mathrm{s=0}$ allows retrieval
of the Wigner-Yanase skew information inherent to the SPSSV state
$\mathrm{\vert\phi\rangle}$, which manifests as

\begin{equation}
\mathrm{W_{s=0}=3\left(\frac{1}{2}+\sinh^{2}(r)\right)}.
\end{equation}

This mathematical relationship implies that skew information exhibits
inherent non-negativity, while its minimal attainable value of $\mathrm{W=0.5}$
emerges uniquely within $\mathrm{coherent\quad states}$. The magnitude
of $\mathrm{W}$ parameter exhibits a direct proportionality to the
degree of quantum non-classicality inherent in a quantum state, thereby
positioning $\mathrm{W}$ as a robust quantifier of its deviation
from classical phase-space behaviour\citep{PhysRevA.100.032116,Turek_2023}.
Furthermore, we quantified the skew information $\mathrm{W}$ of SPSSV
states by computationally evaluating the corresponding parameters
defined in Eq. (\ref{eq:WISI}).

To systematically investigate the influence of postselected von Neumann
measurements on the Wigner-Yanase skew information of the state $\mathrm{\vert\Phi\rangle_{S}}$,
we employ numerical simulations supported by theoretical analysis,
with quantitative results comprehensively presented in Fig. \ref{fig:WYS-I}.
As demonstrated in Fig. \ref{fig:WYS-I} (a), the evolution of $\mathrm{W}$
for the postselected SPSSV state is illustrated under varying parameter
conditions. The investigation specifically focuses on the dependence
of $\mathrm{W}$ on the squeezing parameter r, as a function of the
coupling strength parameter $\mathrm{s}$, for a fixed weak value
of $\mathrm{\alpha=8\pi/9}$. The results demonstrate that the initial
SPSSV state ($\mathrm{\vert\phi\rangle}$,$\mathrm{s=0}$) exhibits
a monotonic increase in $\mathrm{W}$ with rising $\mathrm{r}$. Notably,
the postselection-enhanced state $\mathrm{\vert\Phi\rangle}$ shows
significant advantages in the low-squeezing regime ($\mathrm{0<r<0.27}$),
where the $\mathrm{W}$ value of the enhanced state exceeds that of
the initial state by over $\mathrm{80\%}$ when $\mathrm{s>0}$. This
indicates that appropriately increasing the coupling coefficient under
weak squeezing conditions effectively optimises quantum characteristics.
In Fig. \ref{fig:WYS-I} (b), the context of a fixed coupling coefficient
of $\mathrm{s=0.5}$, the influence of weak values of $\mathrm{\alpha}$
on $\mathrm{W}$ is systematically analysed. The findings indicate
that when $\mathrm{r<0.145}$, the utilisation of $\mathrm{\alpha=8\pi/9}$
results in an enhancement of the $\mathrm{W}$ value by approximately
$\mathrm{60\%}$ in comparison with the $\mathrm{\alpha=\pi/9}$ condition.
The synergistic interplay between weak values and squeezing depth
significantly amplifies non-classical effects in this regime. In Fig.
\ref{fig:WYS-I} (c), Further investigation under a baseline squeezing
parameter $\mathrm{r=0.1}$ elucidates the interaction mechanism between
coupling strength parameter $\mathrm{s}$ and weak value $\mathrm{\alpha}$.
Numerical simulations demonstrate that within the optimised range
of $\mathrm{0.22<s<0.8}$, the $\mathrm{\alpha=8\pi/9}$ configuration
achieves $\mathrm{W=0.07\pm0.03}$ (at $\mathrm{s=0.5}$), representing
a $\mathrm{35\%}$ improvement over the $\mathrm{\alpha=\pi/9}$ case.
These findings demonstrate that larger weak values yield superior
non-classical effects and provide explicit guidance for synergistic
optimisation of experimental parameters.

\subsection{AS squeezing of SPSSV state}

The non-classical phenomenon known as amplitude-squared squeezing
of the field amplitude has been explored, with particular instances
having been investigated in previous studies \citep{PhysRevA.32.974,HILLERY1987135,PhysRevA.36.3796}.
To delineate the concept of AS squeezing, we focus on the real and
imaginary components of the squared field mode amplitude, that is

\begin{align}
\mathrm{Y_{1}} & \mathrm{=\frac{(\hat{A}^{\dagger2}+\hat{A}^{2})}{2}},\\
\mathrm{Y_{2}} & \mathrm{=i\frac{(\hat{A}^{\dagger2}-\hat{A}^{2})}{2}}.
\end{align}

Here, $\mathrm{A}$ and $\mathrm{A^{\dagger}}$ represent slowly varying
operators defined as $\mathrm{\hat{A}=e^{i\vartheta t}\hat{a}}$,
$\mathrm{\hat{A}^{\dagger}=e^{-i\vartheta t}\hat{a}^{\dagger}}$and
they adhere to the canonical commutation relations. The operators
$\mathrm{Y_{1}}$ and $\mathrm{Y_{2}}$, on the other hand, fulfill
a commutation relation given by:

\begin{equation}
\mathrm{\left[Y_{1},Y_{2}\right]=i\left(2\hat{A}\hat{A}^{\dagger}+1\right)}.
\end{equation}

Consequently, the uncertainty principle is upheld by the operators
$\mathrm{Y_{1}}$ and $\mathrm{Y_{2}}$, as they adhere to the following
relation of uncertainty.

\begin{align}
\mathrm{\Delta Y_{1}\vartriangle Y_{2}} & =\mathrm{\sqrt{\langle Y_{1}^{2}\rangle-\langle Y_{1}\rangle^{2}}\sqrt{\langle Y_{2}^{2}\rangle-\langle Y_{2}\rangle^{2}}}\nonumber \\
 & \mathrm{\geqslant\left\langle AA^{\dagger}+\frac{1}{2}\right\rangle }.
\end{align}

Here, $\mathrm{\langle(\Delta Y_{1,2})^{2}\rangle}$ represents the
variance of $\mathrm{Y_{1}}$ and $\mathrm{Y_{2}}$ with respect to
a general state ($\mathrm{\vert\Phi\rangle_{S}}$). We can claim that
AS squeezing occurs in the variable $\mathrm{Y_{i}}$ when the following
condition is met:

\begin{equation}
\mathrm{\left(\triangle Y_{i}\right)^{2}<\left\langle N+\frac{1}{2}\right\rangle \quad\quad i=1,2}.
\end{equation}

Typically, the AS squeezing factor may be expressed as $\mathrm{Y=\langle(Y_{\vartheta}-\langle Y_{\vartheta}\rangle)^{2}\rangle-\left(\langle\hat{a}^{\dagger}\hat{a}\rangle+\frac{1}{2}\right)}$,
here $\mathrm{Y_{\vartheta}=\frac{1}{2}\left(\hat{a}^{\dagger2}e^{i\vartheta}+\hat{a}^{2}e^{-i\vartheta}\right)}$,
Upon examining the minimum value of $Y_{\vartheta}$ as a function
of phase $\vartheta$, we find that

\begin{equation}
\mathrm{Y_{\min}=\langle\hat{a}^{\dagger2}\hat{a}^{2}\rangle-\left|\langle\hat{a}^{2}\rangle\right|^{2}-\left|\langle\hat{a}^{4}\rangle-\langle\hat{a}^{2}\rangle^{2}\right|},
\end{equation}
Whenever $\mathrm{Y_{\min}}$ is negative, AS squeezing is identified.
To compare AS squeezing across states with varying energy levels,
the renormalized factor is employed

\begin{equation}
\mathrm{AS=\frac{\left[\langle\hat{a}^{\dagger2}\hat{a}^{2}\rangle-\left|\langle\hat{a}^{2}\rangle\right|^{2}-\left|\langle\hat{a}^{4}\rangle-\langle\hat{a}^{2}\rangle^{2}\right|\right]}{\frac{1}{2}\langle\hat{a}^{\dagger}\hat{a}\rangle+1}}.
\end{equation}

Consequently, AS squeezing is observed when the parameter Sass falls
within the range of $-1$ to $0$. Here, $\langle\cdots\rangle$ denotes
the expectation values of the corresponding quantum operators for
the state $\mathrm{\mathrm{\vert\Phi_{S}\rangle}}$, with their analytic
expressions rigorously derived in Appendix \ref{sec:A1}. Setting
the coupling coefficient $\mathrm{s=0}$ allows retrieval of the AS
squeezing inherent to the SPSSV state $\mathrm{\vert\phi_{1}\rangle}$,
which manifests as
\begin{align}
\mathrm{AS_{s=0}} & =\mathrm{\mathrm{Sech}(2\mathrm{r})\bigg[\mathrm{\sinh}^{2}(r)\left(3+5\mathrm{\sinh}^{2}(r)\right)}\nonumber \\
 & \quad\mathrm{-\frac{5}{4}e^{-2\theta}\sinh^{2}(2r)\bigg]}.
\end{align}

\begin{figure}[t]
\begin{centering}
\includegraphics[scale=0.355]{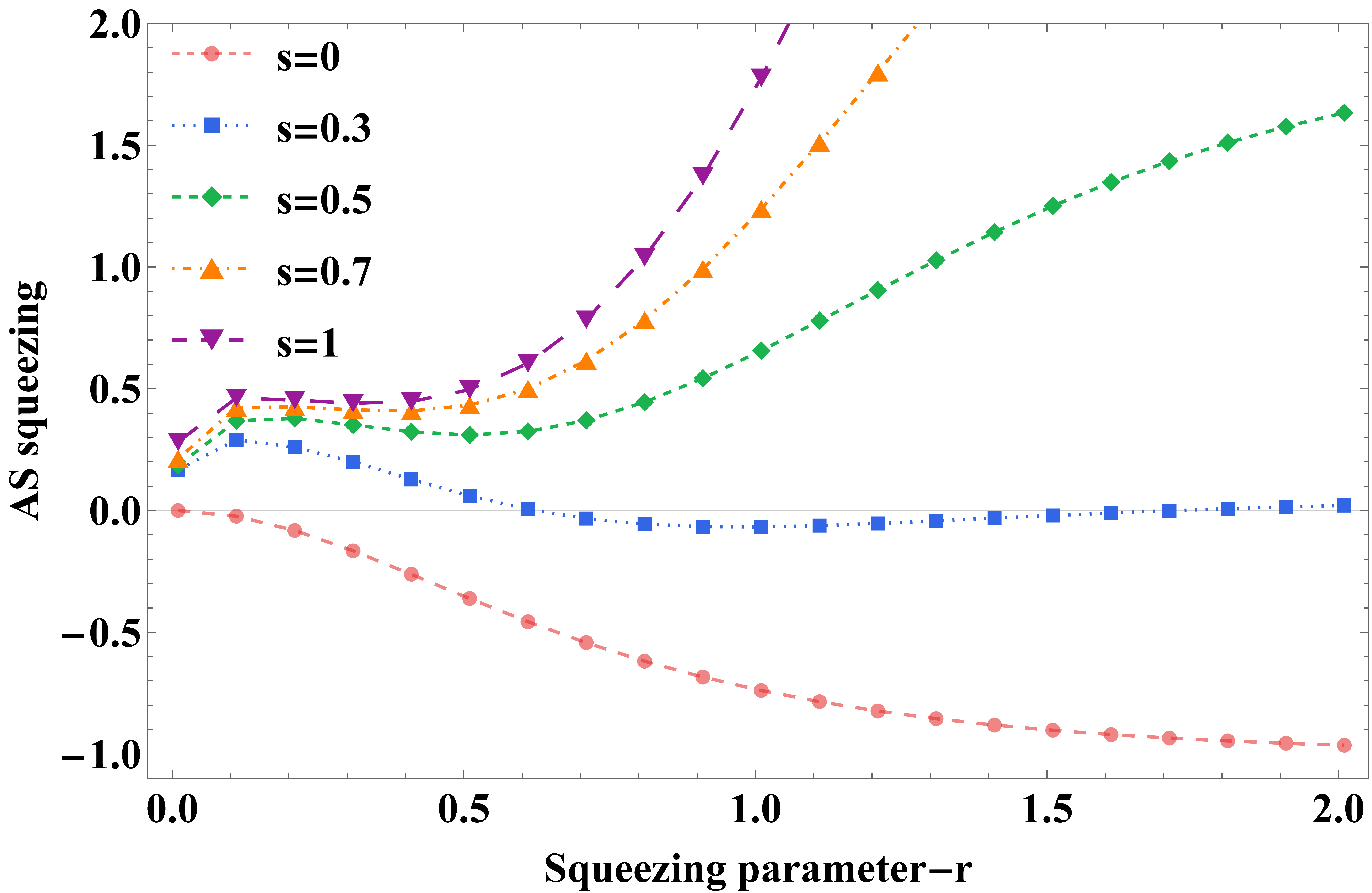}
\par\end{centering}
\caption{AS squeezing of the SPSSV state after psotselected measurement. $\mathrm{AS}$
as a dependent variable on $\mathrm{r}$ for different $\mathrm{s}$.
Here $\mathrm{\alpha=8\pi/9}$ and $\mathrm{\ensuremath{\theta}=\delta=0}$.\label{fig:ASS-I}}
\end{figure}

Theoretical and numerical analyses carried out within the framework
shown in Fig. \ref{fig:ASS-I} systematically reveal the measurement-dependent
properties of AS squeezed in SPSSV state$\mathrm{\mathrm{\vert\Phi\rangle}}$.
Simulation results show that the non-classical properties of AS squeezing
undergo a pronounced non-monotonic evolution as the squeezing parameter
$\mathrm{r}$ increases linearly from $\mathrm{0}$ to $\mathrm{2}$
under a fixed weak value $\mathrm{\langle\hat{\sigma}_{x}\rangle_{w}=5.671}$($\mathrm{\alpha=8\pi/9}$).
In particular, when the coupling strength parameter $\mathrm{s=0}$,
the AS squeezing strength manifests an exponential enhancement of
quantum non-classicality with progressive $\mathrm{r}$-increment.
Paradoxically, As coupling strength parameter $\mathrm{s}$ increases
monotonically from $\mathrm{0}$ to $\mathrm{1}$, the AS squeezing
parameters based on the von Neumann measurement exhibit exponential
decay characteristics, with the decay rate being significantly enhanced
by higher $\mathrm{\alpha}$ weak values $\mathrm{\langle\hat{\sigma}_{x}\rangle_{w}}$,
a phenomenon that progressively reduces the non-classical properties
of the system. This counterintuitive behaviour is in stark contrast
to conventional weak value enhancement paradigms, where parameter
enhancement typically correlates with signal enhancement. The observed
inverse relationship between squeezing parameter evolution and non-classical
property preservation thus represents a fundamental departure from
established weak enhancement principles. Next, we consider the phenomenon
of sum squeezing within the TMSVS state.

\subsection{Sum squeezing of TMSV state}

The multi-mode nonclassical phenomenon known as sum squeezing\citep{PhysRevA.40.3147},
for two modes a and b, is characterized by reduced fluctuations in
a particular two-mode quadrature $\mathrm{V_{\Theta}}$ observable

\begin{equation}
\mathrm{V_{\Theta}=\frac{1}{2}\left(e^{i\Theta}\hat{a}^{\dagger}\hat{b}^{\dagger}+e^{-i\Theta}\hat{a}\hat{b}\right)},\label{eq:quadrature operator}
\end{equation}
where $\mathrm{\Theta}$ is an angle made by $\mathrm{V_{\Theta}}$
with the real axis in the complex plane \citep{NGUYENBAAN199934,PhysRevLett.112.070402,REN2019106,PhysRevA.40.3147}
,A state is said to be sum squeezed for a $\Theta$ if

\begin{equation}
\mathrm{\langle(\Delta V_{\Theta})^{2}\rangle<\frac{1}{4}\langle N_{a}+N_{b}+1}\rangle.
\end{equation}

Where $\mathrm{\langle(\Delta V_{\Theta})^{2}\rangle=\langle V_{\Theta}^{2}\rangle-\langle V_{\Theta}\rangle^{2}}$,
$\mathrm{N_{a}=\hat{a}^{\dagger}\hat{a}}$ and $\mathrm{N_{b}=\hat{b}^{\dagger}\hat{b}}$,
we can define the degree of sum squeezing $\mathrm{S}$ in the following
manner

\begin{equation}
\mathrm{S=\frac{4\left\langle (\Delta V_{\Theta})^{2}\right\rangle }{\langle N_{a}+N_{b}+1\rangle}-1}.\label{eq:sum squeezing}
\end{equation}

The sum squeezing occurs if $\mathrm{S_{2s}<0}$ and a lower bound
of $\mathrm{S_{2s}}$ is equal to $\mathrm{-1}$. Hence, the closer
the value of S to $\mathrm{-1}$ the higher the degree of sum squeezing.
By substituting $\mathrm{V_{\Theta}}$ in Eq.\ref{eq:quadrature operator}
into Eq.\ref{eq:sum squeezing}, we obtain $\mathrm{S}$ in the form
of the normal ordering operators

\begin{equation}
\mathrm{S=\frac{2\left[\operatorname{Re}[e^{-2i\Theta}\left\langle \hat{a}^{2}\hat{b}^{2}\right\rangle ]-2\left(\operatorname{Re}[e^{-i\Theta}\left\langle \hat{a}\hat{b}\right\rangle ]\right)^{2}+\left\langle N_{a}N_{b}\right\rangle \right]}{\left\langle N_{a}\right\rangle +\left\langle N_{b}\right\rangle +1}}.
\end{equation}

Here, $\langle\cdots\rangle$ represents quantum expectation values
(Appendix \ref{sec:A1}). Setting $\mathrm{s=0}$ recovers the TMSV
state $\mathrm{\vert\phi_{2}\rangle}$ inherent sum squeezing

\begin{equation}
\mathrm{S_{s=0}=\frac{\sinh^{2}(2\eta)\left[\cos(2\Theta)-\cos^{2}\Theta\right]}{1+2\sinh^{2}\eta}+2\sinh^{2}\eta}.
\end{equation}

\begin{figure}[t]
\begin{centering}
\includegraphics[scale=0.355]{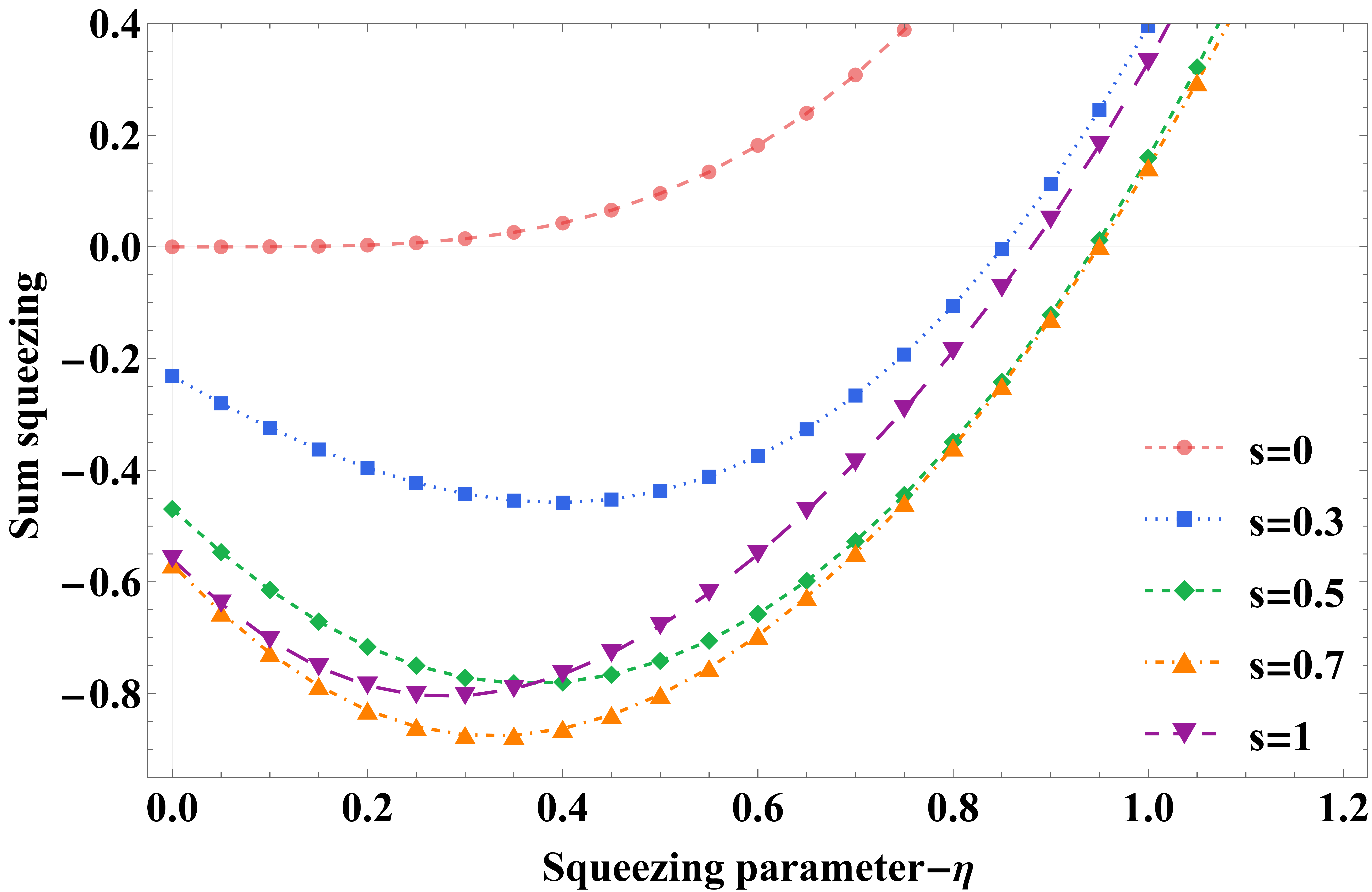}
\par\end{centering}
\caption{sum squeezing of the TMSV state after psotselected measurement. $\mathrm{S}$
as a dependent variable on $\mathrm{\eta}$ for different $\mathrm{s}$.
Here $\mathrm{\alpha=8\pi/9}$ and $\mathrm{\delta=0,\mathrm{\Theta=\pi/4}}$.\label{fig:SUM-SQ}}
\end{figure}

Analysis of Fig. \ref{fig:SUM-SQ} reveals a distinct squeezing behavior
compared to the single-mode case. Specifically, under post-selection
measurement with a weak value of $\mathrm{\langle\hat{\sigma}_{x}\rangle_{w}=5.671}$($\mathrm{\alpha=8\pi/9}$),
the degree of sum squeezing increases with the squeezing parameter
($\mathrm{0<\eta<0.3}$), indicating enhanced non-classical properties.
For coupling strength parameter $\mathrm{s=0}$, the total squeezing
parameter of the TMSV state remains unsqueezed . In contrast, for
$\mathrm{s=0.7}$, coupling strength parameter exhibits significant
squeezing enhancement within a specific range around $\mathrm{\eta=0.3}$,
with the degree of squeezing deepening as $\mathrm{s}$ increases.

\subsection{photon number distribution of SPSSV state }

In this section, the distributions in the phase space of the post-selected
von Neumann measurements on the SPSSV states for the photon number
distribution , respectively, will be studied. The photon number distribution
is one of the important parameters used to characterise quantum states.
The distinguishing characteristic of different quantum states is the
photon number distribution, and this can be measured in order to distinguish
between them.

The photon number distribution is defined as the statistical law that
governs the number of photons in a quantum state, thereby characterising
the particle nature of the light-field quantum state\citep{loudon2000quantum,Int}.
Common light field quantum states manifest distinct photon number
distribution characteristics: The Poisson distribution is employed
for coherent states , whilst the sub-Poisson distribution is utilised
when the variance of the photon number distribution is less than the
mean ($\mathrm{\langle n^{2}\rangle-\langle n\rangle^{2}<\langle n\rangle}$).
The former is typically observed in light fields exhibiting anti-clustering
effects, indicating anti-correlated photon emission that tends to
form pairs or clusters. The hyper-Poisson distribution is characterised
by a variance that exceeds the mean ($\mathrm{\langle n^{2}\rangle-\langle n\rangle^{2}>\langle n\rangle}$).
The thermophoton field exhibits a Bose-Einstein distribution, which,
at low photon numbers, approximates a super-Poisson distribution,
thereby suggesting positive photon emission correlations\citep{loudon2000quantum,Agarwal2013,Turek_2020}.

The probability $\mathrm{\mathrm{P}(n)}$ of measuring $\mathrm{n}$
photons in a postselected measurement is investigated for the quantum
state $\mathrm{\vert\Phi\rangle_{S}}$ is formally defined as

\begin{align}
\mathrm{P(n)}=\mathrm{\left|\langle n|\Phi\rangle_{S}\right|^{2}} & =\mathrm{\sum_{n=0}^{\infty}\left|\lambda C_{n}\left[t_{+}I_{+}+t_{-}I_{-}\right]\right|^{2}},\label{Eq-PN}
\end{align}
where $\mathrm{|n\rangle}$ represents the Fock state (photon number
state) containing exactly $\mathrm{n}$ photons. To calculate parameter
$\mathrm{I_{\pm}}$, we employ the following formula for its characterization\citep{TUREK2023128663}
\begin{align}
\mathrm{\langle m|D(\alpha)|n\rangle} & \mathrm{=e^{-\frac{|\alpha|^{2}}{2}}}\nonumber \\
\times & \mathrm{\begin{cases}
\sqrt{\frac{n!}{m!}}L_{n}^{m-n}\left(|\alpha|^{2}\right)(\alpha)^{m-n}, & m\geq n\\
\sqrt{\frac{m!}{n!}}L_{m}^{n-m}\left(|\alpha|^{2}\right)(\alpha)^{n-m}. & m\leq n
\end{cases}}
\end{align}
with generalized Laguerre polynomials given by

\begin{equation}
\mathrm{L_{n}^{m}(x)=\sum_{k=0}^{n}\left(\begin{array}{c}
n+m\\
n-k
\end{array}\right)\frac{(-1)^{k}}{k!}x^{k}},
\end{equation}
therefore, we conclude that

\begin{align}
\mathrm{I_{\pm}} & \mathrm{=\langle n|\hat{D}\left(\pm\frac{s}{2}\right)|2n+1\rangle}\nonumber \\
 & \mathrm{=\sqrt{\frac{n!}{(2n+1)!}}e^{-\frac{s^{2}}{8}}\left(\mp\frac{s}{2}\right)^{n+1}L_{n}^{n+1}\left(\frac{s^{2}}{4}\right)}.
\end{align}

Therefore, it is imperative that we delve deeper into, setting the
coupling coefficient $\mathrm{s=0}$ allows retrieval of the photon
number distribution inherent to the SPSSV state $\mathrm{\vert\phi_{1}\rangle}$,
which manifests as

\begin{equation}
\mathrm{P(n)_{s=0}}=\mathrm{\left|\langle n|\phi_{1}\rangle\right|^{2}}=0.
\end{equation}

\begin{figure*}
\begin{centering}
\includegraphics[scale=0.355]{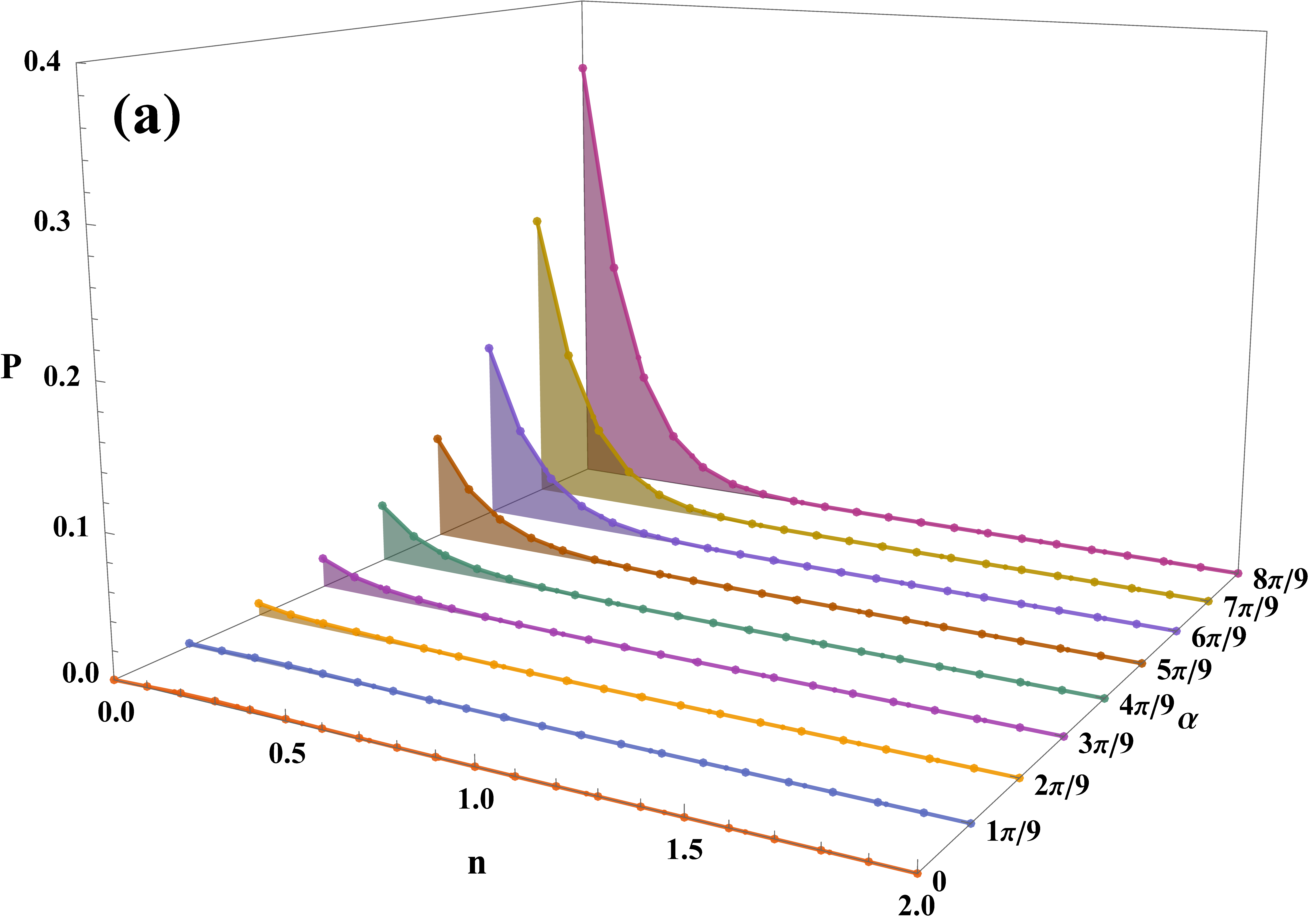}\includegraphics[scale=0.355]{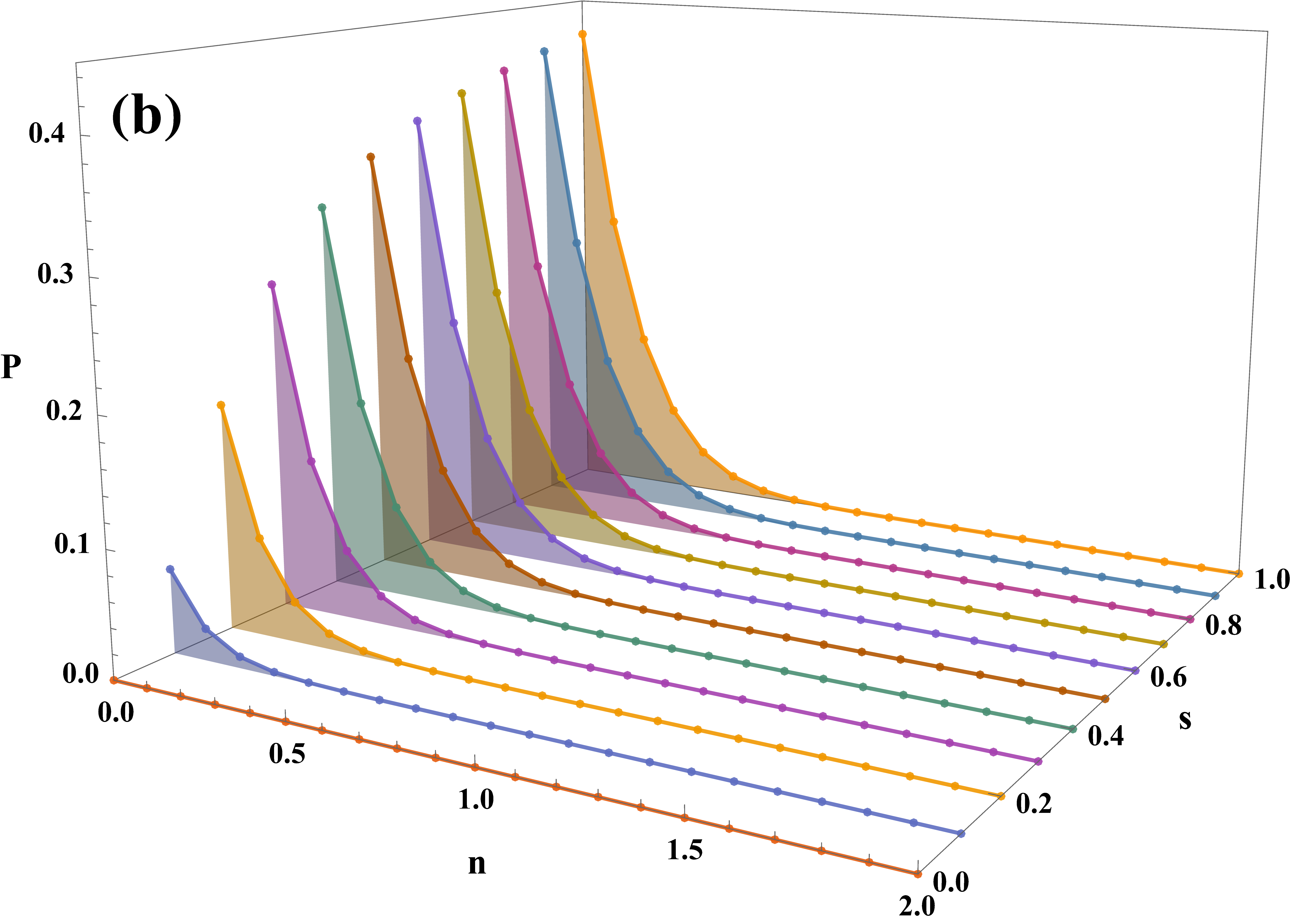}
\par\end{centering}
\caption{Photon statistics $\mathrm{P}$ of the SPSSV state after post-selected
measurement. (a) $\mathrm{P}$ as a dependent variable on $\mathrm{n}$
for different $\mathrm{\alpha}$, with $\mathrm{s=0.5}$. (b) $\mathrm{P}$
as a dependent variable on $\mathrm{n}$ for different $\mathrm{s}$,
with $\mathrm{\alpha=8\pi/9}$. Here $\mathrm{\ensuremath{\theta}=\delta=0}$
and $\mathrm{r=0.1}$.\label{fig:pn}}
\end{figure*}

The probability amplitude for the target outcome being zero in the
initial basis implies that no probability is assigned to this event
prior to measurement. Through the utilisation of numerical simulations
in accordance with Eq. (\ref{Eq-PN}), Fig. \ref{fig:pn} has been
deduced, thereby providing a visual representation of the photon number
distribution $\mathrm{P(n)}$ of the SPSSV state under post-selected
measurement as a function of varying parameters. As demonstrated in
Fig. \ref{fig:pn}(a) demonstrates that, for a fixed $\mathrm{s=0.5}$,
the distribution amplitude in the low photon number region undergoes
a progressive increase as $\mathrm{\alpha}$ is tuned from $\mathrm{4\pi/9}$
to $\mathrm{8\pi/9}$. A sharp single-peak structure emerges at $\mathrm{\alpha=8\pi/9}$,
providing clear evidence of weak value amplification. 

Fig. \ref{fig:pn}(b), when the weak value $\mathrm{\langle\hat{\sigma}_{x}\rangle_{w}=5.671}$($\mathrm{\alpha=8\pi/9}$)
is established, the system with finite coupling strength parameter
($\mathrm{s\neq0}$) manifests elevated photon number distribution
values in the low photon number regime in comparison to the uncoupled
case ($\mathrm{s=0}$). It is noteworthy that the peak amplitude increases
monotonically with the coupling strength parameter $\mathrm{s}$,
thereby demonstrating a characteristic Poissonian distribution profile.
It can be deduced from the evidence presented that postselected measurement
has the capacity to effectively tailor the Poissonian statistical
properties of the optical field by either enhancing the weak value
$\mathrm{\alpha}$ or strengthening the coupling strength parameter
$\mathrm{s}$. 

Theoretical analysis and numerical simulations have been used to confirm
that post-selected measurement enables non-classical manipulation
of photonic statistical properties. The synergistic interplay between
weak value amplification and interaction strength has been shown to
significantly modify the photon statistics of the optical field.

\section{Systematic Framework for Weak-to-Strong Measurement Transitions\label{sec:4}}

\subsection{Measurement transition}

This section formulates the parametric evolution framework for weak-to-strong
measurement transitions and rigorously derives the critical phenomena
associated with quantum state bifurcation under parametric amplification
of measurement operators. The expectation values of the position operator
$\mathrm{\hat{X}}$ and momentum operator $\mathrm{\hat{P}}$ in a
single mode are formally defined as follows

\begin{align}
\mathrm{\delta X} & \mathrm{=\langle\hat{X}\rangle_{\mathrm{\Phi},w}-\langle\hat{X}\rangle_{\mathrm{\phi_{1}},i}}\nonumber \\
 & =\mathrm{_{S}\langle\Phi\vert\hat{X}\vert\Phi\rangle_{S}-\langle\phi_{1}\vert\hat{X}\vert\phi_{1}\rangle},\label{eq:X-shift}
\end{align}
and

\begin{align}
\mathrm{\delta P} & \mathrm{=\langle\hat{P}\rangle_{\mathrm{\Phi},w}-\langle\hat{P}\rangle_{\mathrm{\phi}_{1},i}}\nonumber \\
 & =\mathrm{_{S}\langle\Phi\vert\hat{P}\vert\Phi\rangle_{S}-\langle\phi_{1}\vert\hat{P}\vert\phi_{1}\rangle},\label{eq:P-shift}
\end{align}
where, the corresponding results within the pointer state $\mathrm{\vert\phi\rangle}$
representation are expressed as
\begin{align}
\mathrm{\langle\phi_{1}\vert\hat{X}\vert\phi_{1}\rangle} & \mathrm{\mathrm{=\sigma\langle\phi_{1}\vert(\hat{a}+\hat{a}^{\dagger})\vert\phi_{1}\rangle=0,}}\\
\mathrm{\langle\phi_{1}\vert\hat{P}\vert\phi_{1}\rangle} & =\mathrm{\frac{i}{2\sigma}\langle\phi_{1}\vert(\hat{a}^{\dagger}-\hat{a})\vert\phi_{1}\rangle=0.}
\end{align}

Analogously, within the framework of the final pointer state $\mathrm{|\Phi\rangle}$,
the corresponding results manifest as

\begin{align}
\mathrm{\langle\Phi\vert X\vert\Phi\rangle} & =\mathrm{\sigma\langle\Phi\vert(\hat{a}^{\dagger}+\hat{a})\vert\Phi\rangle}\nonumber \\
 & =\mathrm{2\sigma\operatorname{Re}\left[\langle\Phi\vert\hat{a}\vert\Phi\rangle\right]},
\end{align}
thus

\begin{align}
\mathrm{\langle\Phi\vert\hat{P}\vert\Phi\rangle} & =\mathrm{\frac{i}{2\sigma}\langle\Phi\vert(\hat{a}^{\dagger}-\hat{a})\vert\Phi\rangle}\nonumber \\
 & =\mathrm{\frac{g}{\sigma^{2}}\frac{1}{s}\operatorname{Im}\left[\langle\Phi\vert a\vert\Phi\rangle\right]}.
\end{align}

Provided that the coupling parameter is sufficiently small, the above
position and momentum shifts are asymptotically reduced to their counterparts
in the post-selected weak measurement regime.

\begin{align}
\mathrm{h_{1}(s\to0)} & \mathrm{=3s\bigg[\frac{1}{2}\left(1+e^{i\theta}\sinh(2r)\right)-\cosh^{2}(r)\Bigg],}
\end{align}
building upon the preceding derivations, the final theoretical solution
is obtained, which can be mathematically expressed as

\begin{align}
\mathrm{\frac{\mathrm{\delta X}_{s\to0}}{g}} & =\mathrm{\operatorname{Re}[\langle\hat{\sigma}_{x}\rangle_{w}]+\mathrm{\frac{3}{2}\sin\theta\sinh(2r)\operatorname{Im}[\langle\hat{\sigma}_{x}\rangle_{w}]},}\\
\frac{\mathrm{\mathrm{\delta P}_{s\to0}}\sigma^{2}}{\mathrm{g}} & =\mathrm{\frac{3}{2}\left(\cosh(2r)-\cos\theta\sinh(2r)\right)\operatorname{Im}[\langle\hat{\sigma}_{x}\rangle_{w}].}
\end{align}

Similarly, the obtained results under strong coupling regime are presented
as

\begin{align}
\mathrm{h_{1}(s\to\infty)} & \mathrm{=0},
\end{align}
the obtained results demonstrate

\begin{align}
\frac{\mathrm{\delta X}_{s\to\infty}}{\mathrm{g}} & \mathrm{=\cos\delta\sin\varphi=\sigma_{x}^{c}},\\
\mathrm{\delta P}_{s\to\infty} & \mathrm{=0.}
\end{align}

Following this, we will discuss the representations of the position
and momentum operators in a two-mode framework, for a two mode system
the position and momentum operators can be written as
\begin{align}
\mathrm{\hat{X}_{2}} & =\mathrm{\sigma(\hat{a}+\hat{a}^{\dagger}+\hat{b}+\hat{b}^{\dagger}}),\\
\mathrm{\hat{P}_{2}} & \mathrm{=\frac{i}{2\sigma}(\hat{a}^{\dagger}-\hat{a}+\hat{b}^{\dagger}-\hat{b}).}
\end{align}

The mean displacements $\mathrm{\delta X}_{2}$ and $\mathrm{\delta P_{2}}$
are given by 

\begin{align}
\mathrm{\delta X}_{2} & \mathrm{=\langle\hat{X}_{2}\rangle_{\mathrm{\Phi_{T}},w}-\langle\hat{X}_{2}\rangle_{\mathrm{\phi}_{2},i}}\nonumber \\
 & =\mathrm{_{T}\langle\Phi\vert\hat{X}_{2}\vert\Phi\rangle_{T}}=\mathrm{2\sigma\{\operatorname{Re}[\Lambda_{1}]+\operatorname{Re}[\Lambda_{2}]\},}\label{shift-x2}
\end{align}
and

\begin{align}
\mathrm{\delta P_{2}} & \mathrm{=\langle\hat{P}_{2}\rangle_{\mathrm{\Phi_{T}},w}-\langle\hat{P}_{2}\rangle_{\mathrm{\phi_{2}},i}}\nonumber \\
 & \mathrm{=\mathrm{_{T}\langle\Phi\vert\hat{P}_{2}\vert\Phi\rangle_{T}}=\frac{g}{\sigma^{2}}\frac{1}{s}\Bigg[\operatorname{Im}[\Lambda_{1}]+\operatorname{Im}[\Lambda_{2}]\Bigg]},\label{shift-p2}
\end{align}
among these, $\mathrm{\mathrm{\Lambda_{1}}}$ and $\mathrm{\mathrm{\Lambda_{2}}}$
are defined by

\begin{align}
\mathrm{\Lambda_{1}} & =\mathrm{\frac{s\vert\kappa\vert^{2}}{2}\left\{ \operatorname{Re}\left[\langle\hat{\sigma}_{x}\rangle_{w}\right]-i\operatorname{Im}\left[\langle\hat{\sigma}_{x}\rangle_{w}\right]\cosh(2\eta)K\right\} },\\
\mathrm{\Lambda_{2}} & =\mathrm{\frac{s\vert\kappa\vert^{2}}{2}\left\{ \operatorname{Re}\left[\langle\hat{\sigma}_{x}\rangle_{w}\right]-i\operatorname{Im}\left[\langle\hat{\sigma}_{x}\rangle_{w}\right][\sinh(2\eta)-1]K\right\} }.
\end{align}
Furthermore, we summarise the results obtained for both the weak coupling
and the strong coupling regimes, where the complex parameters $\mathrm{\Lambda_{1}}$
and $\mathrm{\Lambda_{2}}$ for weak coupling ($\mathrm{s\to0}$)
are given by

\begin{align}
\mathrm{\Lambda_{1(s\to0)}} & =\mathrm{\frac{s}{2}\left\{ \operatorname{Re}\left[\langle\hat{\sigma}_{x}\rangle_{w}\right]-i\operatorname{Im}\left[\langle\hat{\sigma}_{x}\rangle_{w}\right]\cosh(2\eta)\right\} },\label{eq:weak-s-1}\\
\mathrm{\Lambda_{2(s\to0)}} & =\mathrm{\frac{s}{2}\left\{ \operatorname{Re}\left[\langle\hat{\sigma}_{x}\rangle_{w}\right]-i\operatorname{Im}\left[\langle\hat{\sigma}_{x}\rangle_{w}\right](\sinh(2\eta)-1)\right\} }.\label{eq:weak-s-2}
\end{align}

Consequently, the weak-coupling position and momentum shifts foll
ow from Eqs. (\ref{eq:weak-s-1})-(\ref{eq:weak-s-2}): 

\begin{align}
\frac{\mathrm{\delta X}_{2}(s\to0)}{g} & =\mathrm{\frac{1}{2}\operatorname{Re}[\langle\hat{\sigma}_{x}\rangle_{w}]},\\
\frac{\mathrm{\delta P_{2}(s\to0)\sigma^{2}}}{g} & \mathrm{=-e^{\eta}\sinh\eta\operatorname{Im}[\langle\hat{\sigma}_{x}\rangle_{w}].}
\end{align}

In the strong coupling ($\mathrm{s\to\infty}$), however, $\mathrm{\Lambda_{1}}$
and $\mathrm{\Lambda_{\ensuremath{2}}}$ coalesce into

\begin{equation}
\mathrm{\Lambda_{1(s\to\infty)}}=\mathrm{\Lambda_{2(s\to\infty)}=\frac{s}{2}\cos\delta\sin\varphi},
\end{equation}
therefore, the conclusion under strong coupling is that

\begin{align}
\frac{\mathrm{\delta X}_{2}(s\to\infty)}{g} & =\mathrm{\cos\delta\sin\varphi=\sigma_{x}^{c},}\\
\mathrm{\delta P}(s\to\infty) & =0.
\end{align}

\begin{figure}
\begin{centering}
\includegraphics[scale=0.37]{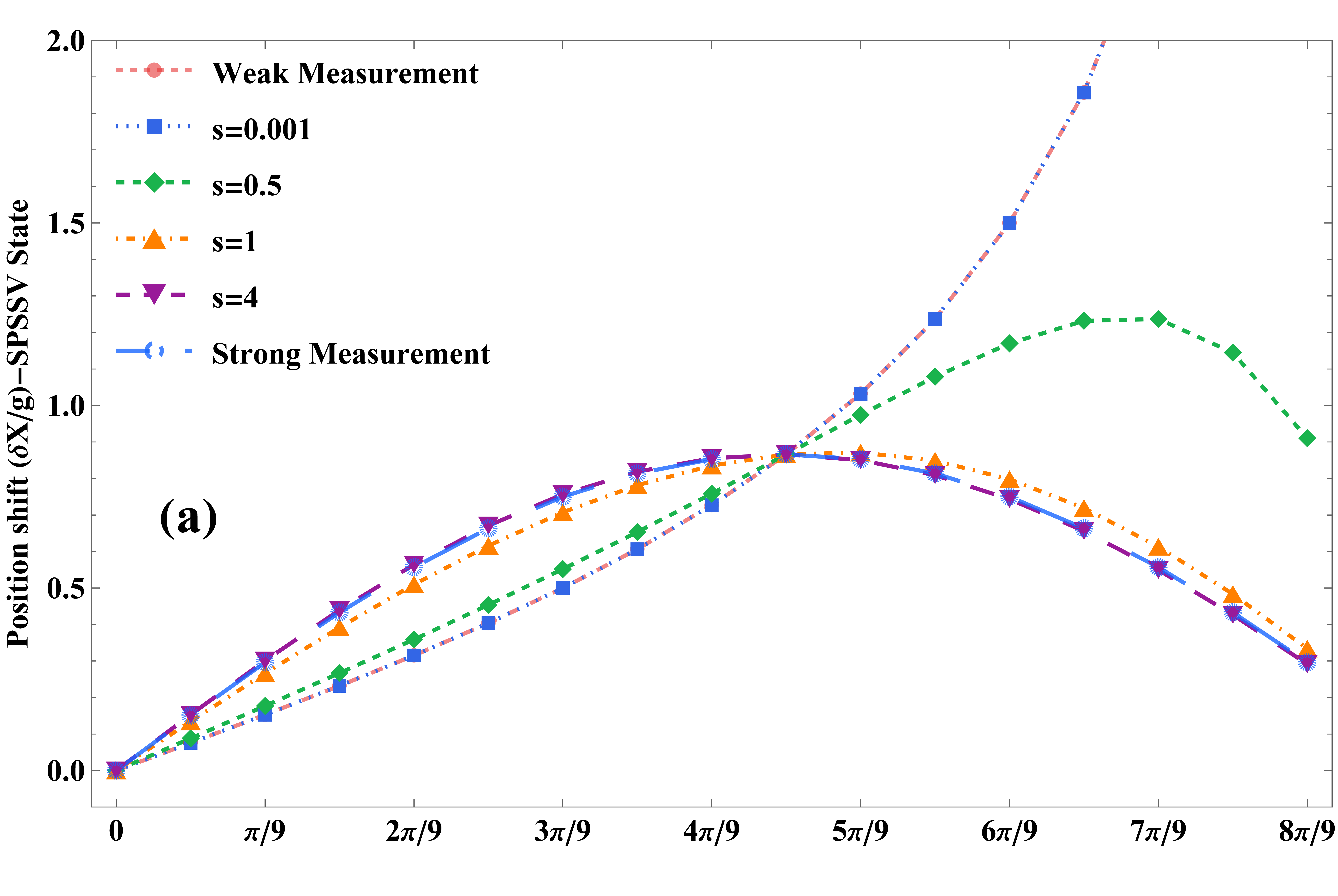}
\par\end{centering}
\begin{centering}
\includegraphics[scale=0.37]{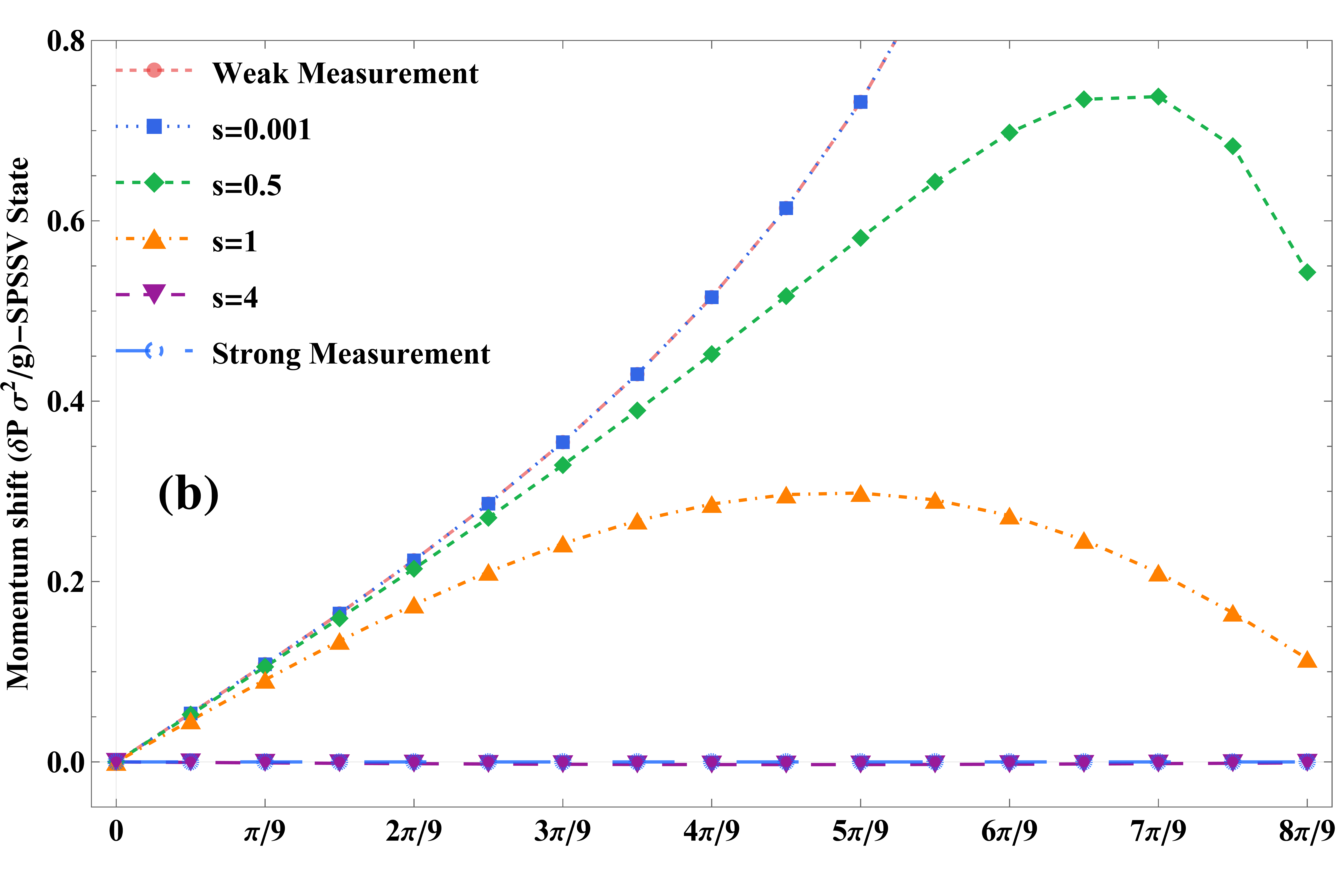}
\par\end{centering}
\begin{centering}
\includegraphics[scale=0.37]{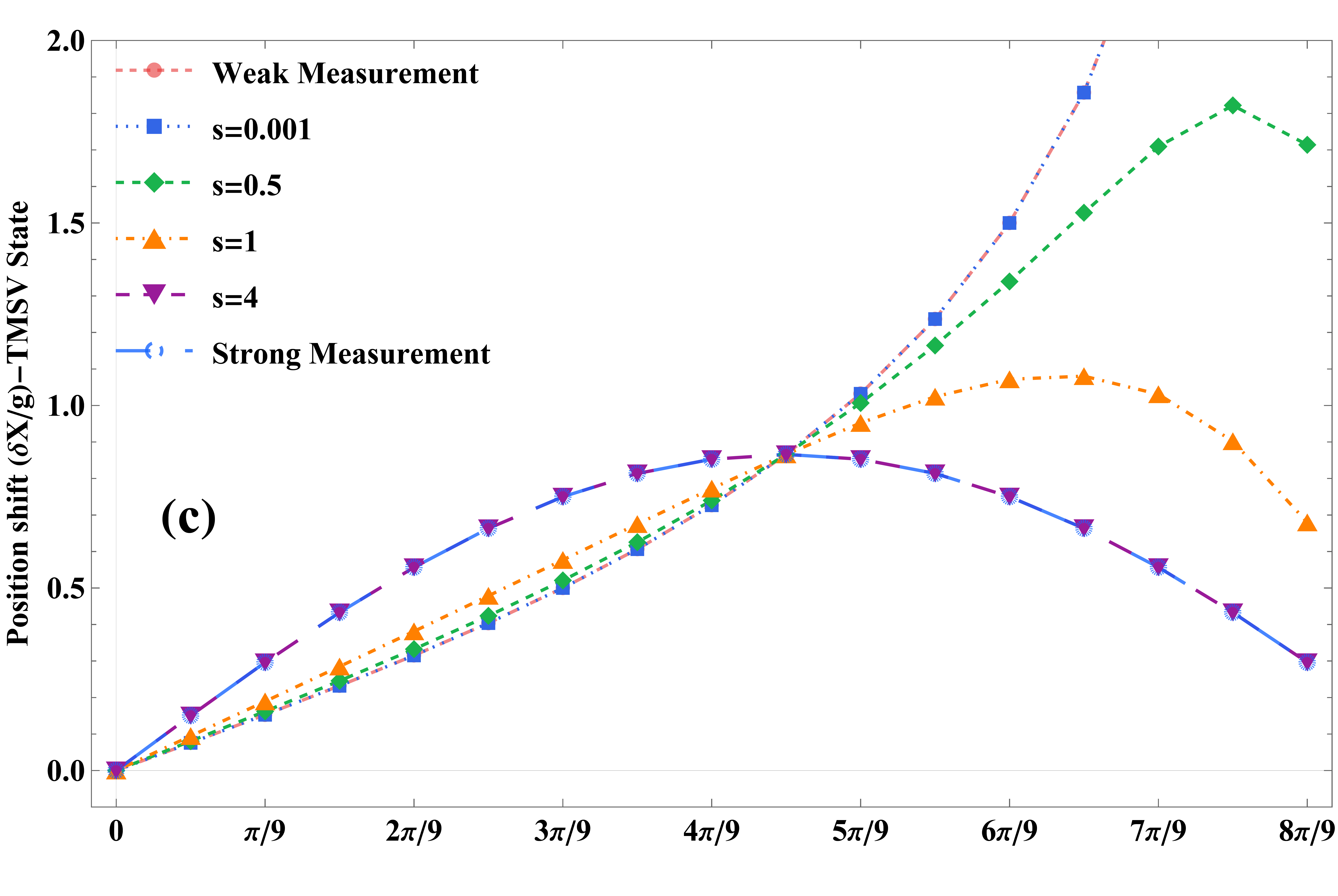}
\par\end{centering}
\begin{centering}
\includegraphics[scale=0.37]{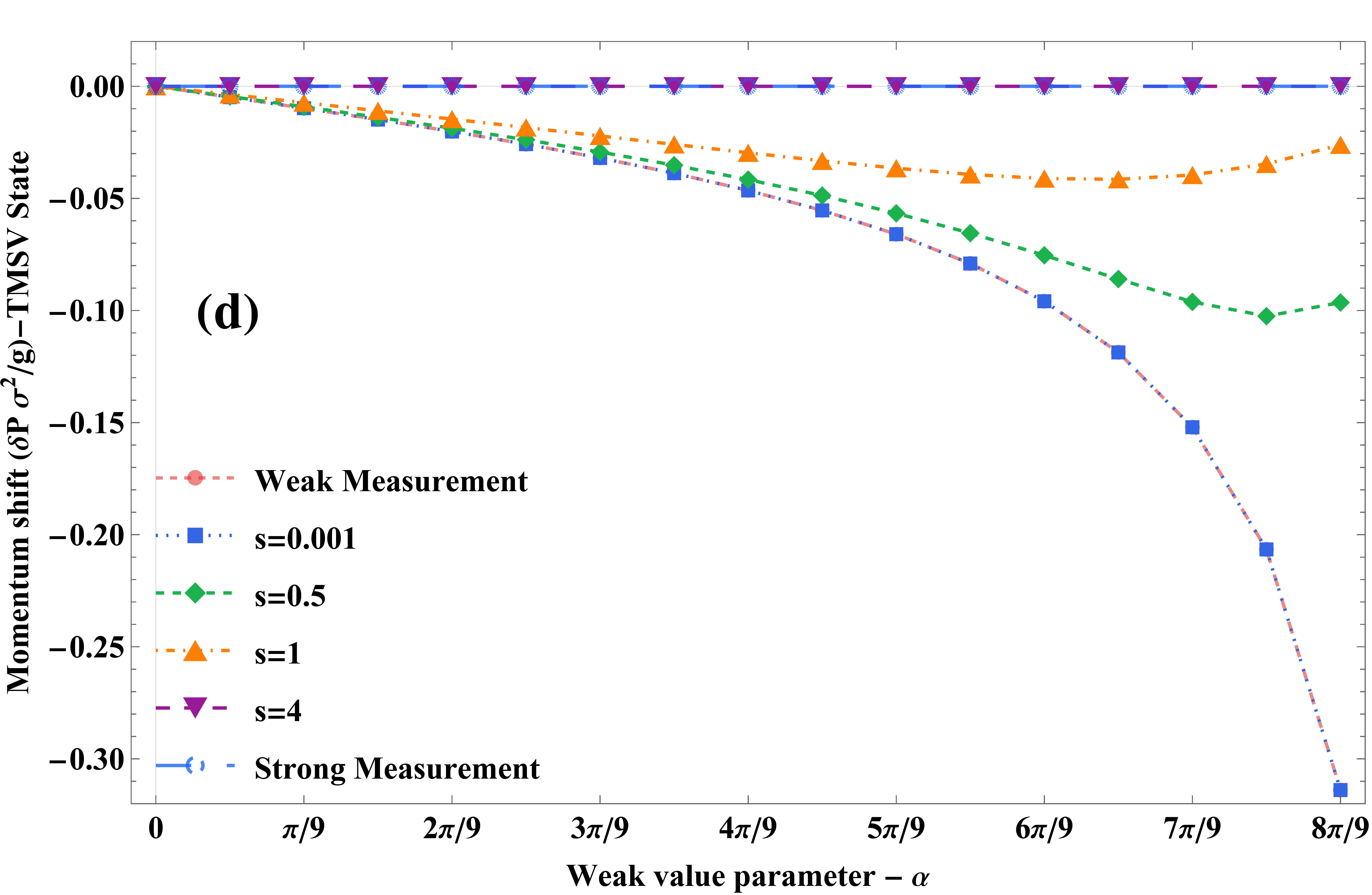}
\par\end{centering}
\caption{Post-selection driven measurement transitions in SPSSV and TMSV state
pointer shifts. (a) and (c) show the position shifts, and panels (b)
and (d) show the momentum shifts for the SPSSV and TMSV states as
a function of $\mathrm{\alpha}$ under varying coupling parameter
$\mathrm{s}$. Here, we set $\mathrm{r=0.1}$, $\mathrm{\eta=0.1}$,
$\mathrm{\ensuremath{\theta=0}}$, and $\mathrm{\ensuremath{\delta=\pi/6}}$.
\label{fig:Shifts}}
\end{figure}

Figure \ref{fig:Shifts} demonstrates the evolution of the pointer
shift during the weak-to-strong measurement transition, where increased
coupling strength $\mathrm{s}$ and weak-value parameter $\mathrm{\mathrm{\alpha}}$
induce significant changes in the pointer displacement. In Figs. \ref{fig:Shifts}
(a) and (b), we present the transition characteristics of SPSSV state
variables from weak to strong quantum measurement regimes as governed
by the theoretical models Eqs. (\ref{eq:X-shift}) and (\ref{eq:P-shift}).
The pointer shift exhibits significant variation with increasing coupling
coefficient $\mathrm{s}$ and weak value parameter $\alpha$. In the
weak coupling limit ($\mathrm{s=0.001}$), the shift significantly
depends on $\alpha$ (e.g., at $\mathrm{\alpha=8\pi/9}$, the weak
value $\mathrm{\langle\hat{\sigma}_{x}\rangle_{w}=5.671}$ corresponds
to the maximum displacement). This behavior aligns with the weak-value-dominated
regime described by Eq. (\ref{eq:weak-approach}), $\mathrm{\delta X/g\propto Re[\langle\hat{\sigma}_{x}\rangle_{w}]}$.
In the strong coupling region ($\mathrm{s=4}$), the displacement
curve flattens and asymptotically approaches the classical expectation
value $\mathrm{\sigma_{x}^{c}=\cos\delta\sin\alpha}$, corresponding
to the strong measurement limit of Eq. (\ref{eq:strong-approach})
and indicative of the complete loss of quantum coherence. We further
observe that the momentum shift in the weak measurement regime is
dominated by the imaginary part of the weak value ($\mathrm{\delta P\cdot\sigma{{}^2}/g\propto Im[\langle\hat{\sigma}_{x}\rangle_{w}]}$),
whereas it asymptotically vanishes in the strong coupling limit. In
figs.\ref{fig:Shifts} (c) and (d) show the weak-to-strong measurement
transition of TMSV state variables according to Eqs. (\ref{shift-x2})
and (\ref{shift-p2}). In fig. (c) demonstrates a halving of displacement
amplitude in the weak measurement regime ($\mathrm{Re[\langle\hat{\sigma}_{x}\rangle_{w}]/2}$),
while asymptotic convergence to $\mathrm{\sigma_{x}^{c}}$ persists
in the strong coupling limit, validating the universality of decoherence
mechanisms. fig. (d) further reveals modulation of the weak-regime
response by the squeezing parameter $\mathrm{\eta}$ ($\mathrm{\propto-e^{\eta}\sinh\eta\operatorname{Im}[\langle\hat{\sigma}_{x}\rangle_{w}]}$),
Reflects the nonclassicality of enhanced bipartite quantum correlations,
whereas it vanishes in the strong coupling correlation region. 

The analysis demonstrates that progressively increasing the system's
coupling strength parameter s induces a continuous evolution in the
measurement dynamics. Notably, in the limiting weak coupling regime
($\mathrm{s\to0}$) and the strongly coupled regime ($\mathrm{s\to\infty}$),
the position shift curve transitions from weak measurement behavior
to strong measurement behavior as the coupling strength $\mathrm{s}$
increases, which closely aligns with our theoretical predictions.
To elucidate the microscopic mechanism of quantum coherence degradation
through the transition from weak to strong measurements, the Husimi-Kano
$\mathrm{Q}$ function is introduced in this section to characterize
the features of quantum-classical boundary phase transitions.

\subsection{Husimi-Kano $\mathrm{Q}$ function}

This section analyzes the weak-to-strong measurement transition for
SPSSV and TMSV states via the phase-space Husimi-Kano $\mathrm{Q}$
function. The Husimi - Kano $\mathrm{Q}$ function is a non-negative
quasi-probability distribution in quantum optics that characterises
the phase-space representation of quantum states, as demonstrated
by Milburn\citep{Schleich1989}, the interference effects in phase
space originate from the properties of the $\mathrm{Q}$ function
for quantum state.

\subsubsection{$\mathrm{Q}$ Function in single mode field}

For a single-mode field, its $\mathrm{Q}$-function can be expressed
in terms of the SPSSV state, as given below:

\begin{figure*}[t]
\begin{centering}
\includegraphics[scale=0.14]{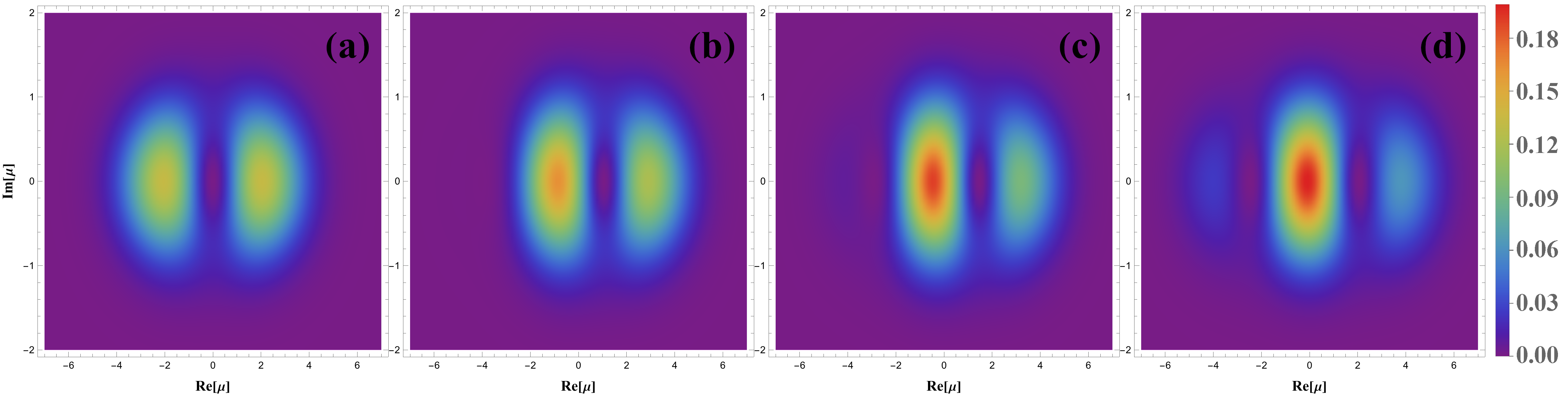}
\par\end{centering}
\caption{$\mathrm{Q}$-function of the post-selected SPSSV state is shown here
for different $\mathrm{s}$. Here, we set $\mathrm{r=1}$, $\mathrm{\ensuremath{\theta}=\delta=0}$
and $\mathrm{\alpha=8\pi/9}$. Columns correspond to different values
of $\mathrm{s}$: (a) non-interacting case ($\mathrm{s=0}$); (b)
$\mathrm{s=0.5}$; (c) $\mathrm{s=1}$; (d) $\mathrm{s=3}$. \label{fig:Q-function}}
\end{figure*}

\begin{align}
\mathrm{Q(\mu)} & =\mathrm{\frac{1}{\pi}\vert\langle\mu\vert\Phi\rangle_{S}\vert^{2}}=\mathrm{\frac{\left|\lambda\right|^{2}}{\pi}\left|t_{+}R_{+}+t_{-}R_{-}\right|^{2}}.\label{eq:Q-fuction}
\end{align}

Where the mathematical expression for $\mathrm{R_{\pm}=\langle\mu\vert\hat{D}\left(\pm\frac{s}{2}\right)\vert\phi_{1}\rangle}$
is concerned, a full derivation and final form is given in Appendix
\ref{sec:B} of this paper, and $\mathrm{\vert\mu\rangle}$ is coherent
states, $\mathrm{\mu}$ be a complex number expressed in the standard
form $\mathrm{\mu=x+iy}$, where $\mathrm{x}$ ($\mathrm{\operatorname{Re}[\mu]}$)
and $\mathrm{y}$ ($\mathrm{\operatorname{Im}[\mu]}$) denote the
real part and imaginary part of the complex number, respectively,
with $\mathrm{x,y\in\mathbb{R}}$, the $\mathrm{Q}$ function of our
initial SPSSV state is defined as

\begin{align}
\mathrm{Q(\phi_{1})} & =\mathrm{\frac{1}{\pi}\vert\langle\mu\vert\phi_{1}\rangle\vert^{2}}\nonumber \\
 & =\mathrm{\frac{e^{\mu^{*2}\mathrm{e}^{\mathrm{i}\theta}\tanh r-|\mu|^{2}}}{\pi\sinh^{2}r\cosh r}\left|\mathrm{e}^{-\mathrm{i}\theta/4}(\mathrm{e}^{\mathrm{i}\theta}\tanh r\mu^{*}-\mu)+\mu\right|^{2}}\nonumber \\
 & =\mathrm{\frac{e^{\mu^{*2}\mathrm{e}^{\mathrm{i}\theta}\tanh r-|\mu|^{2}}}{\pi\sinh^{2}r\cosh r}\bigg(\left|\mathrm{e}^{\mathrm{i}\theta}\tanh r\mu^{*}-\mu\right|^{2}+\left|\mu\right|^{2}}\nonumber \\
 & \quad\mathrm{+2\mathrm{\operatorname{Re}[\mathrm{e}^{-\mathrm{i}\theta/4}(\mathrm{e}^{\mathrm{i}\theta}\tanh r\mu^{*}-\mu)\mu^{*}]\Bigg)}}.
\end{align}

As demonstrated in Fig. (\ref{fig:Q-function}), the $\mathrm{Q}$-function
evolution of SPSSV states across varying measurement strengths ($\mathrm{s=0,0.5,1,3}$)
reveals a continuous quantum-to-classical transition in phase space,
parameterized by the real and imaginary components of the coherent
state $\mathrm{\mu}$. At $\mathrm{s=0}$, the $\mathrm{Q}$ function
exhibits symmetric bimodal Gaussian profiles with prominent interference
fringes between the two peaks, corresponding to the superposition
of Pauli eigenstates and indicating robust quantum coherence in the
weak measurement regime. When $\mathrm{s}$ increases to $1$, decoherence
initiates partial separation of the twin peaks, accompanied by weakened
interference visibility, signifying the onset of measurement-induced
state discrimination. At $\mathrm{s=1}$, the distribution collapses
into two spatially isolated single-mode Gaussian structures, marking
the dominance of projective measurement dynamics that resolve the
pointer states into classical-like eigenstates. Finally, at $\mathrm{s=3}$,
the $\mathrm{Q}$-function fully bifurcates into non-overlapping sharp
peaks with complete suppression of quantum interference, reflecting
a classical statistical mixture where the system irreversibly collapses
into a definite eigenstate. These observations conclusively establish
that the persistence of interference fringes in the weak regime characterizes
quantum coherence, while their disappearance in the strong regime
signals the emergence of classical probability dominance, thereby
providing direct phase-space evidence of measurement-induced quantum
state collapse.

\subsubsection{$\mathrm{Q}$-Function in two mode field}

Similarly, for a two-mode field, its $\mathrm{Q}$ function can be
expressed using the TMSVS state as:

\begin{align}
 & \mathrm{Q(\Psi_{T})}\nonumber \\
 & =\mathrm{\frac{1}{\pi^{2}}\vert\langle\mu_{1},\mu_{2}\vert\Psi\rangle_{T}\vert^{2}}\nonumber \\
 & =\mathrm{\frac{\left|\kappa\right|^{2}}{4\pi^{2}}\Bigg\{\left(1+2\operatorname{Re}\left[\langle\hat{\sigma}_{x}\rangle_{w}\right]+\left|\langle\hat{\sigma}_{x}\rangle_{w}\right|^{2}\right)\left|R_{+}^{\prime}\right|^{2}}\nonumber \\
 & \quad\mathrm{+\operatorname{Re}\left[\left(1-\langle\hat{\sigma}_{x}\rangle_{w}^{*}+\langle\hat{\sigma}_{x}\rangle_{w}+\left|\langle\hat{\sigma}_{x}\rangle_{w}\right|^{2}\right)R_{+}^{\prime}R_{-}^{\prime*}\right]}\nonumber \\
 & \quad\mathrm{+\left(1-2\operatorname{Re}\left[\langle\hat{\sigma}_{x}\rangle_{w}\right]+\left|\langle\hat{\sigma}_{x}\rangle_{w}\right|^{2}\right)\left|R_{-}^{\prime}\right|^{2}\Bigg\},}
\end{align}
where $\mathrm{\vert\mu_{i}\rangle}$ denotes the standard coherent
state for mode $\mathrm{i}$ (with $\mathrm{i=1,2}$), $\mathrm{|\mu_{1},\mu_{2}\rangle\equiv|\mu_{1}\rangle\otimes|\mu_{2}\rangle}$
represents the associated two-mode coherent state, and the mathematical
expression for $\mathrm{R_{\pm}^{\prime}}$ is given by

\begin{align}
\mathrm{R_{\pm}^{\prime}} & =\mathrm{\langle\mu_{1},\mu_{2}\vert\hat{D}\left(\pm\frac{s}{2}\right)\vert\phi_{2}\rangle}\nonumber \\
 & =\mathrm{\frac{1}{\cosh\eta}e^{\frac{1}{2}\left(2\mu_{1}\mu_{2}\tanh\eta\pm\mu_{1}s-\vert\mu_{1}\vert^{2}-\vert\mu_{2}\vert^{2}\right)},}
\end{align}
yielding the squared modulus

\begin{align}
\mathrm{\left|R_{\pm}^{\prime}\right|^{2}} & =\mathrm{\frac{1}{\cosh^{2}\eta}e^{\operatorname{Re}[2\mu_{1}\mu_{2}\tanh\eta\pm\mu_{1}s]-\vert\mu_{1}\vert^{2}-\vert\mu_{2}\vert^{2}},}
\end{align}
and

\begin{align}
 & \mathrm{R_{+}^{\prime}R_{-}^{\prime*}}\nonumber \\
 & =\mathrm{\frac{1}{\cosh^{2}\eta}e^{2\tanh\eta\operatorname{Re}[\mu_{1}\mu_{2}]+s\operatorname{Im}[\mu_{1}]-\vert\mu_{1}\vert^{2}-\vert\mu_{2}\vert^{2}},}
\end{align}
For $\mathrm{s=0}$, the $\mathrm{Q}$ function of our initial TMSV
state reduces to

\begin{align}
\mathrm{Q(\phi_{2})} & =\mathrm{\frac{1}{\pi^{2}}\vert\langle\mu_{1},\mu_{2}\vert\phi_{2}\rangle\vert^{2}}\nonumber \\
 & =\mathrm{\frac{1}{\pi^{2}\cosh^{2}\eta}e^{2\tanh\eta\operatorname{Re}[\mu_{1}\mu_{2}]-|\mu_{1}|^{2}-|\mu_{2}|^{2}}}.
\end{align}

\begin{figure*}[t]
\begin{centering}
\includegraphics[scale=0.27]{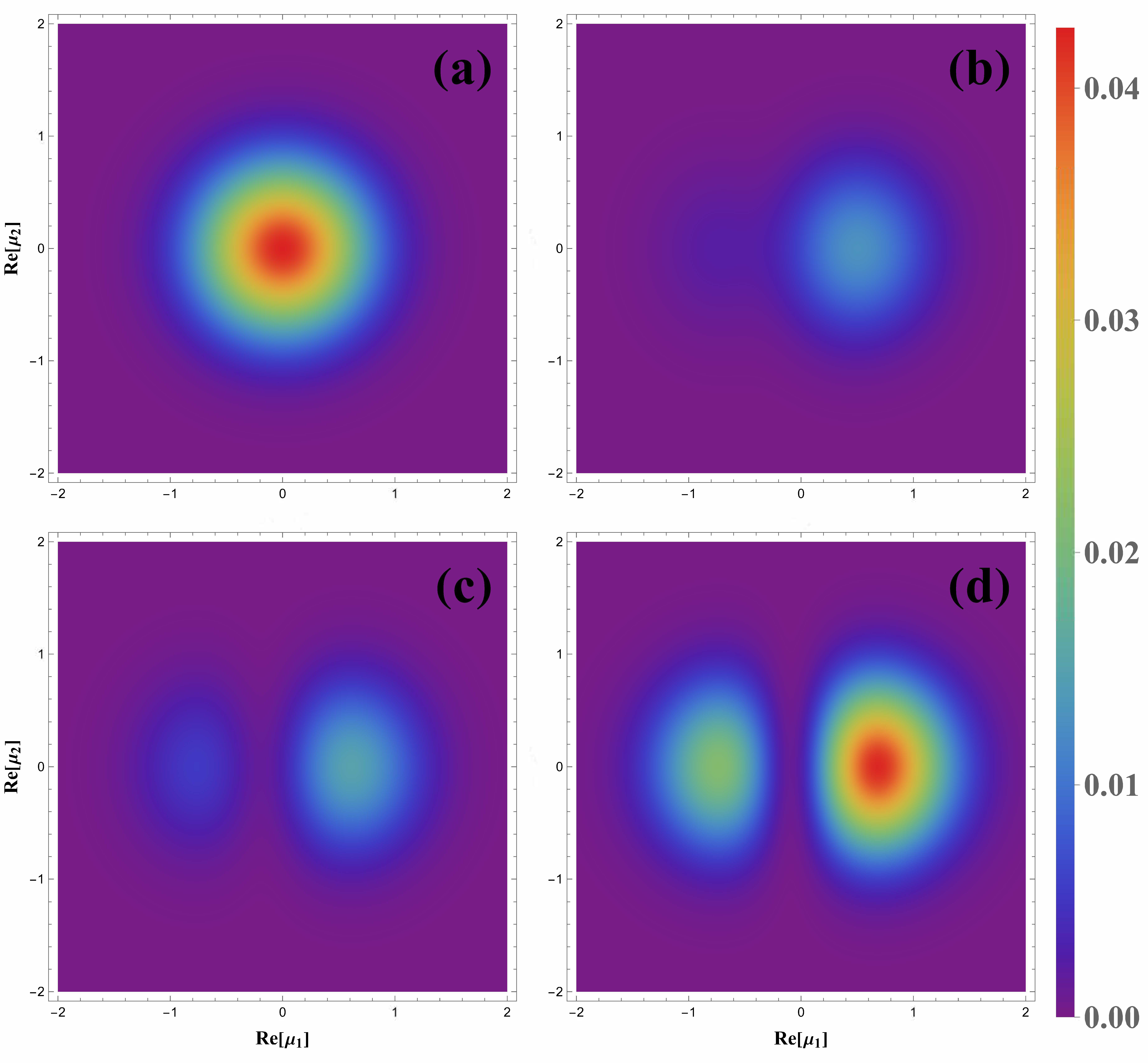}\includegraphics[scale=0.27]{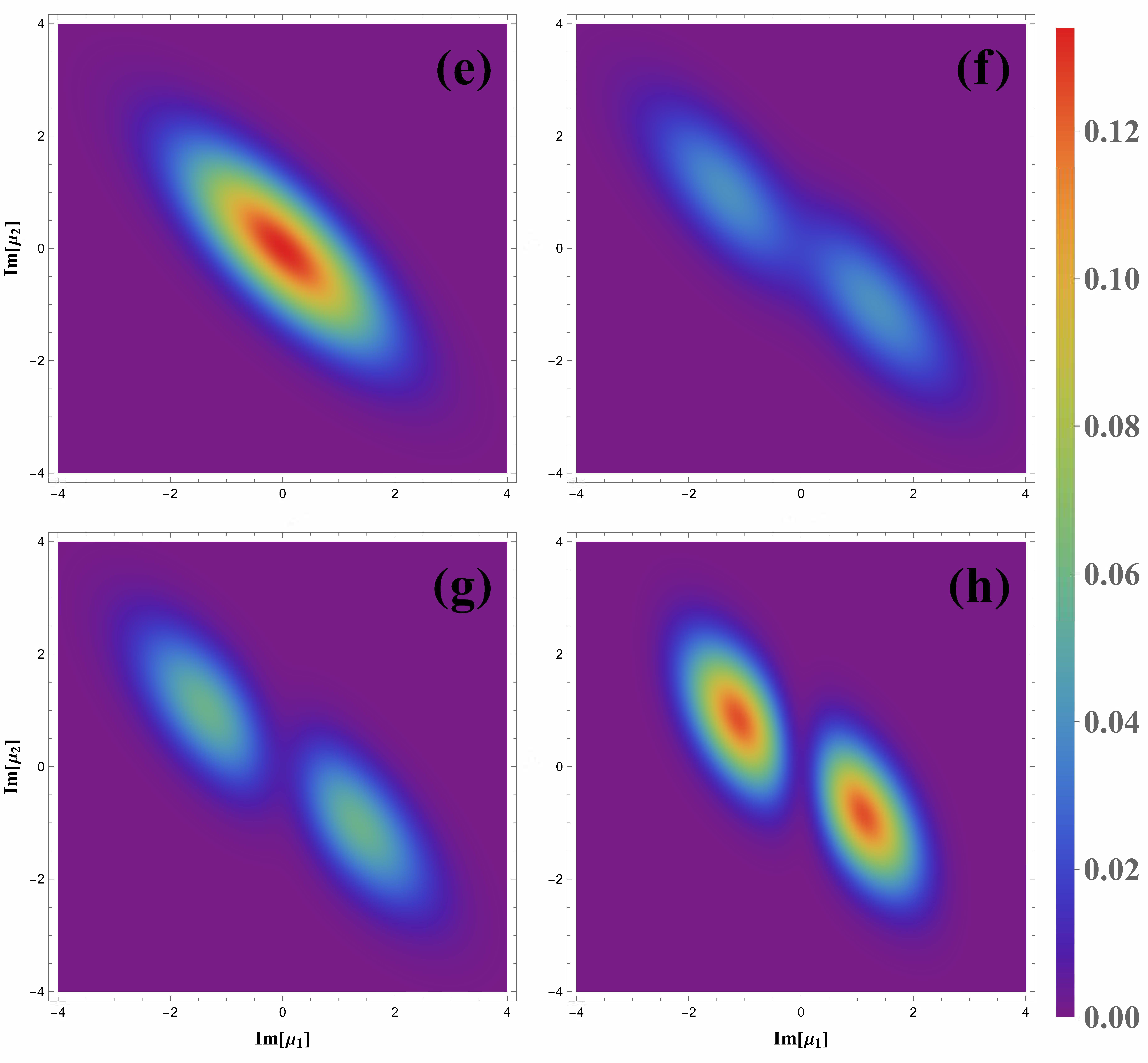}
\par\end{centering}
\caption{$\mathrm{Q}$-function distributions of the TMSV state following post-selected
measurement. Parameters are set with $\mathrm{\mathrm{\eta=1}}$,
$\mathrm{\mathrm{\ensuremath{\theta}=\delta=0}}$, and $\mathrm{\mathrm{\alpha=8\pi/9}}$.
(a-d) Conditional measurement with $\mathrm{\operatorname{Im}[\mu_{1}]=\operatorname{Im}[\mu_{2}]=0}$,
showing dependence on the real components. (e-h) Conditional measurement
with $\mathrm{\operatorname{Re}[\mu_{1}]=\operatorname{Re}[\mu_{2}]=0}$,
showing dependence on the imaginary components. Columns correspond
to different values of $\mathrm{s}$: (a) and (e) are the non-interacting
case ($\mathrm{s=0}$); (b) and (f) $\mathrm{s=0.5}$; (c) and (g)
$\mathrm{s=1}$; (d) and (h) $\mathrm{s=2}$. \label{fig:Q-function-T1}}
\end{figure*}

In order to facilitate an analysis of the transition phenomenon, Figure
\ref{fig:Q-function-T1} presents the distribution of the dual-mode
$\mathrm{Q}$-function as a function of the coupling strength $\mathrm{s}$,
under fixed parameters ($\mathrm{\mathrm{\eta=1}}$, $\mathrm{\mathrm{\theta=\delta=0}}$,
and $\mathrm{\mathrm{\alpha=8\pi/9}}$), at different coupling strengths
($\mathrm{s=0,0.5,1,2}$), the figure displays the real components
of the $\mathrm{Q}$-function under constraint ($\mathrm{\operatorname{Im}[\mu_{1}]=\operatorname{Im}[\mu_{2}]=0}$)
(left: $\mathrm{a\lyxmathsym{–}d}$) and the imaginary components
under constraint ($\mathrm{\operatorname{Re}[\mu_{1}]=\operatorname{Re}[\mu_{2}]=0}$)
(right: $\mathrm{e\lyxmathsym{–}h}$). Initially, at $\mathrm{s=0}$
(non-interacting inital state), as shown in Figs. \ref{fig:Q-function-T1}
(a) and (e), the $\mathrm{Q}$-function exhibits both centrosymmetric
unimodal Gaussian distributions and asymmetrically squeezed unimodal
Gaussian distributions, as $\mathrm{s}$ increases from $\mathrm{0}$
($\mathrm{s\neq0}$) to $\mathrm{1}$, it can be seen from Figs. \ref{fig:Q-function-T1}
(b-c) and (f-g) that the $\mathrm{Q}$-function gradually splits from
a unimodal Gaussian distribution with two overlapping Gaussian wave
packets into a bimodal structure, eventually separating completely
into two independent wave packets at $\mathrm{s=2}$ {[}Figs. \ref{fig:Q-function-T1}
(d) and (h){]}.

As demonstrated in Fig. \ref{fig:Q-function-T1}, an increase in the
coupling parameter ($\mathrm{s}$) results in a transition of the
system from the weak-measurement regime, characterised by a dominance
of quantum coherence, to the strong-measurement regime, which is dominated
by decoherence. In the real part, the structure evolves from unimodal
to a superimposed bimodal structure, thereby reflecting quantum superposition.
As the value of s increases, the bimodal structure undergoes a complete
separation, accompanied by the complete dissipation of interference
(i.e. decoherence is achieved). However, the imaginary part of the
wavefunction displays asymmetric squeezing, thereby revealing phase-sensitive
coherence, a property that is unique to squeezed states. When the
two wave peaks in the imaginary part fully separate, the squeezing
effect significantly enhances, implying persistent quantum correlations
in orthogonal degrees of freedom. It is evident that the two-mode
system further reveals an imbalance in the response to measurement
across varying degrees of freedom. The real part attains projective
strong measurement via prior decoherence, while the imaginary part
preserves the characteristics of a squeezed coherent state (e.g.,
entanglement ). This provides a new experimental dimension for manipulating
many-body quantum measurements.

\section{CONCLUSION AND REMARKS\label{sec:5}}

In this work, we conduct a systematic investigation into the regulation
of quantum characteristics of SPSSV and TMSV states through post-selected
von Neumann measurement. By constructing theoretical models, developing
numerical simulations, and analyzing phase space evolution, we elucidate
the physical mechanism underlying the weak-to-strong measurement transition
and its application potential in quantum precision measurement.

First, we establish a von Neumann framework measurement theory model,
mathematically representing the system-instrument interaction Hamiltonian
through time-dependent coupling terms. Through post-selection protocols,
we derive analytical expressions for normalized pointer states. Subsequently,
we introduce non-classical characteristic evaluation metrics including
Wigner-Yanase skew information, AS squeezing parameters, and photon
statistics.

We then perform numerical simulations systematically investigate the
coupling strength parameter's regulatory effects on SPSSV and TMSV
states quantum features: weak value amplification significantly enhances
non-classical properties in low-squeezing parameter regimes, while
AS squeezing exhibits non-monotonic evolution. Crucially, we demonstrate
that adjusting system-pointer coupling strength enables continuous
transition from Aharonov weak measurement to von Neumann projection
measurement across all coupling regimes.

Finally, we derive an analytical expression for pointer position displacement
in weak measurement regimes, numerical calculations for SPSSV and
TMSV states confirm theoretical predictions of pointer displacements
in both position and momentum. Phase-space analysis via the Husimi-Kano
$\mathrm{Q}$ function further reveals continuous transitions: from
Gaussian weak-measurement distributions to monomodal strong-measurement
structures.

Our results demonstrate that precise control of system-instrument
coupling parameter s enables simultaneous preservation of SPSSV and
TMSV states deterministic single-photon emission advantages with optimized
noise resistance and information capacity. This provides novel methodologies
for designing and controlling quantum resources in quantum precision
measurement and information processing. Future extensions may include
joint measurements of multi-photon entangled states\citep{PhysRevA.110.052611}
and dynamic behavior studies of SPSSV and TMSV states in quantum networks\citep{fan2025quantum}
and quantum sensing\citep{PhysRevA.90.013821,SCHNABEL20171}, advancing
post-selection measurement technology toward practical quantum technological
implementations.
\begin{acknowledgments}
acknowledgments The research is supported by the National Natural Science Foundation of China (No. 12174379, No. E31Q02BG, No. 12365005), the Chinese Academy of Sciences (No. E0SEBB11, No. E27RBB11), the Innovation Program for Quantum Science and Technology (No. 2021ZD0302300) and Chinese Academy of Sciences Project for Young Scientists in Basic Research (YSBR-090). 
\end{acknowledgments}

\appendix

\section{\label{sec:A1}Related expression}
In this study, we rigorously derive closed-form analytical
expressions for all relevant quantities without resorting to approximations.
However, due to their inherent mathematical complexity and length,
these results are not ideally suited for inclusion in the main text
to preserve its readability. For completeness and transparency, we
present the full derivations and detailed formulations in this Appendix.
\begin{widetext} 

1. The expectation values of observables $\mathrm{\hat{a}}$, $\mathrm{\hat{a}^{2}}$,
$\mathrm{\hat{a}^{\dagger}\hat{a}}$, and $\mathrm{\hat{a}^{\dagger2}\hat{a}^{2}}$
in the eigenstates of SPSSV states ($\mathrm{\vert\Phi_{S}\rangle}$)
are rigorously determined through closed-form analytical expressions.

\begin{equation}
\mathrm{\langle\hat{a}\rangle}\mathrm{=2\vert\lambda\vert^{2}\left[\operatorname{Re}[\langle\hat{\sigma}_{x}\rangle_{w}]s-i\operatorname{Im}[\langle\hat{\sigma}_{x}\rangle_{w}]h_{1}(s)\right]},
\end{equation}

\begin{equation}
\mathrm{\langle\hat{a}^{2}\rangle}\mathrm{=2\vert\lambda\vert^{2}\left[\frac{1}{2}\left(1+\vert\langle\hat{\sigma}_{x}\rangle_{w}\vert^{2}\right)\left(3e^{i\theta}\sinh(2r)+\frac{s^{2}}{2}\right)+\left(1-\vert\langle\hat{\sigma}_{x}\rangle_{w}\vert^{2}\right)\operatorname{Re}[h_{2}(s)]\right]},
\end{equation}

\begin{equation}
\mathrm{\langle\hat{a}^{\dagger}\hat{a}\rangle}\mathrm{=2\vert\lambda\vert^{2}\left[\left(1+\vert\langle\hat{\sigma}_{x}\rangle_{w}\vert^{2}\right)\left(1+3\sinh^{2}(r)+\frac{s^{2}}{4}\right)+\left(1-\vert\langle\hat{\sigma}_{x}\rangle_{w}\vert^{2}\right)\operatorname{Re}[h_{3}(s)]\right]},
\end{equation}

\[
\mathrm{\langle a^{\dagger2}a^{2}\rangle}\mathrm{=2\vert\lambda\vert^{2}\left[(1+|\langle\sigma_{x}\rangle_{w}|^{2})h_{4}(s)+2\left(1-\vert\langle\hat{\sigma}_{x}\rangle_{w}\vert^{2}\right)\operatorname{Re}[h_{5}(s)]\right]}.
\]

respectively. Here, the $\mathrm{h_{1}(s)\sim h_{5}(s)}$ given by 

\begin{subequations}

\begin{equation}
\mathrm{h_{1}(s)}\mathrm{=\left\{ \beta\cosh(r)(\beta^{2}e^{-i\theta}\coth(r)+3)+\frac{s}{2}(2\beta^{2}e^{-i\theta}\coth(r)-\vert\beta\vert^{2}+3)\right\} e^{-\frac{1}{2}\vert\beta\vert^{2}}},
\end{equation}

\begin{align}
\mathrm{h_{2}(s)} & \mathrm{=\Bigg\{\beta^{2}\cosh^{2}(r)\left[\beta^{2}e^{-i\theta}\coth(r)+6\right]+\frac{3}{2}e^{i\theta}\sinh(2r)-2s\beta\cosh(r)\left(\beta^{2}e^{-i\theta}\coth(r)+3\right)}\nonumber \\
 & \mathrm{\quad+\frac{s^{2}}{4}\left(2\beta^{2}e^{-i\theta}\coth(r)-\vert\beta\vert^{2}+3\right)\Bigg\} e^{-\frac{1}{2}\vert\beta\vert^{2}}},
\end{align}

\begin{align}
\mathrm{h_{3}(s)} & \mathrm{=\bigg\{\beta^{2}e^{-i\theta}\left[\coth(r)\cosh^{2}(r)+\frac{5}{2}\sinh(2r)+\beta^{2}e^{-i\theta}\cosh^{2}(r)\right]+1+3\sinh^{2}(r)}\nonumber \\
 & \mathrm{\quad+\frac{3s}{2}\beta e^{-i\theta}\left[\left(\ensuremath{\coth}(r)+\beta^{2}e^{-i\theta}\right)\cosh(r)+3\sinh(r)\right]+\frac{s^{2}}{2}e^{-i\theta}\left[\coth(r)+\beta^{2}e^{-i\theta}\right]}\nonumber \\
 & \mathrm{\quad+\frac{s}{2}\beta\cosh(r)\left[e^{-i\theta}\beta^{2}\ensuremath{\coth}(r)+3\right]+\frac{s^{2}}{4}\left(2\beta^{2}e^{-i\theta}\coth(r)-\vert\beta\vert^{2}+3\right)\bigg\} e^{-\frac{1}{2}\vert\beta\vert^{2}}},
\end{align}

\begin{equation}
\mathrm{h_{4}(s)}\mathrm{=\left\{ 3\sinh^{2}(r)\left(3+5\sinh^{2}(r)\right)+\frac{s^{2}}{2}\left[\frac{3}{2}\cos(\theta)\sinh(2r)+2(1+3\sinh^{2}(r))\right]+\frac{s^{4}}{16}\right\} e^{-\frac{1}{2}\vert\beta\vert^{2}}},
\end{equation}

\begin{align}
\mathrm{h_{5}(s)} & \mathrm{=\bigg\{\left(k_{1}+sk_{2}\right)+s\left[\left(k_{2}+sk_{3}\right)+\left(k_{4}+sk_{9}\right)\right]+\frac{s^{2}}{4}\left[\left(k_{3}+sk_{5}\right)+\left(k_{8}+sk_{6}\right)+4\left(k_{9}+sk_{10}\right)\right]}\nonumber \\
 & \quad\mathrm{+\frac{s^{3}}{4}\left[\left(k_{10}+sk_{11}\right)+\left(k_{6}+sk_{7}\right)\right]+\frac{s^{4}}{16}\bigg\} e^{-\frac{1}{2}\vert\beta\vert^{2}}}.
\end{align}

\end{subequations}

where the derived analytical expression for $\mathrm{K_{1}}$to $\mathrm{K_{11}}$
is:

\begin{subequations}

\begin{align}
\mathrm{k_{1}} & =\mathrm{\beta^{2}e^{-i\theta}\cosh(r)}\{\mathrm{\beta^{4}e^{-2i\theta}\sinh(r)\cosh^{2}(r)+3\beta^{2}e^{-i\theta}\cosh(r)\left(1+5\sinh^{2}(r)\right)}\nonumber \\
 & \mathrm{\quad+9\sinh(r)\left(2+5\sinh^{2}(r)\right)}\}\mathrm{+3\sinh^{2}(r)\left(3+5\sinh^{2}(r)\right)},
\end{align}

\begin{align}
\mathrm{k_{2}} & =\mathrm{\beta^{5}e^{-3i\theta}\sinh(r)\cosh^{2}(r)+\beta^{3}e^{-2i\theta}\cosh(r)\left(3+10\sinh^{2}(r)\right)}\\
 & \mathrm{\quad+3\beta e^{-i\theta}\sinh(r)\left(3+5\sinh^{2}(r)\right)},\nonumber 
\end{align}

\begin{equation}
\mathrm{k_{3}}=\mathrm{e^{-i\theta}}\{\mathrm{\beta^{4}\frac{e^{-2i\theta}}{2}\sinh(2r)+3\beta^{2}e^{-i\theta}\cosh(2r)+\frac{3}{2}\sinh(2r)}\},
\end{equation}

\begin{align}
\mathrm{k_{4}} & =\beta^{5}e^{-2i\theta}\cosh^{3}(r)+\beta^{3}e^{-i\theta}\cosh^{2}(r)\left\{ \cosh(r)\coth(r)+9\sinh(r)\right\} \nonumber \\
 & \mathrm{\quad+3\beta\cosh(r)\left(\cosh^{2}(r)+4\sinh^{2}(r)\right)},
\end{align}

\begin{equation}
\mathrm{k_{5}}=\mathrm{\beta e^{-2i\theta}[\beta^{2}e^{-i\theta}\sinh(r)+3\cosh(r)]},
\end{equation}

\begin{equation}
\mathrm{k_{6}}=\mathrm{\beta\cosh(r)[\beta^{2}e^{-i\theta}\coth(r)+3]},
\end{equation}
\begin{equation}
\mathrm{k_{7}}=\mathrm{\beta^{2}e^{-i\theta}\coth(r)+1},
\end{equation}
\begin{equation}
\mathrm{k_{8}}=\mathrm{\beta^{2}\cosh^{2}(r)\left(\beta^{2}e^{-i\theta}\coth(r)+6\right)+\frac{3}{2}e^{i\theta}\sinh(2r)},
\end{equation}
\begin{align}
\mathrm{k_{9}} & =\mathrm{\beta^{2}e^{-i\theta}\left(\coth(r)\cosh^{2}(r)+5\sinh(r)\cosh(r)+\beta^{2}e^{-i\theta}\cosh^{2}(r)\right)+1+3\sinh^{2}(r),}
\end{align}
\begin{equation}
\mathrm{k_{10}}=\mathrm{\beta e^{-i\theta}}\left\{ \mathrm{\sinh^{-1}(r)+3\sinh(r)+\beta^{2}e^{-i\theta}\cosh(r)}\right\} ,
\end{equation}
\begin{equation}
\mathrm{k_{11}}=\mathrm{e^{-i\theta}\left(\coth(r)+\beta^{2}e^{-i\theta}\right)}.
\end{equation}

\end{subequations}

2. Expectation values $\mathrm{\hat{a}^{\dagger}\hat{a}}$, $\mathrm{\hat{b}^{\dagger}\hat{b}}$,
$\mathrm{\hat{a}\hat{b}}$, $\mathrm{\hat{a}^{\dagger}\hat{a}\hat{b}^{\dagger}\hat{b}}$
and $\mathrm{\hat{a}^{2}\hat{b}^{2}}$ for TMSV states ($\mathrm{\vert\Phi\rangle_{T}}$)
eigenstates are analytic

\begin{equation}
\mathrm{\langle\hat{a}^{\dagger}\hat{a}\rangle}=\mathrm{\frac{\vert\kappa\vert^{2}}{2}\Bigg\{\left(1+\vert\langle\hat{\sigma}_{x}\rangle_{w}\vert^{2}\right)\left(\sinh^{2}\eta+\frac{s^{2}}{4}\right)+\left(1-\vert\langle\hat{\sigma}_{x}\rangle_{w}\vert^{2}\right)\left(\sinh^{2}\eta\left(1-s^{2}\cosh^{2}\eta\right)-\frac{s^{2}}{4}\right)K\Bigg\}},
\end{equation}

\begin{equation}
\mathrm{\langle\hat{b}^{\dagger}\hat{b}\rangle=\frac{\vert\kappa\vert^{2}}{2}\Bigg\{\left(1+\vert\langle\hat{\sigma}_{x}\rangle_{w}\vert^{2}\right)\left(\sinh^{2}\eta+\frac{s^{2}}{4}\right)+\left(1-\vert\langle\hat{\sigma}_{x}\rangle_{w}\vert^{2}\right)\left[\sinh^{2}\eta\left(1-s^{2}\cosh^{2}\eta\right)+\frac{s^{2}}{4}\right]K\Bigg\}},
\end{equation}

\begin{align}
\mathrm{\langle\hat{a}\hat{b}\rangle} & =\mathrm{\frac{\vert\kappa\vert^{2}}{4}\Bigg\{\left(1-\vert\langle\hat{\sigma}_{x}\rangle_{w}\vert^{2}\right)\left[\sinh\left(2\eta\right)\left(1-s^{2}\cosh^{2}\eta\right)-s^{2}\left(\cosh^{2}\eta-\frac{1}{2}\sinh(2\eta)+\frac{1}{2}\right)\right]K}\nonumber \\
 & \quad\mathrm{+\left(1+\vert\langle\hat{\sigma}_{x}\rangle_{w}\vert^{2}\right)\left(\sinh(2\eta)+\frac{s^{2}}{2}\right)\Bigg\}},
\end{align}

\begin{align}
\mathrm{\langle\hat{a}^{\dagger}\hat{a}\hat{b}^{\dagger}\hat{b}\rangle} & =\mathrm{\frac{\vert\kappa\vert^{2}}{2}\Bigg\{\left(1+\vert\langle\hat{\sigma}_{x}\rangle_{w}\vert^{2}\right)J_{0}+\left(1-\vert\langle\hat{\sigma}_{x}\rangle_{w}\vert^{2}\right)\bigg[J_{1}-\frac{s}{2}\left(J_{2}+J_{3}+J_{4}+J_{5}\right)}\nonumber \\
 & \quad\mathrm{+\frac{s^{2}}{4}\left[2\sinh^{2}\eta\left(1-s^{2}\cosh^{2}\eta\right)+\sinh(2\eta)\right]-\frac{s^{4}}{16}\bigg]K\Bigg\}},
\end{align}

\begin{align}
\mathrm{\langle\hat{a}^{2}\hat{b}^{2}\rangle} & =\mathrm{\frac{\vert\kappa\vert^{2}}{2}\Bigg\{\left[1+\vert\langle\hat{\sigma}_{x}\rangle_{w}\vert^{2}\right]\frac{1}{2}\left(\sinh(2\eta)\left(1+s^{2}\right)+\frac{s^{4}}{8}\right)}\nonumber \\
 & \quad\mathrm{+\left[1-\vert\langle\hat{\sigma}_{x}\rangle_{w}\vert^{2}\right]\bigg\{ g_{0}-s(g_{1}+J_{5})+\frac{s^{2}}{4}(g_{2}+g_{3}+4g_{4})\mathrm{-\frac{s^{4}}{4}\left[\cosh\eta(\cosh\eta-\sinh\eta)+\frac{1}{4}\right]K\bigg\}\Bigg\}},}
\end{align}

with$\mathrm{g_{0}}$to $\mathrm{g_{4}}$ and $\mathrm{J_{0}}$to
$\mathrm{J_{5}}$analytically expressed as:

\begin{subequations}

\begin{equation}
\mathrm{g_{0}=\frac{\sinh^{2}(2\eta)}{4}\left\{ s^{4}\cosh^{4}\eta-4s^{2}\cosh^{2}\eta+2\right\} K},
\end{equation}

\begin{equation}
\mathrm{g_{1}=s\frac{\sinh(2\eta)\cosh^{2}\eta}{2}\left\{ 2-s^{2}\cosh^{2}\eta\right\} k},
\end{equation}

\begin{equation}
\mathrm{g_{2}=ks^{2}\cosh^{4}\eta},
\end{equation}

\begin{equation}
\mathrm{g_{3}=ks^{2}\frac{\sinh^{2}(2\eta)}{4}},
\end{equation}

\begin{equation}
\mathrm{g_{4}=\frac{\sinh(2\eta)}{2}\left\{ 1-s^{2}\cosh^{2}\eta\right\} k},
\end{equation}

\end{subequations}

and

\begin{subequations}
\begin{equation}
\mathrm{J_{0}=\sinh^{2}\eta\cosh(2\eta)+\frac{s^{4}}{16}+\frac{s^{2}}{2}\sinh\eta(\sinh\eta+\cosh\eta)},
\end{equation}
\begin{equation}
\mathrm{J_{1}=\sinh^{2}\eta\cosh(2\eta)+\frac{s^{2}\sinh^{2}(2\eta)}{4}\left(\frac{s^{2}}{4}\sinh^{2}(2\eta)-4\sinh^{2}\eta-1\right)},
\end{equation}
\begin{equation}
\mathrm{J_{2}=\frac{s\sinh(2\eta)}{8}\left(4\cosh(2\eta)-s^{2}\sinh^{2}(2\eta)\right)},
\end{equation}
\begin{equation}
\mathrm{J_{3}=\frac{s\sinh(2\eta)\sinh^{2}\eta}{2}\left(s^{2}\cosh^{2}\eta-2\right)},
\end{equation}
\begin{equation}
\mathrm{J_{4}=\sinh^{2}\lambda\left(s^{2}\sinh^{2}\eta\cosh^{2}\eta-\cosh(2\eta)\right)},
\end{equation}
\begin{equation}
\mathrm{J_{5}=\frac{s\sinh^{2}(2\eta)}{4}\left(2-s^{2}\cosh^{2}\eta\right)}.
\end{equation}

\end{subequations}

\section{\label{sec:B}Derivations of the Husimi-Kano $\mathrm{Q}$ Function}

To facilitate subsequent calculations, we first derived the expression
for the probability amplitude of finding $\mathrm{n}$ photons in
a squeezed coherent state and expressed it as:

\begin{equation}
\mathrm{\langle n|s,\xi\rangle=\mathrm{i}^{n}\sqrt{\frac{\left(\mathrm{e}^{\mathrm{i}\theta/2}\tanh r\right)^{n}}{2^{n}n!\cosh r}}H_{n}\left[\frac{\mathrm{i}}{2}\mathrm{e}^{-\mathrm{i}\theta/2}\sqrt{\frac{2}{\tanh r}}(s^{*}\mathrm{e}^{\mathrm{i}\theta}\tanh r-s)\right]e^{\frac{1}{2}(s^{*2}\mathrm{e}^{\mathrm{i}\theta}\tanh r-|s|^{2})},}
\end{equation}
where $\mathrm{H_{n}(x)}$ denotes the Hermite polynomial of order
$\mathrm{n}$, Moreover, it is imperative to undertake a thorough
analysis of the underlying transformations in coherent states, with
a particular focus on those characterised by the combination of annihilation
and creation operators\citep{PhysRevA.43.492}
\begin{align}
\mathrm{a^{\dagger m}|\mu\rangle} & =\mathrm{a^{\dagger m}D\left(\mu\right)|0\rangle}\nonumber \\
 & =\mathrm{D(\mu)D^{\dagger}(\mu)a^{\dagger m}D(\mu)|0\rangle}\nonumber \\
 & =\mathrm{D(\mu)(a^{\dagger}+\mu^{*})^{m}|0\rangle},
\end{align}
and

\begin{equation}
\mathrm{\langle\mu|\xi,s\rangle=\frac{1}{\sqrt{\cosh r}}\exp\left[\mathrm{e}^{\mathrm{i}\theta}\frac{(\mu^{*}-s^{*})^{2}}{2}\tanh r\right]\exp\left[-\frac{1}{2}|\mu-s|^{2}-\frac{1}{2}(\mu s^{*}-\mu^{*}s)\right]}.
\end{equation}

Based on the derivation in Sec. \ref{sec:4}(B) (see Eq. (\ref{eq:Q-fuction})),
the relevant computational expressions are presented as follows:

\begin{align}
\mathrm{\langle\mu\vert\hat{D}\left(\frac{s}{2}\right)\vert\phi\rangle} & =\mathrm{\frac{1}{\sinh r}\langle\mu\vert\hat{D}\left(\frac{s}{2}\right)\hat{a}\hat{S}(\xi)\vert0\rangle}\nonumber \\
 & =\mathrm{\frac{1}{\sinh r}e^{-\frac{s}{2}i\operatorname{Im}[\mu]}\langle\mu-\frac{s}{2}|\hat{a}\hat{S}(\xi)\vert0\rangle}\nonumber \\
 & =\mathrm{\frac{1}{\sinh r}e^{-\frac{s}{2}i\operatorname{Im}[\mu]}\left[\langle1|+\langle0|\left(\mu-\frac{s}{2}\right)\right]\hat{D}^{\dagger}\left(\mu-\frac{s}{2}\right)\hat{S}(\xi)\vert0\rangle}\nonumber \\
 & =\mathrm{\frac{1}{\sinh r}e^{-\frac{s}{2}i\operatorname{Im}[\mu]}\left[\langle1|\hat{D}^{\dagger}(\mu-\frac{s}{2})\hat{S}(\xi)\vert0\rangle+\left(\mu-\frac{s}{2}\right)\langle0|\hat{D}^{\dagger}(\mu-\frac{s}{2})\hat{S}(\xi)\vert0\rangle\right]}\nonumber \\
 & =\mathrm{\frac{1}{\sinh r}e^{-\frac{s}{2}i\operatorname{Im}[\mu]}\left[\langle1|\hat{D}(\frac{s}{2}-\mu)\hat{S}(\xi)\vert0\rangle+\left(\mu-\frac{s}{2}\right)\langle0|\hat{D}(\frac{s}{2}-\mu)\hat{S}(\xi)\vert0\rangle\right]},
\end{align}
and

\begin{align}
\mathrm{\langle\mu\vert\hat{D}^{\dagger}\left(\frac{s}{2}\right)\vert\phi\rangle} & =\mathrm{\frac{1}{\sinh r}e^{\frac{s}{2}i\operatorname{Im}[\mu]}\left[\langle1|\hat{D}^{\dagger}(\frac{s}{2}+\mu)\hat{S}(\xi)\vert0\rangle+\left(\mu+\frac{s}{2}\right)\langle0|\hat{D}^{\dagger}(\frac{s}{2}+\mu)\hat{S}(\xi)\vert0\rangle\right]},
\end{align}
here

\begin{align}
 & \mathrm{\langle1|\hat{D}\left(\frac{s}{2}-\mu\right)\hat{S}(\xi)\vert0\rangle}\nonumber \\
 & =\mathrm{\exp\left((\frac{s}{2}-\mu)^{*2}\frac{\mathrm{e}^{\mathrm{i}\theta}}{2}\tanh r-\frac{1}{2}|\frac{s}{2}-\mu|^{2}\right)\mathrm{i}\sqrt{\frac{\tanh r\mathrm{e}^{\mathrm{i}\theta/2}}{2\cosh r}}}\nonumber \\
 & \quad\times\mathrm{H_{1}\left[-\frac{\mathrm{i}}{2}\mathrm{e}^{-\mathrm{i}\theta/2}\sqrt{\frac{2}{\tanh r}}\left[\left(\frac{s}{2}-\mu\right)-\mathrm{e}^{\mathrm{i}\theta}\tanh r\left(\frac{s}{2}-\mu\right)^{*}\right]\right]}\nonumber \\
 & =\mathrm{\exp\left((\frac{s}{2}-\mu)^{*2}\frac{\mathrm{e}^{\mathrm{i}\theta}}{2}\tanh r-\frac{1}{2}|\frac{s}{2}-\mu|^{2}\right)\mathrm{i}\sqrt{\frac{\tanh r\mathrm{e}^{\mathrm{i}\theta/2}}{2\cosh r}}}\times\mathrm{-\mathrm{i}\mathrm{e}^{-\mathrm{i}\theta/2}\sqrt{\frac{2}{\tanh r}}\left[\left(\frac{s}{2}-\mu\right)-\mathrm{e}^{\mathrm{i}\theta}\tanh r\left(\frac{s}{2}-\mu\right)^{*}\right]}\nonumber \\
 & =\mathrm{\exp\left((\frac{s}{2}-\mu)^{*2}\frac{\mathrm{e}^{\mathrm{i}\theta}}{2}\tanh r-\frac{1}{2}|\frac{s}{2}-\mu|^{2}\right)\sqrt{\frac{\tanh r\mathrm{e}^{\mathrm{i}\theta/2}}{2\cosh r}}}\mathrm{\mathrm{e}^{-\mathrm{i}\theta/2}\sqrt{\frac{2}{\tanh r}}\left[\left(\frac{s}{2}-\mu\right)-\mathrm{e}^{\mathrm{i}\theta}\tanh r\left(\frac{s}{2}-\mu\right)^{*}\right]}\nonumber \\
 & =\mathrm{\sqrt{\frac{\tanh r\mathrm{e}^{\mathrm{i}\theta/2}}{2\cosh r}}\times\sqrt{\frac{2\mathrm{e}^{-\mathrm{i}\theta}}{\tanh r}}\left\{ \left(\frac{s}{2}-\mu\right)-\mathrm{e}^{\mathrm{i}\theta}\tanh r\left(\frac{s}{2}-\mu\right)^{*}\right\} \mathrm{\exp\left[\frac{1}{2}\left[(\frac{s}{2}-\mu)^{*2}\mathrm{e}^{\mathrm{i}\theta}\tanh r-|\frac{s}{2}-\mu|^{2}\right]\right]}}\nonumber \\
 & =\mathrm{\sqrt{\frac{\mathrm{e}^{-\mathrm{i}\theta/2}}{\cosh r}}\left[\left(\frac{s}{2}-\mu\right)-\mathrm{e}^{\mathrm{i}\theta}\tanh r\left(\frac{s}{2}-\mu\right)^{*}\right]\exp\left[\frac{1}{2}\left[(\frac{s}{2}-\mu)^{*2}\mathrm{e}^{\mathrm{i}\theta}\tanh r-|\frac{s}{2}-\mu|^{2}\right]\right]},
\end{align}
and

\begin{align}
\mathrm{\langle1|\hat{D}^{\dagger}\left(\frac{s}{2}+\mu\right)\hat{S}(\xi)\vert0\rangle} & =\sqrt{\frac{\mathrm{e}^{-\mathrm{i}\theta/2}}{\cosh r}}\left[\left(-\frac{s}{2}-\mu\right)-\mathrm{e}^{\mathrm{i}\theta}\tanh r\left(-\frac{s}{2}-\mu\right)^{*}\right]\nonumber \\
 & \mathrm{\quad\times\exp\left[\frac{1}{2}\left[(-\frac{s}{2}-\mu)^{*2}\mathrm{e}^{\mathrm{i}\theta}\tanh r-|-\frac{s}{2}-\mu|^{2}\right]\right]},
\end{align}

\begin{equation}
\mathrm{\langle0|\hat{D}\left(\pm\frac{s}{2}-\mu\right)\hat{S}(\xi)\vert0\rangle=\sqrt{\frac{1}{\cosh r}}\exp\left[\frac{1}{2}\left[(\pm\frac{s}{2}-\mu)^{*2}\mathrm{e}^{\mathrm{i}\theta}\tanh r-|\pm\frac{s}{2}-\mu|^{2}\right]\right]}.
\end{equation}

Appendix ends here.

\end{widetext}

\bibliographystyle{apsrev4-1}
\bibliography{ref-SPSSVS-II}

\begin{thebibliography}{100}%
\makeatletter
\providecommand \@ifxundefined [1]{%
 \@ifx{#1\undefined}
}%
\providecommand \@ifnum [1]{%
 \ifnum #1\expandafter \@firstoftwo
 \else \expandafter \@secondoftwo
 \fi
}%
\providecommand \@ifx [1]{%
 \ifx #1\expandafter \@firstoftwo
 \else \expandafter \@secondoftwo
 \fi
}%
\providecommand \natexlab [1]{#1}%
\providecommand \enquote  [1]{``#1''}%
\providecommand \bibnamefont  [1]{#1}%
\providecommand \bibfnamefont [1]{#1}%
\providecommand \citenamefont [1]{#1}%
\providecommand \href@noop [0]{\@secondoftwo}%
\providecommand \href [0]{\begingroup \@sanitize@url \@href}%
\providecommand \@href[1]{\@@startlink{#1}\@@href}%
\providecommand \@@href[1]{\endgroup#1\@@endlink}%
\providecommand \@sanitize@url [0]{\catcode `\\12\catcode `\$12\catcode `\&12\catcode `\#12\catcode `\^12\catcode `\_12\catcode `\%12\relax}%
\providecommand \@@startlink[1]{}%
\providecommand \@@endlink[0]{}%
\providecommand \url  [0]{\begingroup\@sanitize@url \@url }%
\providecommand \@url [1]{\endgroup\@href {#1}{\urlprefix }}%
\providecommand \urlprefix  [0]{URL }%
\providecommand \Eprint [0]{\href }%
\providecommand \doibase [0]{http://dx.doi.org/}%
\providecommand \selectlanguage [0]{\@gobble}%
\providecommand \bibinfo  [0]{\@secondoftwo}%
\providecommand \bibfield  [0]{\@secondoftwo}%
\providecommand \translation [1]{[#1]}%
\providecommand \BibitemOpen [0]{}%
\providecommand \bibitemStop [0]{}%
\providecommand \bibitemNoStop [0]{.\EOS\space}%
\providecommand \EOS [0]{\spacefactor3000\relax}%
\providecommand \BibitemShut  [1]{\csname bibitem#1\endcsname}%
\let\auto@bib@innerbib\@empty
\bibitem [{\citenamefont {Briegel}\ \emph {et~al.}(2009)\citenamefont {Briegel}, \citenamefont {Browne}, \citenamefont {D{\"u}r}, \citenamefont {Raussendorf},\ and\ \citenamefont {Van~den Nest}}]{briegel2009measurement}%
  \BibitemOpen
  \bibfield  {author} {\bibinfo {author} {\bibfnamefont {H.~J.}\ \bibnamefont {Briegel}}, \bibinfo {author} {\bibfnamefont {D.~E.}\ \bibnamefont {Browne}}, \bibinfo {author} {\bibfnamefont {W.}~\bibnamefont {D{\"u}r}}, \bibinfo {author} {\bibfnamefont {R.}~\bibnamefont {Raussendorf}}, \ and\ \bibinfo {author} {\bibfnamefont {M.}~\bibnamefont {Van~den Nest}},\ }\href {\doibase 10.1038/nphys1157} {\bibfield  {journal} {\bibinfo  {journal} {Nat. Phys}\ }\textbf {\bibinfo {volume} {5}},\ \bibinfo {pages} {19} (\bibinfo {year} {2009})}\BibitemShut {NoStop}%
\bibitem [{\citenamefont {Taylor}\ \emph {et~al.}(2013)\citenamefont {Taylor}, \citenamefont {Janousek}, \citenamefont {Daria}, \citenamefont {Knittel}, \citenamefont {Hage}, \citenamefont {Bachor},\ and\ \citenamefont {Bowen}}]{taylor2013biological}%
  \BibitemOpen
  \bibfield  {author} {\bibinfo {author} {\bibfnamefont {M.~A.}\ \bibnamefont {Taylor}}, \bibinfo {author} {\bibfnamefont {J.}~\bibnamefont {Janousek}}, \bibinfo {author} {\bibfnamefont {V.}~\bibnamefont {Daria}}, \bibinfo {author} {\bibfnamefont {J.}~\bibnamefont {Knittel}}, \bibinfo {author} {\bibfnamefont {B.}~\bibnamefont {Hage}}, \bibinfo {author} {\bibfnamefont {H.-A.}\ \bibnamefont {Bachor}}, \ and\ \bibinfo {author} {\bibfnamefont {W.~P.}\ \bibnamefont {Bowen}},\ }\href {\doibase 10.1038/nphoton.2012.346} {\bibfield  {journal} {\bibinfo  {journal} {Nat. Photonics}\ }\textbf {\bibinfo {volume} {7}},\ \bibinfo {pages} {229} (\bibinfo {year} {2013})}\BibitemShut {NoStop}%
\bibitem [{\citenamefont {Zurek}(2003)}]{RevModPhys.75.715}%
  \BibitemOpen
  \bibfield  {author} {\bibinfo {author} {\bibfnamefont {W.~H.}\ \bibnamefont {Zurek}},\ }\href {\doibase 10.1103/RevModPhys.75.715} {\bibfield  {journal} {\bibinfo  {journal} {Rev. Mod. Phys.}\ }\textbf {\bibinfo {volume} {75}},\ \bibinfo {pages} {715} (\bibinfo {year} {2003})}\BibitemShut {NoStop}%
\bibitem [{\citenamefont {Aharonov}\ \emph {et~al.}(1988)\citenamefont {Aharonov}, \citenamefont {Albert},\ and\ \citenamefont {Vaidman}}]{PhysRevLett.60.1351}%
  \BibitemOpen
  \bibfield  {author} {\bibinfo {author} {\bibfnamefont {Y.}~\bibnamefont {Aharonov}}, \bibinfo {author} {\bibfnamefont {D.~Z.}\ \bibnamefont {Albert}}, \ and\ \bibinfo {author} {\bibfnamefont {L.}~\bibnamefont {Vaidman}},\ }\href {\doibase 10.1103/PhysRevLett.60.1351} {\bibfield  {journal} {\bibinfo  {journal} {Phys. Rev. Lett.}\ }\textbf {\bibinfo {volume} {60}},\ \bibinfo {pages} {1351} (\bibinfo {year} {1988})}\BibitemShut {NoStop}%
\bibitem [{\citenamefont {Wiseman}\ and\ \citenamefont {Milburn}(2014)}]{2014Quantum}%
  \BibitemOpen
  \bibfield  {author} {\bibinfo {author} {\bibfnamefont {H.~M.}\ \bibnamefont {Wiseman}}\ and\ \bibinfo {author} {\bibfnamefont {G.~J.}\ \bibnamefont {Milburn}},\ }\href@noop {} {\emph {\bibinfo {title} {Quantum Measurement and Control}}}\ (\bibinfo  {publisher} {Cambridge Universirty Press, Cambridge, England},\ \bibinfo {year} {2014})\BibitemShut {NoStop}%
\bibitem [{\citenamefont {Pang}\ and\ \citenamefont {Brun}(2015)}]{PhysRevA.92.012120}%
  \BibitemOpen
  \bibfield  {author} {\bibinfo {author} {\bibfnamefont {S.}~\bibnamefont {Pang}}\ and\ \bibinfo {author} {\bibfnamefont {T.~A.}\ \bibnamefont {Brun}},\ }\href {\doibase 10.1103/PhysRevA.92.012120} {\bibfield  {journal} {\bibinfo  {journal} {Phys. Rev. A}\ }\textbf {\bibinfo {volume} {92}},\ \bibinfo {pages} {012120} (\bibinfo {year} {2015})}\BibitemShut {NoStop}%
\bibitem [{\citenamefont {Xu}\ \emph {et~al.}(2021)\citenamefont {Xu}, \citenamefont {Xu}, \citenamefont {Jiang}, \citenamefont {Xu}, \citenamefont {Zheng}, \citenamefont {Wang}, \citenamefont {Zhang},\ and\ \citenamefont {Zhang}}]{PhysRevLett.127.180401}%
  \BibitemOpen
  \bibfield  {author} {\bibinfo {author} {\bibfnamefont {L.}~\bibnamefont {Xu}}, \bibinfo {author} {\bibfnamefont {H.}~\bibnamefont {Xu}}, \bibinfo {author} {\bibfnamefont {T.}~\bibnamefont {Jiang}}, \bibinfo {author} {\bibfnamefont {F.}~\bibnamefont {Xu}}, \bibinfo {author} {\bibfnamefont {K.}~\bibnamefont {Zheng}}, \bibinfo {author} {\bibfnamefont {B.}~\bibnamefont {Wang}}, \bibinfo {author} {\bibfnamefont {A.}~\bibnamefont {Zhang}}, \ and\ \bibinfo {author} {\bibfnamefont {L.}~\bibnamefont {Zhang}},\ }\href {\doibase 10.1103/PhysRevLett.127.180401} {\bibfield  {journal} {\bibinfo  {journal} {Phys. Rev. Lett.}\ }\textbf {\bibinfo {volume} {127}},\ \bibinfo {pages} {180401} (\bibinfo {year} {2021})}\BibitemShut {NoStop}%
\bibitem [{\citenamefont {Gillett}\ \emph {et~al.}(2010)\citenamefont {Gillett}, \citenamefont {Dalton}, \citenamefont {Lanyon}, \citenamefont {Almeida}, \citenamefont {Barbieri}, \citenamefont {Pryde}, \citenamefont {O'Brien}, \citenamefont {Resch}, \citenamefont {Bartlett},\ and\ \citenamefont {White}}]{PhysRevLett.104.080503}%
  \BibitemOpen
  \bibfield  {author} {\bibinfo {author} {\bibfnamefont {G.~G.}\ \bibnamefont {Gillett}}, \bibinfo {author} {\bibfnamefont {R.~B.}\ \bibnamefont {Dalton}}, \bibinfo {author} {\bibfnamefont {B.~P.}\ \bibnamefont {Lanyon}}, \bibinfo {author} {\bibfnamefont {M.~P.}\ \bibnamefont {Almeida}}, \bibinfo {author} {\bibfnamefont {M.}~\bibnamefont {Barbieri}}, \bibinfo {author} {\bibfnamefont {G.~J.}\ \bibnamefont {Pryde}}, \bibinfo {author} {\bibfnamefont {J.~L.}\ \bibnamefont {O'Brien}}, \bibinfo {author} {\bibfnamefont {K.~J.}\ \bibnamefont {Resch}}, \bibinfo {author} {\bibfnamefont {S.~D.}\ \bibnamefont {Bartlett}}, \ and\ \bibinfo {author} {\bibfnamefont {A.~G.}\ \bibnamefont {White}},\ }\href {\doibase 10.1103/PhysRevLett.104.080503} {\bibfield  {journal} {\bibinfo  {journal} {Phys. Rev. Lett.}\ }\textbf {\bibinfo {volume} {104}},\ \bibinfo {pages} {080503} (\bibinfo {year} {2010})}\BibitemShut {NoStop}%
\bibitem [{\citenamefont {Hofmann}\ \emph {et~al.}(2012)\citenamefont {Hofmann}, \citenamefont {Goggin}, \citenamefont {Almeida},\ and\ \citenamefont {Barbieri}}]{PhysRevA.86.040102}%
  \BibitemOpen
  \bibfield  {author} {\bibinfo {author} {\bibfnamefont {H.~F.}\ \bibnamefont {Hofmann}}, \bibinfo {author} {\bibfnamefont {M.~E.}\ \bibnamefont {Goggin}}, \bibinfo {author} {\bibfnamefont {M.~P.}\ \bibnamefont {Almeida}}, \ and\ \bibinfo {author} {\bibfnamefont {M.}~\bibnamefont {Barbieri}},\ }\href {\doibase 10.1103/PhysRevA.86.040102} {\bibfield  {journal} {\bibinfo  {journal} {Phys. Rev. A}\ }\textbf {\bibinfo {volume} {86}},\ \bibinfo {pages} {040102} (\bibinfo {year} {2012})}\BibitemShut {NoStop}%
\bibitem [{\citenamefont {Alves}\ \emph {et~al.}(2015)\citenamefont {Alves}, \citenamefont {Escher}, \citenamefont {de~Matos~Filho}, \citenamefont {Zagury},\ and\ \citenamefont {Davidovich}}]{PhysRevA.91.062107}%
  \BibitemOpen
  \bibfield  {author} {\bibinfo {author} {\bibfnamefont {G.~B.}\ \bibnamefont {Alves}}, \bibinfo {author} {\bibfnamefont {B.~M.}\ \bibnamefont {Escher}}, \bibinfo {author} {\bibfnamefont {R.~L.}\ \bibnamefont {de~Matos~Filho}}, \bibinfo {author} {\bibfnamefont {N.}~\bibnamefont {Zagury}}, \ and\ \bibinfo {author} {\bibfnamefont {L.}~\bibnamefont {Davidovich}},\ }\href {\doibase 10.1103/PhysRevA.91.062107} {\bibfield  {journal} {\bibinfo  {journal} {Phys. Rev. A}\ }\textbf {\bibinfo {volume} {91}},\ \bibinfo {pages} {062107} (\bibinfo {year} {2015})}\BibitemShut {NoStop}%
\bibitem [{\citenamefont {Shomroni}\ \emph {et~al.}(2013)\citenamefont {Shomroni}, \citenamefont {Bechler}, \citenamefont {Rosenblum},\ and\ \citenamefont {Dayan}}]{PhysRevLett.111.023604}%
  \BibitemOpen
  \bibfield  {author} {\bibinfo {author} {\bibfnamefont {I.}~\bibnamefont {Shomroni}}, \bibinfo {author} {\bibfnamefont {O.}~\bibnamefont {Bechler}}, \bibinfo {author} {\bibfnamefont {S.}~\bibnamefont {Rosenblum}}, \ and\ \bibinfo {author} {\bibfnamefont {B.}~\bibnamefont {Dayan}},\ }\href {\doibase 10.1103/PhysRevLett.111.023604} {\bibfield  {journal} {\bibinfo  {journal} {Phys. Rev. Lett.}\ }\textbf {\bibinfo {volume} {111}},\ \bibinfo {pages} {023604} (\bibinfo {year} {2013})}\BibitemShut {NoStop}%
\bibitem [{\citenamefont {Kim}\ \emph {et~al.}(2012)\citenamefont {Kim}, \citenamefont {Lee}, \citenamefont {Kwon},\ and\ \citenamefont {Kim}}]{kim2012protecting}%
  \BibitemOpen
  \bibfield  {author} {\bibinfo {author} {\bibfnamefont {Y.-S.}\ \bibnamefont {Kim}}, \bibinfo {author} {\bibfnamefont {J.-C.}\ \bibnamefont {Lee}}, \bibinfo {author} {\bibfnamefont {O.}~\bibnamefont {Kwon}}, \ and\ \bibinfo {author} {\bibfnamefont {Y.-H.}\ \bibnamefont {Kim}},\ }\href {\doibase 10.1038/nphys2178} {\bibfield  {journal} {\bibinfo  {journal} {Nat. Phys}\ }\textbf {\bibinfo {volume} {8}},\ \bibinfo {pages} {117} (\bibinfo {year} {2012})}\BibitemShut {NoStop}%
\bibitem [{\citenamefont {Pang}\ \emph {et~al.}(2014)\citenamefont {Pang}, \citenamefont {Brun}, \citenamefont {Wu},\ and\ \citenamefont {Chen}}]{PhysRevA.90.012108}%
  \BibitemOpen
  \bibfield  {author} {\bibinfo {author} {\bibfnamefont {S.}~\bibnamefont {Pang}}, \bibinfo {author} {\bibfnamefont {T.~A.}\ \bibnamefont {Brun}}, \bibinfo {author} {\bibfnamefont {S.}~\bibnamefont {Wu}}, \ and\ \bibinfo {author} {\bibfnamefont {Z.-B.}\ \bibnamefont {Chen}},\ }\href {\doibase 10.1103/PhysRevA.90.012108} {\bibfield  {journal} {\bibinfo  {journal} {Phys. Rev. A}\ }\textbf {\bibinfo {volume} {90}},\ \bibinfo {pages} {012108} (\bibinfo {year} {2014})}\BibitemShut {NoStop}%
\bibitem [{\citenamefont {Ritchie}\ \emph {et~al.}(1991)\citenamefont {Ritchie}, \citenamefont {Story},\ and\ \citenamefont {Hulet}}]{PhysRevLett.66.1107}%
  \BibitemOpen
  \bibfield  {author} {\bibinfo {author} {\bibfnamefont {N.~W.~M.}\ \bibnamefont {Ritchie}}, \bibinfo {author} {\bibfnamefont {J.~G.}\ \bibnamefont {Story}}, \ and\ \bibinfo {author} {\bibfnamefont {R.~G.}\ \bibnamefont {Hulet}},\ }\href {\doibase 10.1103/PhysRevLett.66.1107} {\bibfield  {journal} {\bibinfo  {journal} {Phys. Rev. Lett.}\ }\textbf {\bibinfo {volume} {66}},\ \bibinfo {pages} {1107} (\bibinfo {year} {1991})}\BibitemShut {NoStop}%
\bibitem [{\citenamefont {Anisimov}\ \emph {et~al.}(2010)\citenamefont {Anisimov}, \citenamefont {Raterman}, \citenamefont {Chiruvelli}, \citenamefont {Plick}, \citenamefont {Huver}, \citenamefont {Lee},\ and\ \citenamefont {Dowling}}]{PhysRevLett.104.103602}%
  \BibitemOpen
  \bibfield  {author} {\bibinfo {author} {\bibfnamefont {P.~M.}\ \bibnamefont {Anisimov}}, \bibinfo {author} {\bibfnamefont {G.~M.}\ \bibnamefont {Raterman}}, \bibinfo {author} {\bibfnamefont {A.}~\bibnamefont {Chiruvelli}}, \bibinfo {author} {\bibfnamefont {W.~N.}\ \bibnamefont {Plick}}, \bibinfo {author} {\bibfnamefont {S.~D.}\ \bibnamefont {Huver}}, \bibinfo {author} {\bibfnamefont {H.}~\bibnamefont {Lee}}, \ and\ \bibinfo {author} {\bibfnamefont {J.~P.}\ \bibnamefont {Dowling}},\ }\href {\doibase 10.1103/PhysRevLett.104.103602} {\bibfield  {journal} {\bibinfo  {journal} {Phys. Rev. Lett.}\ }\textbf {\bibinfo {volume} {104}},\ \bibinfo {pages} {103602} (\bibinfo {year} {2010})}\BibitemShut {NoStop}%
\bibitem [{\citenamefont {Ouyang}\ \emph {et~al.}(2016)\citenamefont {Ouyang}, \citenamefont {Wang},\ and\ \citenamefont {Zhang}}]{Ouyang:16}%
  \BibitemOpen
  \bibfield  {author} {\bibinfo {author} {\bibfnamefont {Y.}~\bibnamefont {Ouyang}}, \bibinfo {author} {\bibfnamefont {S.}~\bibnamefont {Wang}}, \ and\ \bibinfo {author} {\bibfnamefont {L.}~\bibnamefont {Zhang}},\ }\href {\doibase 10.1364/JOSAB.33.001373} {\bibfield  {journal} {\bibinfo  {journal} {J. Opt. Soc. Am. B}\ }\textbf {\bibinfo {volume} {33}},\ \bibinfo {pages} {1373} (\bibinfo {year} {2016})}\BibitemShut {NoStop}%
\bibitem [{\citenamefont {Yonezawa}\ and\ \citenamefont {Furusawa}(2010)}]{RN1927}%
  \BibitemOpen
  \bibfield  {author} {\bibinfo {author} {\bibfnamefont {H.}~\bibnamefont {Yonezawa}}\ and\ \bibinfo {author} {\bibfnamefont {A.}~\bibnamefont {Furusawa}},\ }\href {\doibase 10.1134/S0030400X10020189} {\bibfield  {journal} {\bibinfo  {journal} {Opt Spectrosc+}\ }\textbf {\bibinfo {volume} {108}},\ \bibinfo {pages} {288} (\bibinfo {year} {2010})}\BibitemShut {NoStop}%
\bibitem [{\citenamefont {Zhang}\ and\ \citenamefont {Duan}(2014)}]{Zhang2014}%
  \BibitemOpen
  \bibfield  {author} {\bibinfo {author} {\bibfnamefont {Z.}~\bibnamefont {Zhang}}\ and\ \bibinfo {author} {\bibfnamefont {L.~M.}\ \bibnamefont {Duan}},\ }\href {\doibase 10.1088/1367-2630/16/10/103037} {\bibfield  {journal} {\bibinfo  {journal} {New. J. Phys}\ }\textbf {\bibinfo {volume} {16}},\ \bibinfo {pages} {103037} (\bibinfo {year} {2014})}\BibitemShut {NoStop}%
\bibitem [{\citenamefont {Blais}\ \emph {et~al.}(2020)\citenamefont {Blais}, \citenamefont {Girvin},\ and\ \citenamefont {Oliver}}]{RN1928}%
  \BibitemOpen
  \bibfield  {author} {\bibinfo {author} {\bibfnamefont {A.}~\bibnamefont {Blais}}, \bibinfo {author} {\bibfnamefont {S.~M.}\ \bibnamefont {Girvin}}, \ and\ \bibinfo {author} {\bibfnamefont {W.~D.}\ \bibnamefont {Oliver}},\ }\href {\doibase 10.1038/s41567-020-0806-z} {\bibfield  {journal} {\bibinfo  {journal} {Nat. Phys}\ }\textbf {\bibinfo {volume} {16}},\ \bibinfo {pages} {247} (\bibinfo {year} {2020})}\BibitemShut {NoStop}%
\bibitem [{\citenamefont {A.Nielsen}\ and\ \citenamefont {L.Chuang}(2010)}]{Chuang2010}%
  \BibitemOpen
  \bibfield  {author} {\bibinfo {author} {\bibfnamefont {M.}~\bibnamefont {A.Nielsen}}\ and\ \bibinfo {author} {\bibfnamefont {I.}~\bibnamefont {L.Chuang}},\ }\href@noop {} {\emph {\bibinfo {title} {Quantum Computation and Quantum Information}}}\ (\bibinfo  {publisher} {Cambridge Universirty Press, Cambridge, England},\ \bibinfo {year} {2010})\BibitemShut {NoStop}%
\bibitem [{\citenamefont {Schumacher}(1996)}]{PhysRevA.54.2614}%
  \BibitemOpen
  \bibfield  {author} {\bibinfo {author} {\bibfnamefont {B.}~\bibnamefont {Schumacher}},\ }\href {\doibase 10.1103/PhysRevA.54.2614} {\bibfield  {journal} {\bibinfo  {journal} {Phys. Rev. A}\ }\textbf {\bibinfo {volume} {54}},\ \bibinfo {pages} {2614} (\bibinfo {year} {1996})}\BibitemShut {NoStop}%
\bibitem [{\citenamefont {Schumacher}\ and\ \citenamefont {Nielsen}(1996)}]{PhysRevA.54.2629}%
  \BibitemOpen
  \bibfield  {author} {\bibinfo {author} {\bibfnamefont {B.}~\bibnamefont {Schumacher}}\ and\ \bibinfo {author} {\bibfnamefont {M.~A.}\ \bibnamefont {Nielsen}},\ }\href {\doibase 10.1103/PhysRevA.54.2629} {\bibfield  {journal} {\bibinfo  {journal} {Phys. Rev. A}\ }\textbf {\bibinfo {volume} {54}},\ \bibinfo {pages} {2629} (\bibinfo {year} {1996})}\BibitemShut {NoStop}%
\bibitem [{\citenamefont {Milburn}\ and\ \citenamefont {Braunstein}(1999)}]{PhysRevA.60.937}%
  \BibitemOpen
  \bibfield  {author} {\bibinfo {author} {\bibfnamefont {G.~J.}\ \bibnamefont {Milburn}}\ and\ \bibinfo {author} {\bibfnamefont {S.~L.}\ \bibnamefont {Braunstein}},\ }\href {\doibase 10.1103/PhysRevA.60.937} {\bibfield  {journal} {\bibinfo  {journal} {Phys. Rev. A}\ }\textbf {\bibinfo {volume} {60}},\ \bibinfo {pages} {937} (\bibinfo {year} {1999})}\BibitemShut {NoStop}%
\bibitem [{\citenamefont {Zhang}\ \emph {et~al.}(2003)\citenamefont {Zhang}, \citenamefont {Goh}, \citenamefont {Chou}, \citenamefont {Lodahl},\ and\ \citenamefont {Kimble}}]{PhysRevA.67.033802}%
  \BibitemOpen
  \bibfield  {author} {\bibinfo {author} {\bibfnamefont {T.~C.}\ \bibnamefont {Zhang}}, \bibinfo {author} {\bibfnamefont {K.~W.}\ \bibnamefont {Goh}}, \bibinfo {author} {\bibfnamefont {C.~W.}\ \bibnamefont {Chou}}, \bibinfo {author} {\bibfnamefont {P.}~\bibnamefont {Lodahl}}, \ and\ \bibinfo {author} {\bibfnamefont {H.~J.}\ \bibnamefont {Kimble}},\ }\href {\doibase 10.1103/PhysRevA.67.033802} {\bibfield  {journal} {\bibinfo  {journal} {Phys. Rev. A}\ }\textbf {\bibinfo {volume} {67}},\ \bibinfo {pages} {033802} (\bibinfo {year} {2003})}\BibitemShut {NoStop}%
\bibitem [{\citenamefont {Kraus}\ \emph {et~al.}(2003)\citenamefont {Kraus}, \citenamefont {Hammerer}, \citenamefont {Giedke},\ and\ \citenamefont {Cirac}}]{PhysRevA.67.042314}%
  \BibitemOpen
  \bibfield  {author} {\bibinfo {author} {\bibfnamefont {B.}~\bibnamefont {Kraus}}, \bibinfo {author} {\bibfnamefont {K.}~\bibnamefont {Hammerer}}, \bibinfo {author} {\bibfnamefont {G.}~\bibnamefont {Giedke}}, \ and\ \bibinfo {author} {\bibfnamefont {J.~I.}\ \bibnamefont {Cirac}},\ }\href {\doibase 10.1103/PhysRevA.67.042314} {\bibfield  {journal} {\bibinfo  {journal} {Phys. Rev. A}\ }\textbf {\bibinfo {volume} {67}},\ \bibinfo {pages} {042314} (\bibinfo {year} {2003})}\BibitemShut {NoStop}%
\bibitem [{\citenamefont {Doli\ifmmode~\acute{n}\else \'{n}\fi{}ska}\ \emph {et~al.}(2003)\citenamefont {Doli\ifmmode~\acute{n}\else \'{n}\fi{}ska}, \citenamefont {Buchler}, \citenamefont {Bowen}, \citenamefont {Ralph},\ and\ \citenamefont {Lam}}]{PhysRevA.68.052308}%
  \BibitemOpen
  \bibfield  {author} {\bibinfo {author} {\bibfnamefont {A.}~\bibnamefont {Doli\ifmmode~\acute{n}\else \'{n}\fi{}ska}}, \bibinfo {author} {\bibfnamefont {B.~C.}\ \bibnamefont {Buchler}}, \bibinfo {author} {\bibfnamefont {W.~P.}\ \bibnamefont {Bowen}}, \bibinfo {author} {\bibfnamefont {T.~C.}\ \bibnamefont {Ralph}}, \ and\ \bibinfo {author} {\bibfnamefont {P.~K.}\ \bibnamefont {Lam}},\ }\href {\doibase 10.1103/PhysRevA.68.052308} {\bibfield  {journal} {\bibinfo  {journal} {Phys. Rev. A}\ }\textbf {\bibinfo {volume} {68}},\ \bibinfo {pages} {052308} (\bibinfo {year} {2003})}\BibitemShut {NoStop}%
\bibitem [{\citenamefont {Caves}(1981)}]{PhysRevD.23.1693}%
  \BibitemOpen
  \bibfield  {author} {\bibinfo {author} {\bibfnamefont {C.~M.}\ \bibnamefont {Caves}},\ }\href {\doibase 10.1103/PhysRevD.23.1693} {\bibfield  {journal} {\bibinfo  {journal} {Phys. Rev. D}\ }\textbf {\bibinfo {volume} {23}},\ \bibinfo {pages} {1693} (\bibinfo {year} {1981})}\BibitemShut {NoStop}%
\bibitem [{\citenamefont {Hacker}\ \emph {et~al.}(2018)\citenamefont {Hacker}, \citenamefont {Welte}, \citenamefont {Daiss}, \citenamefont {Shaukat}, \citenamefont {Ritter}, \citenamefont {Li},\ and\ \citenamefont {Rempe}}]{Hacker2018DeterministicCO}%
  \BibitemOpen
  \bibfield  {author} {\bibinfo {author} {\bibfnamefont {B.}~\bibnamefont {Hacker}}, \bibinfo {author} {\bibfnamefont {S.}~\bibnamefont {Welte}}, \bibinfo {author} {\bibfnamefont {S.}~\bibnamefont {Daiss}}, \bibinfo {author} {\bibfnamefont {A.}~\bibnamefont {Shaukat}}, \bibinfo {author} {\bibfnamefont {S.}~\bibnamefont {Ritter}}, \bibinfo {author} {\bibfnamefont {L.}~\bibnamefont {Li}}, \ and\ \bibinfo {author} {\bibfnamefont {G.}~\bibnamefont {Rempe}},\ }\href {https://api.semanticscholar.org/CorpusID:92988275} {\bibfield  {journal} {\bibinfo  {journal} {Nat. Photonics}\ }\textbf {\bibinfo {volume} {13}},\ \bibinfo {pages} {110 } (\bibinfo {year} {2018})}\BibitemShut {NoStop}%
\bibitem [{\citenamefont {guo Meng}\ \emph {et~al.}(2012)\citenamefont {guo Meng}, \citenamefont {Wang}, \citenamefont {yi~Fan},\ and\ \citenamefont {suo Wang}}]{Meng:12}%
  \BibitemOpen
  \bibfield  {author} {\bibinfo {author} {\bibfnamefont {X.}~\bibnamefont {guo Meng}}, \bibinfo {author} {\bibfnamefont {Z.}~\bibnamefont {Wang}}, \bibinfo {author} {\bibfnamefont {H.}~\bibnamefont {yi~Fan}}, \ and\ \bibinfo {author} {\bibfnamefont {J.}~\bibnamefont {suo Wang}},\ }\href {\doibase 10.1364/JOSAB.29.003141} {\bibfield  {journal} {\bibinfo  {journal} {J. Opt. Soc. Am. B}\ }\textbf {\bibinfo {volume} {29}},\ \bibinfo {pages} {3141} (\bibinfo {year} {2012})}\BibitemShut {NoStop}%
\bibitem [{\citenamefont {Ra}\ \emph {et~al.}(2017)\citenamefont {Ra}, \citenamefont {Jacquard}, \citenamefont {Dufour}, \citenamefont {Fabre},\ and\ \citenamefont {Treps}}]{PhysRevX.7.031012}%
  \BibitemOpen
  \bibfield  {author} {\bibinfo {author} {\bibfnamefont {Y.-S.}\ \bibnamefont {Ra}}, \bibinfo {author} {\bibfnamefont {C.}~\bibnamefont {Jacquard}}, \bibinfo {author} {\bibfnamefont {A.}~\bibnamefont {Dufour}}, \bibinfo {author} {\bibfnamefont {C.}~\bibnamefont {Fabre}}, \ and\ \bibinfo {author} {\bibfnamefont {N.}~\bibnamefont {Treps}},\ }\href {\doibase 10.1103/PhysRevX.7.031012} {\bibfield  {journal} {\bibinfo  {journal} {Phys. Rev. X}\ }\textbf {\bibinfo {volume} {7}},\ \bibinfo {pages} {031012} (\bibinfo {year} {2017})}\BibitemShut {NoStop}%
\bibitem [{\citenamefont {Wu}\ \emph {et~al.}(2019)\citenamefont {Wu}, \citenamefont {Zhang}, \citenamefont {Xie}, \citenamefont {Ou}, \citenamefont {Chen}, \citenamefont {Wu},\ and\ \citenamefont {Chen}}]{PhysRevA.100.062111}%
  \BibitemOpen
  \bibfield  {author} {\bibinfo {author} {\bibfnamefont {C.-W.}\ \bibnamefont {Wu}}, \bibinfo {author} {\bibfnamefont {J.}~\bibnamefont {Zhang}}, \bibinfo {author} {\bibfnamefont {Y.}~\bibnamefont {Xie}}, \bibinfo {author} {\bibfnamefont {B.-Q.}\ \bibnamefont {Ou}}, \bibinfo {author} {\bibfnamefont {T.}~\bibnamefont {Chen}}, \bibinfo {author} {\bibfnamefont {W.}~\bibnamefont {Wu}}, \ and\ \bibinfo {author} {\bibfnamefont {P.-X.}\ \bibnamefont {Chen}},\ }\href {\doibase 10.1103/PhysRevA.100.062111} {\bibfield  {journal} {\bibinfo  {journal} {Phys. Rev. A}\ }\textbf {\bibinfo {volume} {100}},\ \bibinfo {pages} {062111} (\bibinfo {year} {2019})}\BibitemShut {NoStop}%
\bibitem [{\citenamefont {Pan}\ \emph {et~al.}(2020)\citenamefont {Pan}, \citenamefont {Zhang}, \citenamefont {Cohen}, \citenamefont {Wu}, \citenamefont {Chen},\ and\ \citenamefont {Davidson}}]{pan2020weak}%
  \BibitemOpen
  \bibfield  {author} {\bibinfo {author} {\bibfnamefont {Y.}~\bibnamefont {Pan}}, \bibinfo {author} {\bibfnamefont {J.}~\bibnamefont {Zhang}}, \bibinfo {author} {\bibfnamefont {E.}~\bibnamefont {Cohen}}, \bibinfo {author} {\bibfnamefont {C.-w.}\ \bibnamefont {Wu}}, \bibinfo {author} {\bibfnamefont {P.-X.}\ \bibnamefont {Chen}}, \ and\ \bibinfo {author} {\bibfnamefont {N.}~\bibnamefont {Davidson}},\ }\href {\doibase 10.1038/s41567-020-0973-y} {\bibfield  {journal} {\bibinfo  {journal} {Nat. Phys}\ }\textbf {\bibinfo {volume} {16}},\ \bibinfo {pages} {1206} (\bibinfo {year} {2020})}\BibitemShut {NoStop}%
\bibitem [{\citenamefont {Zhang}\ \emph {et~al.}(2015)\citenamefont {Zhang}, \citenamefont {Datta},\ and\ \citenamefont {Walmsley}}]{PhysRevLett.114.210801}%
  \BibitemOpen
  \bibfield  {author} {\bibinfo {author} {\bibfnamefont {L.}~\bibnamefont {Zhang}}, \bibinfo {author} {\bibfnamefont {A.}~\bibnamefont {Datta}}, \ and\ \bibinfo {author} {\bibfnamefont {I.~A.}\ \bibnamefont {Walmsley}},\ }\href {\doibase 10.1103/PhysRevLett.114.210801} {\bibfield  {journal} {\bibinfo  {journal} {Phys. Rev. Lett.}\ }\textbf {\bibinfo {volume} {114}},\ \bibinfo {pages} {210801} (\bibinfo {year} {2015})}\BibitemShut {NoStop}%
\bibitem [{\citenamefont {Jebli}\ \emph {et~al.}(2020)\citenamefont {Jebli}, \citenamefont {Amzioug}, \citenamefont {Ennadifi}, \citenamefont {Habiballah},\ and\ \citenamefont {Nassik}}]{Jebli_2020}%
  \BibitemOpen
  \bibfield  {author} {\bibinfo {author} {\bibfnamefont {L.}~\bibnamefont {Jebli}}, \bibinfo {author} {\bibfnamefont {M.}~\bibnamefont {Amzioug}}, \bibinfo {author} {\bibfnamefont {S.~E.}\ \bibnamefont {Ennadifi}}, \bibinfo {author} {\bibfnamefont {N.}~\bibnamefont {Habiballah}}, \ and\ \bibinfo {author} {\bibfnamefont {M.}~\bibnamefont {Nassik}},\ }\href {\doibase 10.1088/1674-1056/aba5fa} {\bibfield  {journal} {\bibinfo  {journal} {Chin. Phys. B}\ }\textbf {\bibinfo {volume} {29}},\ \bibinfo {pages} {110301} (\bibinfo {year} {2020})}\BibitemShut {NoStop}%
\bibitem [{\citenamefont {Turek}\ \emph {et~al.}(2023{\natexlab{a}})\citenamefont {Turek}, \citenamefont {Yuanbek},\ and\ \citenamefont {Abliz}}]{TUREK2023128663}%
  \BibitemOpen
  \bibfield  {author} {\bibinfo {author} {\bibfnamefont {Y.}~\bibnamefont {Turek}}, \bibinfo {author} {\bibfnamefont {J.}~\bibnamefont {Yuanbek}}, \ and\ \bibinfo {author} {\bibfnamefont {A.}~\bibnamefont {Abliz}},\ }\href {\doibase https://doi.org/10.1016/j.physleta.2023.128663} {\bibfield  {journal} {\bibinfo  {journal} {Phys. Lett. A}\ }\textbf {\bibinfo {volume} {462}},\ \bibinfo {pages} {128663} (\bibinfo {year} {2023}{\natexlab{a}})}\BibitemShut {NoStop}%
\bibitem [{\citenamefont {Glauber}(1963)}]{PhysRev.131.2766}%
  \BibitemOpen
  \bibfield  {author} {\bibinfo {author} {\bibfnamefont {R.~J.}\ \bibnamefont {Glauber}},\ }\href {\doibase 10.1103/PhysRev.131.2766} {\bibfield  {journal} {\bibinfo  {journal} {Phys. Rev.}\ }\textbf {\bibinfo {volume} {131}},\ \bibinfo {pages} {2766} (\bibinfo {year} {1963})}\BibitemShut {NoStop}%
\bibitem [{\citenamefont {Titulaer}\ and\ \citenamefont {Glauber}(1965)}]{PhysRev.140.B676}%
  \BibitemOpen
  \bibfield  {author} {\bibinfo {author} {\bibfnamefont {U.~M.}\ \bibnamefont {Titulaer}}\ and\ \bibinfo {author} {\bibfnamefont {R.~J.}\ \bibnamefont {Glauber}},\ }\href {\doibase 10.1103/PhysRev.140.B676} {\bibfield  {journal} {\bibinfo  {journal} {Phys. Rev.}\ }\textbf {\bibinfo {volume} {140}},\ \bibinfo {pages} {B676} (\bibinfo {year} {1965})}\BibitemShut {NoStop}%
\bibitem [{\citenamefont {Stoler}(1971)}]{PhysRevD.4.2309}%
  \BibitemOpen
  \bibfield  {author} {\bibinfo {author} {\bibfnamefont {D.}~\bibnamefont {Stoler}},\ }\href {\doibase 10.1103/PhysRevD.4.2309} {\bibfield  {journal} {\bibinfo  {journal} {Phys. Rev. D}\ }\textbf {\bibinfo {volume} {4}},\ \bibinfo {pages} {2309} (\bibinfo {year} {1971})}\BibitemShut {NoStop}%
\bibitem [{\citenamefont {Carranza}\ and\ \citenamefont {Gerry}(2012)}]{Carranza:12}%
  \BibitemOpen
  \bibfield  {author} {\bibinfo {author} {\bibfnamefont {R.}~\bibnamefont {Carranza}}\ and\ \bibinfo {author} {\bibfnamefont {C.~C.}\ \bibnamefont {Gerry}},\ }\href {\doibase 10.1364/JOSAB.29.002581} {\bibfield  {journal} {\bibinfo  {journal} {J. Opt. Soc. Am. B}\ }\textbf {\bibinfo {volume} {29}},\ \bibinfo {pages} {2581} (\bibinfo {year} {2012})}\BibitemShut {NoStop}%
\bibitem [{\citenamefont {Andersen}\ \emph {et~al.}(2015)\citenamefont {Andersen}, \citenamefont {Gehring}, \citenamefont {Marquardt},\ and\ \citenamefont {Leuchs}}]{Andersen201530YO}%
  \BibitemOpen
  \bibfield  {author} {\bibinfo {author} {\bibfnamefont {U.~L.}\ \bibnamefont {Andersen}}, \bibinfo {author} {\bibfnamefont {T.}~\bibnamefont {Gehring}}, \bibinfo {author} {\bibfnamefont {C.}~\bibnamefont {Marquardt}}, \ and\ \bibinfo {author} {\bibfnamefont {G.}~\bibnamefont {Leuchs}},\ }\href {https://api.semanticscholar.org/CorpusID:118439273} {\bibfield  {journal} {\bibinfo  {journal} {Phys. Scr}\ }\textbf {\bibinfo {volume} {91}} (\bibinfo {year} {2015})}\BibitemShut {NoStop}%
\bibitem [{\citenamefont {Breitenbach}\ \emph {et~al.}(1997)\citenamefont {Breitenbach}, \citenamefont {Schiller},\ and\ \citenamefont {Mlynek}}]{RN1932}%
  \BibitemOpen
  \bibfield  {author} {\bibinfo {author} {\bibfnamefont {G.}~\bibnamefont {Breitenbach}}, \bibinfo {author} {\bibfnamefont {S.}~\bibnamefont {Schiller}}, \ and\ \bibinfo {author} {\bibfnamefont {J.}~\bibnamefont {Mlynek}},\ }\href {\doibase 10.1038/387471a0} {\bibfield  {journal} {\bibinfo  {journal} {Nature}\ }\textbf {\bibinfo {volume} {387}},\ \bibinfo {pages} {471} (\bibinfo {year} {1997})}\BibitemShut {NoStop}%
\bibitem [{\citenamefont {Hong}\ and\ \citenamefont {Mandel}(1986)}]{PhysRevLett.56.58}%
  \BibitemOpen
  \bibfield  {author} {\bibinfo {author} {\bibfnamefont {C.~K.}\ \bibnamefont {Hong}}\ and\ \bibinfo {author} {\bibfnamefont {L.}~\bibnamefont {Mandel}},\ }\href {\doibase 10.1103/PhysRevLett.56.58} {\bibfield  {journal} {\bibinfo  {journal} {Phys. Rev. Lett.}\ }\textbf {\bibinfo {volume} {56}},\ \bibinfo {pages} {58} (\bibinfo {year} {1986})}\BibitemShut {NoStop}%
\bibitem [{\citenamefont {Krause}\ \emph {et~al.}(1987)\citenamefont {Krause}, \citenamefont {Scully},\ and\ \citenamefont {Walther}}]{PhysRevA.36.4547}%
  \BibitemOpen
  \bibfield  {author} {\bibinfo {author} {\bibfnamefont {J.}~\bibnamefont {Krause}}, \bibinfo {author} {\bibfnamefont {M.~O.}\ \bibnamefont {Scully}}, \ and\ \bibinfo {author} {\bibfnamefont {H.}~\bibnamefont {Walther}},\ }\href {\doibase 10.1103/PhysRevA.36.4547} {\bibfield  {journal} {\bibinfo  {journal} {Phys. Rev. A}\ }\textbf {\bibinfo {volume} {36}},\ \bibinfo {pages} {4547} (\bibinfo {year} {1987})}\BibitemShut {NoStop}%
\bibitem [{\citenamefont {Cummings}\ and\ \citenamefont {Rajagopal}(1989)}]{PhysRevA.39.3414}%
  \BibitemOpen
  \bibfield  {author} {\bibinfo {author} {\bibfnamefont {F.~W.}\ \bibnamefont {Cummings}}\ and\ \bibinfo {author} {\bibfnamefont {A.~K.}\ \bibnamefont {Rajagopal}},\ }\href {\doibase 10.1103/PhysRevA.39.3414} {\bibfield  {journal} {\bibinfo  {journal} {Phys. Rev. A}\ }\textbf {\bibinfo {volume} {39}},\ \bibinfo {pages} {3414} (\bibinfo {year} {1989})}\BibitemShut {NoStop}%
\bibitem [{\citenamefont {Varcoe}\ \emph {et~al.}(2000)\citenamefont {Varcoe}, \citenamefont {Brattke}, \citenamefont {Weidinger},\ and\ \citenamefont {Walther}}]{RN1930}%
  \BibitemOpen
  \bibfield  {author} {\bibinfo {author} {\bibfnamefont {B.~T.~H.}\ \bibnamefont {Varcoe}}, \bibinfo {author} {\bibfnamefont {S.}~\bibnamefont {Brattke}}, \bibinfo {author} {\bibfnamefont {M.}~\bibnamefont {Weidinger}}, \ and\ \bibinfo {author} {\bibfnamefont {H.}~\bibnamefont {Walther}},\ }\href {\doibase 10.1038/35001526} {\bibfield  {journal} {\bibinfo  {journal} {Nature}\ }\textbf {\bibinfo {volume} {403}},\ \bibinfo {pages} {743} (\bibinfo {year} {2000})}\BibitemShut {NoStop}%
\bibitem [{\citenamefont {xi~Liu}\ \emph {et~al.}(2004)\citenamefont {xi~Liu}, \citenamefont {Wei},\ and\ \citenamefont {Nori}}]{Liu_2004}%
  \BibitemOpen
  \bibfield  {author} {\bibinfo {author} {\bibfnamefont {Y.}~\bibnamefont {xi~Liu}}, \bibinfo {author} {\bibfnamefont {L.~F.}\ \bibnamefont {Wei}}, \ and\ \bibinfo {author} {\bibfnamefont {F.}~\bibnamefont {Nori}},\ }\href {\doibase 10.1209/epl/i2004-10144-3} {\bibfield  {journal} {\bibinfo  {journal} {Europhys. Lett ({EPL})}\ }\textbf {\bibinfo {volume} {67}},\ \bibinfo {pages} {941} (\bibinfo {year} {2004})}\BibitemShut {NoStop}%
\bibitem [{\citenamefont {Waks}\ \emph {et~al.}(2006)\citenamefont {Waks}, \citenamefont {Diamanti},\ and\ \citenamefont {Yamamoto}}]{Waks2006}%
  \BibitemOpen
  \bibfield  {author} {\bibinfo {author} {\bibfnamefont {E.}~\bibnamefont {Waks}}, \bibinfo {author} {\bibfnamefont {E.}~\bibnamefont {Diamanti}}, \ and\ \bibinfo {author} {\bibfnamefont {Y.}~\bibnamefont {Yamamoto}},\ }\href {\doibase 10.1088/1367-2630/8/1/004} {\bibfield  {journal} {\bibinfo  {journal} {New. J. Phys}\ }\textbf {\bibinfo {volume} {8}},\ \bibinfo {pages} {4} (\bibinfo {year} {2006})}\BibitemShut {NoStop}%
\bibitem [{\citenamefont {Houck}\ \emph {et~al.}(2007)\citenamefont {Houck}, \citenamefont {Schuster}, \citenamefont {Gambetta}, \citenamefont {Schreier}, \citenamefont {Johnson}, \citenamefont {Chow}, \citenamefont {Frunzio}, \citenamefont {Majer}, \citenamefont {Devoret}, \citenamefont {Girvin},\ and\ \citenamefont {Schoelkopf}}]{RN1931}%
  \BibitemOpen
  \bibfield  {author} {\bibinfo {author} {\bibfnamefont {A.~A.}\ \bibnamefont {Houck}}, \bibinfo {author} {\bibfnamefont {D.~I.}\ \bibnamefont {Schuster}}, \bibinfo {author} {\bibfnamefont {J.~M.}\ \bibnamefont {Gambetta}}, \bibinfo {author} {\bibfnamefont {J.~A.}\ \bibnamefont {Schreier}}, \bibinfo {author} {\bibfnamefont {B.~R.}\ \bibnamefont {Johnson}}, \bibinfo {author} {\bibfnamefont {J.~M.}\ \bibnamefont {Chow}}, \bibinfo {author} {\bibfnamefont {L.}~\bibnamefont {Frunzio}}, \bibinfo {author} {\bibfnamefont {J.}~\bibnamefont {Majer}}, \bibinfo {author} {\bibfnamefont {M.~H.}\ \bibnamefont {Devoret}}, \bibinfo {author} {\bibfnamefont {S.~M.}\ \bibnamefont {Girvin}}, \ and\ \bibinfo {author} {\bibfnamefont {R.~J.}\ \bibnamefont {Schoelkopf}},\ }\href {\doibase 10.1038/nature06126} {\bibfield  {journal} {\bibinfo  {journal} {Nature}\ }\textbf {\bibinfo {volume} {449}},\ \bibinfo {pages} {328} (\bibinfo {year} {2007})}\BibitemShut {NoStop}%
\bibitem [{\citenamefont {Monroe}\ \emph {et~al.}(1995)\citenamefont {Monroe}, \citenamefont {Meekhof}, \citenamefont {King}, \citenamefont {Jefferts}, \citenamefont {Itano}, \citenamefont {Wineland},\ and\ \citenamefont {Gould}}]{PhysRevLett.75.4011}%
  \BibitemOpen
  \bibfield  {author} {\bibinfo {author} {\bibfnamefont {C.}~\bibnamefont {Monroe}}, \bibinfo {author} {\bibfnamefont {D.~M.}\ \bibnamefont {Meekhof}}, \bibinfo {author} {\bibfnamefont {B.~E.}\ \bibnamefont {King}}, \bibinfo {author} {\bibfnamefont {S.~R.}\ \bibnamefont {Jefferts}}, \bibinfo {author} {\bibfnamefont {W.~M.}\ \bibnamefont {Itano}}, \bibinfo {author} {\bibfnamefont {D.~J.}\ \bibnamefont {Wineland}}, \ and\ \bibinfo {author} {\bibfnamefont {P.}~\bibnamefont {Gould}},\ }\href {\doibase 10.1103/PhysRevLett.75.4011} {\bibfield  {journal} {\bibinfo  {journal} {Phys. Rev. Lett.}\ }\textbf {\bibinfo {volume} {75}},\ \bibinfo {pages} {4011} (\bibinfo {year} {1995})}\BibitemShut {NoStop}%
\bibitem [{\citenamefont {Yuen}(1976)}]{PhysRevA.13.2226}%
  \BibitemOpen
  \bibfield  {author} {\bibinfo {author} {\bibfnamefont {H.~P.}\ \bibnamefont {Yuen}},\ }\href {\doibase 10.1103/PhysRevA.13.2226} {\bibfield  {journal} {\bibinfo  {journal} {Phys. Rev. A}\ }\textbf {\bibinfo {volume} {13}},\ \bibinfo {pages} {2226} (\bibinfo {year} {1976})}\BibitemShut {NoStop}%
\bibitem [{\citenamefont {Parigi}\ \emph {et~al.}(2007)\citenamefont {Parigi}, \citenamefont {Zavatta}, \citenamefont {Kim},\ and\ \citenamefont {Bellini}}]{doi:10.1126/science.1146204}%
  \BibitemOpen
  \bibfield  {author} {\bibinfo {author} {\bibfnamefont {V.}~\bibnamefont {Parigi}}, \bibinfo {author} {\bibfnamefont {A.}~\bibnamefont {Zavatta}}, \bibinfo {author} {\bibfnamefont {M.}~\bibnamefont {Kim}}, \ and\ \bibinfo {author} {\bibfnamefont {M.}~\bibnamefont {Bellini}},\ }\href {\doibase 10.1126/science.1146204} {\bibfield  {journal} {\bibinfo  {journal} {Science}\ }\textbf {\bibinfo {volume} {317}},\ \bibinfo {pages} {1890} (\bibinfo {year} {2007})}\BibitemShut {NoStop}%
\bibitem [{\citenamefont {{Ashfaq Ahmad}}\ \emph {et~al.}(2011)\citenamefont {{Ashfaq Ahmad}}, \citenamefont {{Hamad Bukhari}}, \citenamefont {{Naeem Khan}}, \citenamefont {{Ran}}, \citenamefont {{Liao}},\ and\ \citenamefont {{Liu}}}]{2011JMOp...58..890A}%
  \BibitemOpen
  \bibfield  {author} {\bibinfo {author} {\bibfnamefont {M.}~\bibnamefont {{Ashfaq Ahmad}}}, \bibinfo {author} {\bibfnamefont {S.}~\bibnamefont {{Hamad Bukhari}}}, \bibinfo {author} {\bibfnamefont {S.}~\bibnamefont {{Naeem Khan}}}, \bibinfo {author} {\bibfnamefont {Z.}~\bibnamefont {{Ran}}}, \bibinfo {author} {\bibfnamefont {Q.}~\bibnamefont {{Liao}}}, \ and\ \bibinfo {author} {\bibfnamefont {S.}~\bibnamefont {{Liu}}},\ }\href {\doibase 10.1080/09500340.2011.577915} {\bibfield  {journal} {\bibinfo  {journal} {J. Mod. Opt.}\ }\textbf {\bibinfo {volume} {58}},\ \bibinfo {pages} {890} (\bibinfo {year} {2011})}\BibitemShut {NoStop}%
\bibitem [{\citenamefont {Andersen}\ \emph {et~al.}(2016)\citenamefont {Andersen}, \citenamefont {Gehring}, \citenamefont {Marquardt},\ and\ \citenamefont {Leuchs}}]{Andersen_2016}%
  \BibitemOpen
  \bibfield  {author} {\bibinfo {author} {\bibfnamefont {U.~L.}\ \bibnamefont {Andersen}}, \bibinfo {author} {\bibfnamefont {T.}~\bibnamefont {Gehring}}, \bibinfo {author} {\bibfnamefont {C.}~\bibnamefont {Marquardt}}, \ and\ \bibinfo {author} {\bibfnamefont {G.}~\bibnamefont {Leuchs}},\ }\href {\doibase 10.1088/0031-8949/91/5/053001} {\bibfield  {journal} {\bibinfo  {journal} {Phys. Scr.}\ }\textbf {\bibinfo {volume} {91}},\ \bibinfo {pages} {053001} (\bibinfo {year} {2016})}\BibitemShut {NoStop}%
\bibitem [{\citenamefont {Agarwal}(2013)}]{Agarwal2013}%
  \BibitemOpen
  \bibfield  {author} {\bibinfo {author} {\bibfnamefont {G.}~\bibnamefont {Agarwal}},\ }\href@noop {} {\emph {\bibinfo {title} {Quantum Optics}}}\ (\bibinfo  {publisher} {Cambridge University Press, Cambridge, England},\ \bibinfo {year} {2013})\BibitemShut {NoStop}%
\bibitem [{\citenamefont {Biswas}\ and\ \citenamefont {Agarwal}(2007)}]{PhysRevA.75.032104}%
  \BibitemOpen
  \bibfield  {author} {\bibinfo {author} {\bibfnamefont {A.}~\bibnamefont {Biswas}}\ and\ \bibinfo {author} {\bibfnamefont {G.~S.}\ \bibnamefont {Agarwal}},\ }\href {\doibase 10.1103/PhysRevA.75.032104} {\bibfield  {journal} {\bibinfo  {journal} {Phys. Rev. A}\ }\textbf {\bibinfo {volume} {75}},\ \bibinfo {pages} {032104} (\bibinfo {year} {2007})}\BibitemShut {NoStop}%
\bibitem [{\citenamefont {Agarwal}\ and\ \citenamefont {Tara}(1991)}]{PhysRevA.43.492}%
  \BibitemOpen
  \bibfield  {author} {\bibinfo {author} {\bibfnamefont {G.~S.}\ \bibnamefont {Agarwal}}\ and\ \bibinfo {author} {\bibfnamefont {K.}~\bibnamefont {Tara}},\ }\href {\doibase 10.1103/PhysRevA.43.492} {\bibfield  {journal} {\bibinfo  {journal} {Phys. Rev. A}\ }\textbf {\bibinfo {volume} {43}},\ \bibinfo {pages} {492} (\bibinfo {year} {1991})}\BibitemShut {NoStop}%
\bibitem [{\citenamefont {Riabinin}\ \emph {et~al.}(2021)\citenamefont {Riabinin}, \citenamefont {Sharapova}, \citenamefont {Bartley},\ and\ \citenamefont {Meier}}]{Riabinin_2021}%
  \BibitemOpen
  \bibfield  {author} {\bibinfo {author} {\bibfnamefont {M.}~\bibnamefont {Riabinin}}, \bibinfo {author} {\bibfnamefont {P.~R.}\ \bibnamefont {Sharapova}}, \bibinfo {author} {\bibfnamefont {T.~J.}\ \bibnamefont {Bartley}}, \ and\ \bibinfo {author} {\bibfnamefont {T.}~\bibnamefont {Meier}},\ }\href {\doibase 10.1088/2399-6528/abeec2} {\bibfield  {journal} {\bibinfo  {journal} {J. Phys. Commun}\ }\textbf {\bibinfo {volume} {5}},\ \bibinfo {pages} {045002} (\bibinfo {year} {2021})}\BibitemShut {NoStop}%
\bibitem [{\citenamefont {Dao-ming}(2015)}]{Daoming2015QuantumPO}%
  \BibitemOpen
  \bibfield  {author} {\bibinfo {author} {\bibfnamefont {L.}~\bibnamefont {Dao-ming}},\ }\href {https://api.semanticscholar.org/CorpusID:120006318} {\bibfield  {journal} {\bibinfo  {journal} {Int. J. Theor. Phys}\ }\textbf {\bibinfo {volume} {54}},\ \bibinfo {pages} {2289} (\bibinfo {year} {2015})}\BibitemShut {NoStop}%
\bibitem [{\citenamefont {Akhtar}\ \emph {et~al.}(2023)\citenamefont {Akhtar}, \citenamefont {Wu}, \citenamefont {Peng}, \citenamefont {Liu},\ and\ \citenamefont {Xianlong}}]{PhysRevA.107.052614}%
  \BibitemOpen
  \bibfield  {author} {\bibinfo {author} {\bibfnamefont {N.}~\bibnamefont {Akhtar}}, \bibinfo {author} {\bibfnamefont {J.}~\bibnamefont {Wu}}, \bibinfo {author} {\bibfnamefont {J.-X.}\ \bibnamefont {Peng}}, \bibinfo {author} {\bibfnamefont {W.-M.}\ \bibnamefont {Liu}}, \ and\ \bibinfo {author} {\bibfnamefont {G.}~\bibnamefont {Xianlong}},\ }\href {\doibase 10.1103/PhysRevA.107.052614} {\bibfield  {journal} {\bibinfo  {journal} {Phys. Rev. A}\ }\textbf {\bibinfo {volume} {107}},\ \bibinfo {pages} {052614} (\bibinfo {year} {2023})}\BibitemShut {NoStop}%
\bibitem [{\citenamefont {Park}\ \emph {et~al.}(2024)\citenamefont {Park}, \citenamefont {Stokowski}, \citenamefont {Ansari}, \citenamefont {Gyger}, \citenamefont {Multani}, \citenamefont {Celik}, \citenamefont {Hwang}, \citenamefont {Dean}, \citenamefont {Mayor}, \citenamefont {McKenna}, \citenamefont {Fejer},\ and\ \citenamefont {Safavi-Naeini}}]{doi:10.1126/sciadv.adl1814}%
  \BibitemOpen
  \bibfield  {author} {\bibinfo {author} {\bibfnamefont {T.}~\bibnamefont {Park}}, \bibinfo {author} {\bibfnamefont {H.}~\bibnamefont {Stokowski}}, \bibinfo {author} {\bibfnamefont {V.}~\bibnamefont {Ansari}}, \bibinfo {author} {\bibfnamefont {S.}~\bibnamefont {Gyger}}, \bibinfo {author} {\bibfnamefont {K.~K.~S.}\ \bibnamefont {Multani}}, \bibinfo {author} {\bibfnamefont {O.~T.}\ \bibnamefont {Celik}}, \bibinfo {author} {\bibfnamefont {A.~Y.}\ \bibnamefont {Hwang}}, \bibinfo {author} {\bibfnamefont {D.~J.}\ \bibnamefont {Dean}}, \bibinfo {author} {\bibfnamefont {F.}~\bibnamefont {Mayor}}, \bibinfo {author} {\bibfnamefont {T.~P.}\ \bibnamefont {McKenna}}, \bibinfo {author} {\bibfnamefont {M.~M.}\ \bibnamefont {Fejer}}, \ and\ \bibinfo {author} {\bibfnamefont {A.}~\bibnamefont {Safavi-Naeini}},\ }\href {\doibase 10.1126/sciadv.adl1814} {\bibfield  {journal} {\bibinfo  {journal} {Sci. Adv.}\ }\textbf {\bibinfo {volume} {10}},\ \bibinfo {pages} {eadl1814} (\bibinfo {year} {2024})}\BibitemShut {NoStop}%
\bibitem [{\citenamefont {Young}\ and\ \citenamefont {Soh}(2025)}]{PhysRevResearch.7.013130}%
  \BibitemOpen
  \bibfield  {author} {\bibinfo {author} {\bibfnamefont {S.~M.}\ \bibnamefont {Young}}\ and\ \bibinfo {author} {\bibfnamefont {D.}~\bibnamefont {Soh}},\ }\href {\doibase 10.1103/PhysRevResearch.7.013130} {\bibfield  {journal} {\bibinfo  {journal} {Phys. Rev. Res.}\ }\textbf {\bibinfo {volume} {7}},\ \bibinfo {pages} {013130} (\bibinfo {year} {2025})}\BibitemShut {NoStop}%
\bibitem [{\citenamefont {Xu}\ and\ \citenamefont {Yuan}(2020)}]{xu2020conditional}%
  \BibitemOpen
  \bibfield  {author} {\bibinfo {author} {\bibfnamefont {X.-x.}\ \bibnamefont {Xu}}\ and\ \bibinfo {author} {\bibfnamefont {H.-c.}\ \bibnamefont {Yuan}},\ }\href {\doibase 10.1007/s11128-020-02823-1} {\bibfield  {journal} {\bibinfo  {journal} {Quantum Inf. Process}\ }\textbf {\bibinfo {volume} {19}},\ \bibinfo {pages} {1} (\bibinfo {year} {2020})}\BibitemShut {NoStop}%
\bibitem [{\citenamefont {Liu}\ \emph {et~al.}(2017)\citenamefont {Liu}, \citenamefont {Ye}, \citenamefont {Zhou}, \citenamefont {Zhang}, \citenamefont {Huang},\ and\ \citenamefont {Hu}}]{liu2017entanglement}%
  \BibitemOpen
  \bibfield  {author} {\bibinfo {author} {\bibfnamefont {C.-J.}\ \bibnamefont {Liu}}, \bibinfo {author} {\bibfnamefont {W.}~\bibnamefont {Ye}}, \bibinfo {author} {\bibfnamefont {W.-D.}\ \bibnamefont {Zhou}}, \bibinfo {author} {\bibfnamefont {H.-L.}\ \bibnamefont {Zhang}}, \bibinfo {author} {\bibfnamefont {J.-H.}\ \bibnamefont {Huang}}, \ and\ \bibinfo {author} {\bibfnamefont {L.-Y.}\ \bibnamefont {Hu}},\ }\href {\doibase 10.1007/s11467-017-0694-6} {\bibfield  {journal} {\bibinfo  {journal} {Frontiers of Physics}\ }\textbf {\bibinfo {volume} {12}},\ \bibinfo {pages} {1} (\bibinfo {year} {2017})}\BibitemShut {NoStop}%
\bibitem [{\citenamefont {Ho}(2024)}]{PhysRevResearch.6.033292}%
  \BibitemOpen
  \bibfield  {author} {\bibinfo {author} {\bibfnamefont {L.~B.}\ \bibnamefont {Ho}},\ }\href {\doibase 10.1103/PhysRevResearch.6.033292} {\bibfield  {journal} {\bibinfo  {journal} {Phys. Rev. Res.}\ }\textbf {\bibinfo {volume} {6}},\ \bibinfo {pages} {033292} (\bibinfo {year} {2024})}\BibitemShut {NoStop}%
\bibitem [{\citenamefont {Gao}\ \emph {et~al.}(2024)\citenamefont {Gao}, \citenamefont {Zheng}, \citenamefont {Lu}, \citenamefont {Shi}, \citenamefont {Tian},\ and\ \citenamefont {Zheng}}]{gao2024generation}%
  \BibitemOpen
  \bibfield  {author} {\bibinfo {author} {\bibfnamefont {L.}~\bibnamefont {Gao}}, \bibinfo {author} {\bibfnamefont {L.-a.}\ \bibnamefont {Zheng}}, \bibinfo {author} {\bibfnamefont {B.}~\bibnamefont {Lu}}, \bibinfo {author} {\bibfnamefont {S.}~\bibnamefont {Shi}}, \bibinfo {author} {\bibfnamefont {L.}~\bibnamefont {Tian}}, \ and\ \bibinfo {author} {\bibfnamefont {Y.}~\bibnamefont {Zheng}},\ }\href {\doibase 10.1038/s41377-024-01606-y} {\bibfield  {journal} {\bibinfo  {journal} {LIGHT-SCI. APPL.}\ }\textbf {\bibinfo {volume} {13}},\ \bibinfo {pages} {294} (\bibinfo {year} {2024})}\BibitemShut {NoStop}%
\bibitem [{\citenamefont {Oelker}\ \emph {et~al.}(2014)\citenamefont {Oelker}, \citenamefont {Barsotti}, \citenamefont {Dwyer}, \citenamefont {Sigg},\ and\ \citenamefont {Mavalvala}}]{Oelker:14}%
  \BibitemOpen
  \bibfield  {author} {\bibinfo {author} {\bibfnamefont {E.}~\bibnamefont {Oelker}}, \bibinfo {author} {\bibfnamefont {L.}~\bibnamefont {Barsotti}}, \bibinfo {author} {\bibfnamefont {S.}~\bibnamefont {Dwyer}}, \bibinfo {author} {\bibfnamefont {D.}~\bibnamefont {Sigg}}, \ and\ \bibinfo {author} {\bibfnamefont {N.}~\bibnamefont {Mavalvala}},\ }\href {\doibase 10.1364/OE.22.021106} {\bibfield  {journal} {\bibinfo  {journal} {Opt. Express}\ }\textbf {\bibinfo {volume} {22}},\ \bibinfo {pages} {21106} (\bibinfo {year} {2014})}\BibitemShut {NoStop}%
\bibitem [{\citenamefont {Braunstein}\ and\ \citenamefont {Kimble}(1998)}]{PhysRevLett.80.869}%
  \BibitemOpen
  \bibfield  {author} {\bibinfo {author} {\bibfnamefont {S.~L.}\ \bibnamefont {Braunstein}}\ and\ \bibinfo {author} {\bibfnamefont {H.~J.}\ \bibnamefont {Kimble}},\ }\href {\doibase 10.1103/PhysRevLett.80.869} {\bibfield  {journal} {\bibinfo  {journal} {Phys. Rev. Lett.}\ }\textbf {\bibinfo {volume} {80}},\ \bibinfo {pages} {869} (\bibinfo {year} {1998})}\BibitemShut {NoStop}%
\bibitem [{\citenamefont {Yamamoto}\ and\ \citenamefont {Haus}(1986)}]{RevModPhys.58.1001}%
  \BibitemOpen
  \bibfield  {author} {\bibinfo {author} {\bibfnamefont {Y.}~\bibnamefont {Yamamoto}}\ and\ \bibinfo {author} {\bibfnamefont {H.~A.}\ \bibnamefont {Haus}},\ }\href {\doibase 10.1103/RevModPhys.58.1001} {\bibfield  {journal} {\bibinfo  {journal} {Rev. Mod. Phys.}\ }\textbf {\bibinfo {volume} {58}},\ \bibinfo {pages} {1001} (\bibinfo {year} {1986})}\BibitemShut {NoStop}%
\bibitem [{\citenamefont {Yuen}\ and\ \citenamefont {Shapiro}(1980)}]{1056132}%
  \BibitemOpen
  \bibfield  {author} {\bibinfo {author} {\bibfnamefont {H.}~\bibnamefont {Yuen}}\ and\ \bibinfo {author} {\bibfnamefont {J.}~\bibnamefont {Shapiro}},\ }\href {\doibase 10.1109/TIT.1980.1056132} {\bibfield  {journal} {\bibinfo  {journal} {IEEE Transactions on Information Theory}\ }\textbf {\bibinfo {volume} {26}},\ \bibinfo {pages} {78} (\bibinfo {year} {1980})}\BibitemShut {NoStop}%
\bibitem [{\citenamefont {Lo~Franco}\ \emph {et~al.}(2005)\citenamefont {Lo~Franco}, \citenamefont {Compagno}, \citenamefont {Messina},\ and\ \citenamefont {Napoli}}]{PhysRevA.72.053806}%
  \BibitemOpen
  \bibfield  {author} {\bibinfo {author} {\bibfnamefont {R.}~\bibnamefont {Lo~Franco}}, \bibinfo {author} {\bibfnamefont {G.}~\bibnamefont {Compagno}}, \bibinfo {author} {\bibfnamefont {A.}~\bibnamefont {Messina}}, \ and\ \bibinfo {author} {\bibfnamefont {A.}~\bibnamefont {Napoli}},\ }\href {\doibase 10.1103/PhysRevA.72.053806} {\bibfield  {journal} {\bibinfo  {journal} {Phys. Rev. A}\ }\textbf {\bibinfo {volume} {72}},\ \bibinfo {pages} {053806} (\bibinfo {year} {2005})}\BibitemShut {NoStop}%
\bibitem [{\citenamefont {Lo~Franco}\ \emph {et~al.}(2007)\citenamefont {Lo~Franco}, \citenamefont {Compagno}, \citenamefont {Messina},\ and\ \citenamefont {Napoli}}]{PhysRevA.76.011804}%
  \BibitemOpen
  \bibfield  {author} {\bibinfo {author} {\bibfnamefont {R.}~\bibnamefont {Lo~Franco}}, \bibinfo {author} {\bibfnamefont {G.}~\bibnamefont {Compagno}}, \bibinfo {author} {\bibfnamefont {A.}~\bibnamefont {Messina}}, \ and\ \bibinfo {author} {\bibfnamefont {A.}~\bibnamefont {Napoli}},\ }\href {\doibase 10.1103/PhysRevA.76.011804} {\bibfield  {journal} {\bibinfo  {journal} {Phys. Rev. A}\ }\textbf {\bibinfo {volume} {76}},\ \bibinfo {pages} {011804} (\bibinfo {year} {2007})}\BibitemShut {NoStop}%
\bibitem [{\citenamefont {Braunstein}\ and\ \citenamefont {Kimble}(2000)}]{PhysRevA.61.042302}%
  \BibitemOpen
  \bibfield  {author} {\bibinfo {author} {\bibfnamefont {S.~L.}\ \bibnamefont {Braunstein}}\ and\ \bibinfo {author} {\bibfnamefont {H.~J.}\ \bibnamefont {Kimble}},\ }\href {\doibase 10.1103/PhysRevA.61.042302} {\bibfield  {journal} {\bibinfo  {journal} {Phys. Rev. A}\ }\textbf {\bibinfo {volume} {61}},\ \bibinfo {pages} {042302} (\bibinfo {year} {2000})}\BibitemShut {NoStop}%
\bibitem [{\citenamefont {Petersen}\ \emph {et~al.}(2005)\citenamefont {Petersen}, \citenamefont {Madsen},\ and\ \citenamefont {M\o{}lmer}}]{PhysRevA.72.053812}%
  \BibitemOpen
  \bibfield  {author} {\bibinfo {author} {\bibfnamefont {V.}~\bibnamefont {Petersen}}, \bibinfo {author} {\bibfnamefont {L.~B.}\ \bibnamefont {Madsen}}, \ and\ \bibinfo {author} {\bibfnamefont {K.}~\bibnamefont {M\o{}lmer}},\ }\href {\doibase 10.1103/PhysRevA.72.053812} {\bibfield  {journal} {\bibinfo  {journal} {Phys. Rev. A}\ }\textbf {\bibinfo {volume} {72}},\ \bibinfo {pages} {053812} (\bibinfo {year} {2005})}\BibitemShut {NoStop}%
\bibitem [{\citenamefont {Kempe}(1999)}]{PhysRevA.60.910}%
  \BibitemOpen
  \bibfield  {author} {\bibinfo {author} {\bibfnamefont {J.}~\bibnamefont {Kempe}},\ }\href {\doibase 10.1103/PhysRevA.60.910} {\bibfield  {journal} {\bibinfo  {journal} {Phys. Rev. A}\ }\textbf {\bibinfo {volume} {60}},\ \bibinfo {pages} {910} (\bibinfo {year} {1999})}\BibitemShut {NoStop}%
\bibitem [{\citenamefont {Yuanbek}\ \emph {et~al.}(2024{\natexlab{a}})\citenamefont {Yuanbek}, \citenamefont {Islam}, \citenamefont {Abliz},\ and\ \citenamefont {Turek}}]{yuanbek2024single}%
  \BibitemOpen
  \bibfield  {author} {\bibinfo {author} {\bibfnamefont {J.}~\bibnamefont {Yuanbek}}, \bibinfo {author} {\bibfnamefont {A.}~\bibnamefont {Islam}}, \bibinfo {author} {\bibfnamefont {A.}~\bibnamefont {Abliz}}, \ and\ \bibinfo {author} {\bibfnamefont {Y.}~\bibnamefont {Turek}},\ }\href {\doibase 10.1140/epjp/s13360-024-05850-4} {\bibfield  {journal} {\bibinfo  {journal} {Eur. Phys. J. Plus}\ }\textbf {\bibinfo {volume} {139}},\ \bibinfo {pages} {1} (\bibinfo {year} {2024}{\natexlab{a}})}\BibitemShut {NoStop}%
\bibitem [{\citenamefont {von Neumann}(2018)}]{vonNeumann+2018}%
  \BibitemOpen
  \bibfield  {author} {\bibinfo {author} {\bibfnamefont {J.}~\bibnamefont {von Neumann}},\ }\href {\doibase doi:10.1515/9781400889921} {\emph {\bibinfo {title} {Mathematical Foundations of Quantum Mechanics}}},\ edited by\ \bibinfo {editor} {\bibfnamefont {N.~A.}\ \bibnamefont {Wheeler}}\ (\bibinfo  {publisher} {Princeton University Press},\ \bibinfo {address} {Princeton},\ \bibinfo {year} {2018})\BibitemShut {NoStop}%
\bibitem [{\citenamefont {Hong-yi}(1990)}]{PhysRevA.41.1526}%
  \BibitemOpen
  \bibfield  {author} {\bibinfo {author} {\bibfnamefont {F.}~\bibnamefont {Hong-yi}},\ }\href {\doibase 10.1103/PhysRevA.41.1526} {\bibfield  {journal} {\bibinfo  {journal} {Phys. Rev. A}\ }\textbf {\bibinfo {volume} {41}},\ \bibinfo {pages} {1526} (\bibinfo {year} {1990})}\BibitemShut {NoStop}%
\bibitem [{\citenamefont {Jozsa}(2007)}]{PhysRevA.76.044103}%
  \BibitemOpen
  \bibfield  {author} {\bibinfo {author} {\bibfnamefont {R.}~\bibnamefont {Jozsa}},\ }\href {\doibase 10.1103/PhysRevA.76.044103} {\bibfield  {journal} {\bibinfo  {journal} {Phys. Rev. A}\ }\textbf {\bibinfo {volume} {76}},\ \bibinfo {pages} {044103} (\bibinfo {year} {2007})}\BibitemShut {NoStop}%
\bibitem [{\citenamefont {Gerry}\ and\ \citenamefont {Knight}(2004)}]{Int}%
  \BibitemOpen
  \bibfield  {author} {\bibinfo {author} {\bibfnamefont {C.}~\bibnamefont {Gerry}}\ and\ \bibinfo {author} {\bibfnamefont {P.}~\bibnamefont {Knight}},\ }\href@noop {} {\emph {\bibinfo {title} {Introductory Quantum Optics}}}\ (\bibinfo  {publisher} {Cambridge Universirty Press, Cambridge, England},\ \bibinfo {year} {2004})\BibitemShut {NoStop}%
\bibitem [{\citenamefont {Araya-Sossa}\ and\ \citenamefont {Orszag}(2021)}]{PhysRevA.103.052215}%
  \BibitemOpen
  \bibfield  {author} {\bibinfo {author} {\bibfnamefont {K.}~\bibnamefont {Araya-Sossa}}\ and\ \bibinfo {author} {\bibfnamefont {M.}~\bibnamefont {Orszag}},\ }\href {\doibase 10.1103/PhysRevA.103.052215} {\bibfield  {journal} {\bibinfo  {journal} {Phys. Rev. A}\ }\textbf {\bibinfo {volume} {103}},\ \bibinfo {pages} {052215} (\bibinfo {year} {2021})}\BibitemShut {NoStop}%
\bibitem [{\citenamefont {Turek}\ \emph {et~al.}(2023{\natexlab{b}})\citenamefont {Turek}, \citenamefont {Islam},\ and\ \citenamefont {Abliz}}]{turek2023single}%
  \BibitemOpen
  \bibfield  {author} {\bibinfo {author} {\bibfnamefont {Y.}~\bibnamefont {Turek}}, \bibinfo {author} {\bibfnamefont {A.}~\bibnamefont {Islam}}, \ and\ \bibinfo {author} {\bibfnamefont {A.}~\bibnamefont {Abliz}},\ }\href {\doibase 10.1140/epjp/s13360-023-03659-1} {\bibfield  {journal} {\bibinfo  {journal} {Eur. Phys. J. Plus}\ }\textbf {\bibinfo {volume} {138}},\ \bibinfo {pages} {1} (\bibinfo {year} {2023}{\natexlab{b}})}\BibitemShut {NoStop}%
\bibitem [{\citenamefont {Zhu}\ \emph {et~al.}(2025)\citenamefont {Zhu}, \citenamefont {Ye}, \citenamefont {Wang}, \citenamefont {Liu}, \citenamefont {Jiang}, \citenamefont {Wang}, \citenamefont {Wu},\ and\ \citenamefont {Zhang}}]{10.1063/5.0230512}%
  \BibitemOpen
  \bibfield  {author} {\bibinfo {author} {\bibfnamefont {J.}~\bibnamefont {Zhu}}, \bibinfo {author} {\bibfnamefont {L.}~\bibnamefont {Ye}}, \bibinfo {author} {\bibfnamefont {Y.}~\bibnamefont {Wang}}, \bibinfo {author} {\bibfnamefont {Y.}~\bibnamefont {Liu}}, \bibinfo {author} {\bibfnamefont {Y.}~\bibnamefont {Jiang}}, \bibinfo {author} {\bibfnamefont {A.}~\bibnamefont {Wang}}, \bibinfo {author} {\bibfnamefont {J.}~\bibnamefont {Wu}}, \ and\ \bibinfo {author} {\bibfnamefont {Z.}~\bibnamefont {Zhang}},\ }\href {\doibase 10.1063/5.0230512} {\bibfield  {journal} {\bibinfo  {journal} {Appl. Phys. Rev.}\ }\textbf {\bibinfo {volume} {12}},\ \bibinfo {pages} {021315} (\bibinfo {year} {2025})}\BibitemShut {NoStop}%
\bibitem [{\citenamefont {Aharonov}\ \emph {et~al.}(1964)\citenamefont {Aharonov}, \citenamefont {Bergmann},\ and\ \citenamefont {Lebowitz}}]{PhysRev.134.B1410}%
  \BibitemOpen
  \bibfield  {author} {\bibinfo {author} {\bibfnamefont {Y.}~\bibnamefont {Aharonov}}, \bibinfo {author} {\bibfnamefont {P.~G.}\ \bibnamefont {Bergmann}}, \ and\ \bibinfo {author} {\bibfnamefont {J.~L.}\ \bibnamefont {Lebowitz}},\ }\href {\doibase 10.1103/PhysRev.134.B1410} {\bibfield  {journal} {\bibinfo  {journal} {Phys. Rev.}\ }\textbf {\bibinfo {volume} {134}},\ \bibinfo {pages} {B1410} (\bibinfo {year} {1964})}\BibitemShut {NoStop}%
\bibitem [{\citenamefont {Turek}\ \emph {et~al.}(2023{\natexlab{c}})\citenamefont {Turek}, \citenamefont {Aishan},\ and\ \citenamefont {Islam}}]{Turek_2023}%
  \BibitemOpen
  \bibfield  {author} {\bibinfo {author} {\bibfnamefont {Y.}~\bibnamefont {Turek}}, \bibinfo {author} {\bibfnamefont {N.}~\bibnamefont {Aishan}}, \ and\ \bibinfo {author} {\bibfnamefont {A.}~\bibnamefont {Islam}},\ }\href {\doibase 10.1088/1402-4896/acdcca} {\bibfield  {journal} {\bibinfo  {journal} {Phys. Scr}\ }\textbf {\bibinfo {volume} {98}},\ \bibinfo {pages} {075103} (\bibinfo {year} {2023}{\natexlab{c}})}\BibitemShut {NoStop}%
\bibitem [{\citenamefont {Wigner}\ and\ \citenamefont {Yanase}(1963)}]{doi:10.1073/pnas.49.6.910}%
  \BibitemOpen
  \bibfield  {author} {\bibinfo {author} {\bibfnamefont {E.~P.}\ \bibnamefont {Wigner}}\ and\ \bibinfo {author} {\bibfnamefont {M.~M.}\ \bibnamefont {Yanase}},\ }\href {\doibase 10.1073/pnas.49.6.910} {\bibfield  {journal} {\bibinfo  {journal} {Proc. Natl. Acad. Sci. USA}\ }\textbf {\bibinfo {volume} {49}},\ \bibinfo {pages} {910} (\bibinfo {year} {1963})}\BibitemShut {NoStop}%
\bibitem [{\citenamefont {Luo}\ and\ \citenamefont {Zhang}(2019)}]{PhysRevA.100.032116}%
  \BibitemOpen
  \bibfield  {author} {\bibinfo {author} {\bibfnamefont {S.}~\bibnamefont {Luo}}\ and\ \bibinfo {author} {\bibfnamefont {Y.}~\bibnamefont {Zhang}},\ }\href {\doibase 10.1103/PhysRevA.100.032116} {\bibfield  {journal} {\bibinfo  {journal} {Phys. Rev. A}\ }\textbf {\bibinfo {volume} {100}},\ \bibinfo {pages} {032116} (\bibinfo {year} {2019})}\BibitemShut {NoStop}%
\bibitem [{\citenamefont {Hong}\ and\ \citenamefont {Mandel}(1985)}]{PhysRevA.32.974}%
  \BibitemOpen
  \bibfield  {author} {\bibinfo {author} {\bibfnamefont {C.~K.}\ \bibnamefont {Hong}}\ and\ \bibinfo {author} {\bibfnamefont {L.}~\bibnamefont {Mandel}},\ }\href {\doibase 10.1103/PhysRevA.32.974} {\bibfield  {journal} {\bibinfo  {journal} {Phys. Rev. A}\ }\textbf {\bibinfo {volume} {32}},\ \bibinfo {pages} {974} (\bibinfo {year} {1985})}\BibitemShut {NoStop}%
\bibitem [{\citenamefont {Hillery}(1987{\natexlab{a}})}]{HILLERY1987135}%
  \BibitemOpen
  \bibfield  {author} {\bibinfo {author} {\bibfnamefont {M.}~\bibnamefont {Hillery}},\ }\href {\doibase https://doi.org/10.1016/0030-4018(87)90097-6} {\bibfield  {journal} {\bibinfo  {journal} {Opt. Commun}\ }\textbf {\bibinfo {volume} {62}},\ \bibinfo {pages} {135} (\bibinfo {year} {1987}{\natexlab{a}})}\BibitemShut {NoStop}%
\bibitem [{\citenamefont {Hillery}(1987{\natexlab{b}})}]{PhysRevA.36.3796}%
  \BibitemOpen
  \bibfield  {author} {\bibinfo {author} {\bibfnamefont {M.}~\bibnamefont {Hillery}},\ }\href {\doibase 10.1103/PhysRevA.36.3796} {\bibfield  {journal} {\bibinfo  {journal} {Phys. Rev. A}\ }\textbf {\bibinfo {volume} {36}},\ \bibinfo {pages} {3796} (\bibinfo {year} {1987}{\natexlab{b}})}\BibitemShut {NoStop}%
\bibitem [{\citenamefont {Hillery}(1989)}]{PhysRevA.40.3147}%
  \BibitemOpen
  \bibfield  {author} {\bibinfo {author} {\bibfnamefont {M.}~\bibnamefont {Hillery}},\ }\href {\doibase 10.1103/PhysRevA.40.3147} {\bibfield  {journal} {\bibinfo  {journal} {Phys. Rev. A}\ }\textbf {\bibinfo {volume} {40}},\ \bibinfo {pages} {3147} (\bibinfo {year} {1989})}\BibitemShut {NoStop}%
\bibitem [{\citenamefont {{Nguyen Ba An}}\ and\ \citenamefont {{Vo Tinh}}(1999)}]{NGUYENBAAN199934}%
  \BibitemOpen
  \bibfield  {author} {\bibinfo {author} {\bibnamefont {{Nguyen Ba An}}}\ and\ \bibinfo {author} {\bibnamefont {{Vo Tinh}}},\ }\href {\doibase https://doi.org/10.1016/S0375-9601(99)00602-7} {\bibfield  {journal} {\bibinfo  {journal} {Physics Letters A}\ }\textbf {\bibinfo {volume} {261}},\ \bibinfo {pages} {34} (\bibinfo {year} {1999})}\BibitemShut {NoStop}%
\bibitem [{\citenamefont {Kurochkin}\ \emph {et~al.}(2014)\citenamefont {Kurochkin}, \citenamefont {Prasad},\ and\ \citenamefont {Lvovsky}}]{PhysRevLett.112.070402}%
  \BibitemOpen
  \bibfield  {author} {\bibinfo {author} {\bibfnamefont {Y.}~\bibnamefont {Kurochkin}}, \bibinfo {author} {\bibfnamefont {A.~S.}\ \bibnamefont {Prasad}}, \ and\ \bibinfo {author} {\bibfnamefont {A.~I.}\ \bibnamefont {Lvovsky}},\ }\href {\doibase 10.1103/PhysRevLett.112.070402} {\bibfield  {journal} {\bibinfo  {journal} {Phys. Rev. Lett.}\ }\textbf {\bibinfo {volume} {112}},\ \bibinfo {pages} {070402} (\bibinfo {year} {2014})}\BibitemShut {NoStop}%
\bibitem [{\citenamefont {Ren}\ \emph {et~al.}(2019)\citenamefont {Ren}, \citenamefont {hai Zhang},\ and\ \citenamefont {jun Xu}}]{REN2019106}%
  \BibitemOpen
  \bibfield  {author} {\bibinfo {author} {\bibfnamefont {G.}~\bibnamefont {Ren}}, \bibinfo {author} {\bibfnamefont {W.}~\bibnamefont {hai Zhang}}, \ and\ \bibinfo {author} {\bibfnamefont {Y.}~\bibnamefont {jun Xu}},\ }\href {\doibase https://doi.org/10.1016/j.physa.2019.01.008} {\bibfield  {journal} {\bibinfo  {journal} {PHYSICA A.}\ }\textbf {\bibinfo {volume} {520}},\ \bibinfo {pages} {106} (\bibinfo {year} {2019})}\BibitemShut {NoStop}%
\bibitem [{\citenamefont {Loudon}(2000)}]{loudon2000quantum}%
  \BibitemOpen
  \bibfield  {author} {\bibinfo {author} {\bibfnamefont {R.}~\bibnamefont {Loudon}},\ }\href@noop {} {\emph {\bibinfo {title} {The quantum theory of light}}}\ (\bibinfo  {publisher} {OUP Oxford},\ \bibinfo {year} {2000})\BibitemShut {NoStop}%
\bibitem [{\citenamefont {Turek}(2020)}]{Turek_2020}%
  \BibitemOpen
  \bibfield  {author} {\bibinfo {author} {\bibfnamefont {Y.}~\bibnamefont {Turek}},\ }\href {\doibase 10.1088/1674-1056/ab9f23} {\bibfield  {journal} {\bibinfo  {journal} {Chin. Phys. B}\ }\textbf {\bibinfo {volume} {29}},\ \bibinfo {pages} {090302} (\bibinfo {year} {2020})}\BibitemShut {NoStop}%
\bibitem [{\citenamefont {Schleich}(1989)}]{Schleich1989}%
  \BibitemOpen
  \bibfield  {author} {\bibinfo {author} {\bibfnamefont {W.~P.}\ \bibnamefont {Schleich}},\ }\enquote {\bibinfo {title} {Phase space, correspondence principle and dynamical phases: Photon count probabilities of coherent and squeezed states via interfering areas in phase space},}\ in\ \href {\doibase 10.1007/978-1-4757-6574-8_10} {\emph {\bibinfo {booktitle} {Squeezed and Nonclassical Light}}},\ \bibinfo {editor} {edited by\ \bibinfo {editor} {\bibfnamefont {P.}~\bibnamefont {Tombesi}}\ and\ \bibinfo {editor} {\bibfnamefont {E.~R.}\ \bibnamefont {Pike}}}\ (\bibinfo  {publisher} {Springer US},\ \bibinfo {address} {Boston, MA},\ \bibinfo {year} {1989})\ pp.\ \bibinfo {pages} {129--149}\BibitemShut {NoStop}%
\bibitem [{\citenamefont {Yuanbek}\ \emph {et~al.}(2024{\natexlab{b}})\citenamefont {Yuanbek}, \citenamefont {Ren}, \citenamefont {Abliz},\ and\ \citenamefont {Turek}}]{PhysRevA.110.052611}%
  \BibitemOpen
  \bibfield  {author} {\bibinfo {author} {\bibfnamefont {J.}~\bibnamefont {Yuanbek}}, \bibinfo {author} {\bibfnamefont {Y.-F.}\ \bibnamefont {Ren}}, \bibinfo {author} {\bibfnamefont {A.}~\bibnamefont {Abliz}}, \ and\ \bibinfo {author} {\bibfnamefont {Y.}~\bibnamefont {Turek}},\ }\href {\doibase 10.1103/PhysRevA.110.052611} {\bibfield  {journal} {\bibinfo  {journal} {Phys. Rev. A}\ }\textbf {\bibinfo {volume} {110}},\ \bibinfo {pages} {052611} (\bibinfo {year} {2024}{\natexlab{b}})}\BibitemShut {NoStop}%
\bibitem [{\citenamefont {Fan}\ \emph {et~al.}(2025)\citenamefont {Fan}, \citenamefont {Luo}, \citenamefont {Guo}, \citenamefont {Wu}, \citenamefont {Zeng}, \citenamefont {Deng}, \citenamefont {Wang}, \citenamefont {Song}, \citenamefont {Wang}, \citenamefont {You} \emph {et~al.}}]{fan2025quantum}%
  \BibitemOpen
  \bibfield  {author} {\bibinfo {author} {\bibfnamefont {Y.-R.}\ \bibnamefont {Fan}}, \bibinfo {author} {\bibfnamefont {Y.}~\bibnamefont {Luo}}, \bibinfo {author} {\bibfnamefont {K.}~\bibnamefont {Guo}}, \bibinfo {author} {\bibfnamefont {J.-P.}\ \bibnamefont {Wu}}, \bibinfo {author} {\bibfnamefont {H.}~\bibnamefont {Zeng}}, \bibinfo {author} {\bibfnamefont {G.-W.}\ \bibnamefont {Deng}}, \bibinfo {author} {\bibfnamefont {Y.}~\bibnamefont {Wang}}, \bibinfo {author} {\bibfnamefont {H.-Z.}\ \bibnamefont {Song}}, \bibinfo {author} {\bibfnamefont {Z.}~\bibnamefont {Wang}}, \bibinfo {author} {\bibfnamefont {L.-X.}\ \bibnamefont {You}},  \emph {et~al.},\ }\href {\doibase 10.1038/s41377-025-01805-1} {\bibfield  {journal} {\bibinfo  {journal} {LIGHT-SCI. APPL.}\ }\textbf {\bibinfo {volume} {14}},\ \bibinfo {pages} {1} (\bibinfo {year} {2025})}\BibitemShut {NoStop}%
\bibitem [{\citenamefont {Braun}\ \emph {et~al.}(2014)\citenamefont {Braun}, \citenamefont {Jian}, \citenamefont {Pinel},\ and\ \citenamefont {Treps}}]{PhysRevA.90.013821}%
  \BibitemOpen
  \bibfield  {author} {\bibinfo {author} {\bibfnamefont {D.}~\bibnamefont {Braun}}, \bibinfo {author} {\bibfnamefont {P.}~\bibnamefont {Jian}}, \bibinfo {author} {\bibfnamefont {O.}~\bibnamefont {Pinel}}, \ and\ \bibinfo {author} {\bibfnamefont {N.}~\bibnamefont {Treps}},\ }\href {\doibase 10.1103/PhysRevA.90.013821} {\bibfield  {journal} {\bibinfo  {journal} {Phys. Rev. A}\ }\textbf {\bibinfo {volume} {90}},\ \bibinfo {pages} {013821} (\bibinfo {year} {2014})}\BibitemShut {NoStop}%
\bibitem [{\citenamefont {Schnabel}(2017)}]{SCHNABEL20171}%
  \BibitemOpen
  \bibfield  {author} {\bibinfo {author} {\bibfnamefont {R.}~\bibnamefont {Schnabel}},\ }\href {\doibase https://doi.org/10.1016/j.physrep.2017.04.001} {\bibfield  {journal} {\bibinfo  {journal} {Phys. Rep.}\ }\textbf {\bibinfo {volume} {684}},\ \bibinfo {pages} {1} (\bibinfo {year} {2017})}\BibitemShut {NoStop}%
\end{thebibliography}%

\end{document}